\documentstyle[12pt]{article}
\input tcilatex
\begin{document}
\title{ Superstrings in Higher Order Extensions
 of Finsler Superspaces}
\author{Sergiu I. Vacaru}
\date{}
\maketitle
\centerline{\it Institute of Applied Physics, Academy of Sciences,}
\centerline{\it 5 Academy str., Chi\c sin\v au 2028, Republic of Moldova}
\vskip7pt
\centerline{Fax: 011-3732-738149, E-mail: lises@cc.acad.md}

\begin{abstract}

The work proposes a general  background of the theory of field
interactions and strings in spaces with higher order anisotro\-py.  Our
approach proceeds by  developing the concept of higher order anisotropic
 superspace which unifies the logical and mathematical aspects of modern
Kaluza--Klein theories and generalized Lagrange and Finsler geometry and
 leads to modelling of physical processes on higher order fiber bundles
provided with nonlinear and distingushed connections and metric structures.
 The view adopted here is that a general field theory should incorporate
 all possible anisotropic and stochastic manifestations of classical and
 quantum interactions and, in consequence, a corresponding modification
 of basic principles and mathematical methods in formulation of physical
 theories.

The presentation is divided into two parts. The first five sections cover
the  higher order anisotropic superspaces.  We focus  on the geometry
 of distinguished by nonlinear connection  vector superbundles, consider
 different supersymmetric extensions of Finsler and Lagrange spaces and
 analyze the structure of basic geometric objects on such superspaces.
 The remaining five sections are devoted to the theory of higher order
 anisotropic superstrings. In the framework of supersymmetric nonlinear
sigma models in Finser extended backgrounds we prove that the low--energy
 dynamics of such strings contains motion equations for locally anisotropic
 field interactions.

 Our work is to be compared with important
 previous variants of extension of Finsler geometry and gravity
(see, for instance, \cite{asa,ma94,mat,bej}). There are substantial
 differences, because we rely on  modeling of higher order
 anisotropic interactions on superbundle spaces and do not propose
 some "exotic" Finsler models but a general approach which for trivial
 or corresponding parametization of
 nonlinear connection stuctures reduces to Kaluza--Klein and another
 variants of compactified higher--dimension space--times. The geometry
 of nonlinear connections (not being confused with connections for nonlinear
 realizations of gauge supergroups) is firstly considered for superspaces and
 possible cosequences on nonlinear connection field for compatible
 propagations  of strings in anisotropic backgrounds are analyzed.

Finally, we note that the developed computation methods are general
(in some line very similar to those for Einstein--Cartan--Weyl spaces which
 is a priority comparing with another combersome calculations in Finsler
 geometry) and admit extension to various Clifford and spinor bundles.

\copyright{\bf \ Sergiu I. Vacaru, 1996}

\end{abstract}
\tableofcontents

\section{Introduction}

The differential supergeometry has been formulated with the aim of getting
a geometric framework for the supersymmetric field theories (see the theory
of graded manifolds \cite{ber,lei,leim,kon}, the theory of supermanifolds
\cite{dew,rog80,bar,jad} and, for detailed considerations of geometric and
topological aspects of supermanifolds and formulation of superanalysis, \cite
{cia,bru,man,hoy,vla,vol}). In this work we apply the supergeometric
formalism for a study of a new class of (higher order anisotropic)
superspaces.

The concept of local anisotropy is largely used in some divisions of
theoretical and mathematical physics \cite{vlas,in,ish,mk} (see also
 possible applications in physics and biology in \cite{az94,am}). The first
models of locally anisotropic (la) spaces (la--spaces) have been proposed by
P.Finsler \cite{fin} and E.Cartan \cite{car35} (early approaches and modern
treatments of Finsler geometry and its extensions can be found, for
instance, in \cite{run,asa,asa88,mat}). In our works
\cite{vjmp,vb295,v295a,vlasg,vsp96,vg,vodg,voa} we try to formulate the
geometry of la--spaces in a manner as to include both variants of Finsler and
Lagrange, in general supersymmetric, extensions and higher dimensional
Kaluza--Klein (super)spaces as well to propose general principles and
methods of construction of models of classical and quantum field interactions
and stochastic processes on spaces with generic anisotropy.

We cite here the works \cite{bej90,bej91} by A. Bejancu where a new
viewpoint on differential geometry of supermanifolds is considered. The
author introduced the nonlinear connection (N--connection) structure and
developed a corresponding distinguished by N--connection supertensor
covariant differential calculus in the frame of De Witt \cite{dew} approach
to supermanifolds in the framework of the geometry of superbundles with
typical fibres parametrized by noncommutative coordinates. This was the
first example of superspace with local anisotropy. In our turn we have given
a general definition of locally anisotropic superspaces (la--superspaces)
\cite{vlasg}. It should be noted here that in our supersymmetric
generalizations we were inspired by the R.Miron, M. Anastasiei and Gh.
Atanasiu works on the geometry of nonlinear connections in vector bundles
and higher order La\-gran\-ge spaces \cite{ma87,ma94,mirata}.
In this work we shall formulate the theory of higher order vector and
tangent superbundles provided with nonlinear and distinguished connections
and metric structures (a generalized model of la--superspaces). Such
superbundles contain as particular cases the supersymmetric extensions and
various higher order prolongations of Riemann, Finsler and Lagrange spaces.

The superstring theory holds the greatest promise as the unification theory
of all fundamental interactions. The superstring models contains a lot a
characteristic features of Kaluza--Klein approaches, supersymmetry and
supergravity, local field theory and dual models. We note that in the string
theories the nonlocal one dimensional quantum objects (strings) mutually
interacting by linking and separating together are considered as fundamental
values. Perturbations of the quantized string are identified with quantum
particles. Symmetry and conservation laws in the string and superstring
theory can be considered as sweeping generalizations of gauge principles
which consists the basis of quantum field models. The new physical concepts
are formulated in the framework a ''new'' for physicists mathematical
formalism of the algebraic geometry and topology \cite{dhoker}.

The relationship between two dimensional $\sigma $-models and strings has
been considered \cite{lov,fts,cfmp,sen,alw} in order to discuss the
effective low energy field equations for the massless models of strings.
Nonlinear $\sigma $-models makes up a class of quantum field systems for
which the fields are also treated as coordinates of some manifolds.
Interactions are introduced in a geometric manner and admit a lot of
applications and generalizations in classical and quantum field and string
theories. The geometric structure of nonlinear sigma models manifests the
existence of topological nontrivial configuration, admits a geometric
interpretation of conterterms and points to a substantial interrelation
between extended supersymmetry and differential supergeometry. In connection
to this a new approach based on nonlocal, in general, higher order
anisotropic constructions seem to be emerging \cite{vstrod,vlags,vlagsr}. We
consider the reader to be familiar with basic results from supergeometry
(see, for instance, \cite{cia,leim,dew,rog80}), supergravity theories \cite
{gates1,sal,west,wes,wesz} and superstrings \cite{hul,sch,ket,ketn}.

In this work we shall present an introduction into the theory of higher
order anisotropic superstrings being a natural generalization to locally
anisotropic (la) backgrounds (we shall write in brief la-backgrounds,
la-spaces and la-geometry) of the Polyakov's covariant functional-integral
approach to string theory \cite{poly}. Our aim is to show that a
corresponding low-energy string dynamics contains the motion equations for
field equations on higher order anisotropic superspaces and to analyze the
geometry of the perturbation theory of the locally anisotropic
supersymmetric sigma models. We note that this work is devoted to
supersymmetric models of locally anisotropic superstrings.

The work is organized as follows: Section 2 contains a brief review on
supermanifolds and superbundles and an introduction into the geometry of
higher order distinguished vector superbundles. Section 3 deals with the
geometry of nonlinear and linear distinguished connections in vector
superbundles and distinguished vector superbundles. The geometry of the
total space of distinguished vector superbundles is studied in section 4;
distinguished connection and metric structures,  their torsions, curvatures
and structure equations are considered. Generalized Lagrange and Finsler
superspaces and thears higher order prolongations are  defined in section 5.
 Section 6 contains
an introduction into the geometry of two dimensional higher order
anisotropic sigma models and an locally anisotropic approach to heterotic
strings. In section 7 the background field method for $\sigma $-models is
generalized for a distinguished calculus locally adapted to the
N--connection structure in higher order anisotropic superspaces. Section 8
is devoted to a study of Green--Schwartz action in distinguished vector
superbundles. Fermi strings in higher order anisotropic spaces are considered
in section 9. An example of one--loop and two--loop calculus for anomalies
of locally anisotropic strings is presented in section 10. A discussion and
conclusions are drawn in section 11.\

\section{ Distinguished Superbundles}

In this section we establish the necessary terminology on supermanifolds
( s--manifolds ) \cite{dew,rog80,rog81,jad,vla,hoy,man,bar,bru,cia,hoy} and
present an introduction into the geometry of distinguished vector
superbundles (dvs--bundles) \cite{v96jpa1}. Here we note that a number of
different approaches to supermanifolds are broadly equivalent for local
considerations. For simplicity, we shall restrict our study only with
geometric constructions on locally trivial superspaces.

\subsection{Supermanifolds and superbundles}

To build up s--manifolds (see \cite{rog80,jad,vla}) one uses as basic
structures \ Grassmann algebra and Banach space. A Grassmann algebra is
introduced as a real associative algebra $\Lambda $ (with unity) possessing
a finite (canonical) set of anticommutative generators $\beta _{\hat A}$, ${{%
[{\beta _{\hat A}},{\beta _{\hat B}}]}_{+}}={{\beta _{\hat A}}}{{\beta
_{\hat C}}}+{{\beta _{\hat C}}}{{\beta _{\hat A}}}=0$, where ${{\hat A},{%
\hat B},...}=1,2,...,{\hat L}$. In this case it is also defined a ${Z_2}$%
-graded commutative algebra ${{\Lambda }_0}+{{\Lambda }_1}$, whose even part
${{\Lambda }_0}$ (odd part ${{\Lambda }_1}$) is a ${2^{{\hat L}-1}}$%
--dimensional real vector space of even (odd) products of generators ${\beta
}_{\hat A}$.After setting ${{\Lambda }_0}={\cal R}+{{\Lambda }_0}^{\prime }$%
, where ${\cal R}$ is the real number field and ${{\Lambda }_0}^{\prime }$
is the subspace of ${\Lambda }$ consisting of nilpotent elements, the
projections ${\sigma }:{\Lambda }\to {\cal R}$ and $s:{\Lambda }\to {{%
\Lambda }_0}^{\prime }$ are called, respectively, the body and soul maps.

A Grassmann algebra can be provided with both structures of a Banach algebra
and Euclidean topological space by the norm \cite{rog80}
$$
{\Vert }{\xi }{\Vert }={{\Sigma }_{{\hat A}_i}}{|}a^{{{\hat A}_1}...{{\hat A}%
_k}}{|},{\xi }={{\Sigma }_{r=0}^{\hat L}}a^{{{\hat A}_1}...{{\hat A}_r}}{{%
\beta }_{{\hat A}_1}}...{{\beta }_{{\hat A}_r}}.
$$
A superspace is introduced as a product
$$
{\Lambda }^{n,k}={\underbrace{{{\Lambda }_0}{\times }...{\times }{{\Lambda }%
_0}}_n{\times }{\underbrace{{{\Lambda }_1}{\times }...{\times }{{\Lambda }_1}%
}_k}}
$$
which is the $\Lambda $-envelope of a $Z_2$-graded vector space ${V^{n,k}}={%
V_0}{\otimes }{V_1}={{\cal R}^n}\oplus {{\cal R}^k}$ is obtained by
multiplication of even (odd) vectors of $V$ on even (odd) elements of ${%
\Lambda }$. The superspace (as the ${\Lambda }$-envelope) posses $(n+k)$
basis vectors $\{{\hat {{\beta }_i}},{\quad }i=0,1,...,n-1,$ and ${\quad }{{%
\beta }_{\hat i}},{\quad }{\hat i}=1,2,...k\}$. Coordinates of even (odd)
elements of $V^{n,k}$ are even (odd) elements of $\Lambda $. We can consider
equivalently a superspace $V^{n,k}$ as a $({2^{{\hat L}-1}})(n+k)$%
-dimensional real vector spaces with a basis $\{{{\hat \beta }_{i({\Lambda }%
)}},{{\beta }_{{\hat i}({\Lambda })}}\}$.

Functions of superspaces, differentiation with respect to Grassmann
coordinates, supersmooth (superanalytic) functions and mappings are
introduced by analogy with the ordinary case, but with a glance to certain
specificity caused by changing of real (or complex) number field into
Grassmann algebra $\Lambda $. Here we remark that functions on a superspace $%
{\Lambda }^{n,k}$ which takes values in Grassmann algebra can be considered
as mappings of the space ${\cal R}^{{({2^{({\hat L}-1)}})}{(n+k)}}$ into the
space ${\cal R}^{2{\hat L}}$. Functions differentiable on Grassmann
coordinates can be rewritten via derivatives on real coordinates, which obey
a generalized form of Cauchy-Riemann conditions.

A $(n,k)$-dimensional s--manifold $\tilde M$ can be defined as a Banach
manifold (see, for example, \cite{len}) modeled on ${\Lambda }^{n,k}$
endowed with an atlas ${\psi }={\{}{U_{(i)}},{{\psi }_{(i)}}:{U_{(i)}}\to {{%
\Lambda }^{n,k}},(i)\in J{\}}$ whose transition functions ${\psi }_{(i)}$
are supersmooth \cite{rog80,jad}. Instead of supersmooth functions we can
use $G^\infty $-functions \cite{rog80} and introduce $G^\infty $%
-supermanifolds ($G^\infty $ denotes the class of superdifferentiable
functions). The local structure of a $G^\infty $--supermanifold is built very
much as on a $C^\infty $--manifold. Just as a vector field on a $n$%
-dimensional $C^\infty $-manifold written locally as%
$$
{\Sigma }_{i=0}^{n-1}{\quad }{f_i}{(x^j)}{\frac{{\partial }}{{{\partial }x^i}%
}},
$$
where $f_i$ are $C^\infty $-functions, a vector field on an $(n,k)$%
--dimensional $G^\infty $--su\-per\-ma\-ni\-fold $\tilde M$ can be expressed
locally on an open region $U{\subset }\tilde M$ as
$$
{\Sigma }_{I=0}^{n-1+k}{\quad }{f_I}{(x^J)}{\frac{{\partial }}{{{\partial }%
x^I}}}= {\Sigma }_{i=0}^{n-1}{\quad }{f_i}{(x^j,{{\theta }^{\hat j}})}{\frac{%
{\partial }}{{{\partial }x^i}}}+{\Sigma }_{{\hat i}=1}^k{\quad }{f_{\hat i}}{%
(x^j,{{\theta }^{\hat j}})}{\frac \partial {\partial {{\theta }^{\hat i}}}},
$$
where $x=({\hat x},{\theta })=\{{x^I}=({{\hat x}^i},{\theta }^{\hat i})\}$
are local (even, odd) coordinates. We shall use indices $I=(i,{\hat i}),J=(j,%
{\hat j}),K=(k,{\hat k}),...$ for geometric objects on $\tilde M$. A vector
field on $U$ is an element $X{\subset }End[{G^\infty }(U)]$ (we can also
consider supersmooth functions instead of $G^\infty $-functions) such that
$$
X(fg)=(Xf)g+{(-)}^{{\mid }f{\mid }{\mid }X{\mid }}fXg,
$$
for all $f,g$ in $G^\infty (U)$, and
$$
X(af)={(-)}^{{\mid }X{\mid }{\mid }a{\mid }}aXf,
$$
where ${\mid }X{\mid }$ and ${\mid }a{\mid }$ denote correspondingly the
parity $(=0,1)$ of values $X$ and $a$ and for simplicity in this work we
shall write ${(-)}^{{\mid }f{\mid }{\mid }X{\mid }}$ instead of ${(-1)}^{{%
\mid }f{\mid }{\mid }X{\mid }}.$

A super Lie group (sl--group) \cite{rog81} is both an abstract group and a
s--manifold, provided that the group composition law fulfills a suitable
smoothness condition (i.e. to be superanalytic, for short, $sa$ \cite{jad}).

In our further considerations we shall use the group of automorphisms of ${%
\Lambda}^{(n,k)}$, denoted as $GL(n,k,{\Lambda})$, which can be parametrized
as the super Lie group of invertible matrices
$$
Q={\left(
\begin{array}{cc}
A & B \\
C & D
\end{array}
\right) } ,%
$$
where A and D are respectively $(n{\times}n)$ and $(k{\times}k)$ matrices
consisting of even Grassmann elements and B and C are rectangular matrices
consisting of odd Grassmann elements. A matrix Q is invertible as soon as
maps ${\sigma}A$ and ${\sigma}D$ are invertible matrices. A sl-group
represents an ordinary Lie group included in the group of linear transforms $%
GL(2^{{\hat L}-1}(n+k),{\cal R})$. For matrices of type Q one defines \cite
{ber,lei,leim} the superdeterminant, $sdetQ$, supertrace, $strQ$, and
superrank, $srankQ$.

A Lie superalgebra (sl--algebra) is a
$Z_2$-graded algebra $A={A_0}\oplus A_1$
endowed with product $[,\}$ satisfying the following properties:
$$
[I,I^{\prime }\}=-{(-)}^{{\mid }I{\mid }{\mid }I^{\prime }{\mid }}[I^{\prime
},I\},
$$
$$
[I,[I^{\prime },I^{\prime \prime }\}\}=[[I,I^{\prime }\},I^{\prime \prime
}\}+{(-)}^{{\mid }I{\mid }{\mid }I^{\prime }{\mid }}[I^{\prime }[I,I^{\prime
\prime }\}\},
$$
$I{\in }A_{{\mid }I{\mid }},{\quad }I^{\prime }{\in }A_{{\mid }I^{\prime }{%
\mid }}$, where ${\mid }I{\mid },{\mid }I^{\prime }{\mid }=0,1$ enumerates,
respectively, the possible parity of elements $I,I^{\prime }$. The even part
$A_0$ of a sl-algebra is a usual Lie algebra and the odd part $A_1$ is a
representation of this Lie algebra. This enables us to classify sl--algebras
following the Lie algebra classification \cite{kac}. We also point out that
irreducible linear representations of Lie superalgebra A are realized in $%
Z_2 $-graded vector spaces by matrices $\left(
\begin{array}{cc}
A & 0 \\
0 & D
\end{array}
\right) $ for even elements and $\left(
\begin{array}{cc}
0 & B \\
C & 0
\end{array}
\right) $ for odd elements and that, roughly speaking, A is a superalgebra
of generators of a sl--group.

A sl--module $W$ (graded Lie module) \cite{rog80} is introduced as a
$Z_2$--graded left $\Lambda $-module endowed with a product
 $[,\}$ which satisfies
the graded Jacobi identity and makes $W$ into a graded-anticommuta\-ti\-ve
Banach algebra over $\Lambda $. One calls the Lie module {\cal G} the set of
the left-invariant derivatives of a sl--group $G$.

The tangent superbundle (ts-bundle) $T\tilde M$ over a s-manifold $\tilde M$%
, ${\pi }:T\tilde M\to {\tilde M}$ is constructed in a usual manner (see,
for instance, \cite{len}) by taking as the typical fibre the superspace ${%
\Lambda }^{n,k}$ and as the structure group the group of automorphisms, i.e.
the sl-group $GL(n,k,{\Lambda }).$

Let us denote by ${\cal F}$ a vector superspace ( vs--space ) of dimension $%
(m,l)$ (with respect to a chosen base we parametrize an element $y\in {\cal E%
}$ as $y=({\hat y},\zeta )=\{{y^A}=({\hat {y^a}},{\zeta }^{\hat a})\}$,
where $a=1,2,...,m$ and ${\hat a}=1,2,...,l$). We shall use indices $A=(a,{%
\hat a}),B=(b,{\hat b}),...$ for objects on vs--spaces. A vector superbundle
( vs--bundle ) $\tilde {{\cal E}}$ over base $\tilde M$ with total superspace 
$\tilde E$, standard fibre ${\hat {{\cal F}}}$ and surjective projection ${{%
\pi }_E}:\tilde E{\to }\tilde M$ is defined (see details and variants in
\cite{bru,vol}) as in the case of ordinary manifolds (see, for instance,
\cite{len,ma87,ma94}). A section of $\tilde {{\cal E}}$ is a supersmooth map
$s:U{\to }\tilde E$ such that ${{\pi }_E}{\cdot }s=id_U.$

A subbundle of ${\tilde {{\cal E}}}$ is a triple $(\tilde {{\cal B}%
},f,f^{\prime})$, where $\tilde {{\cal B}}$ is a vs--bundle on $\tilde M$,
maps $f: \tilde {{\cal B}} {\to} \tilde {{\cal E}}$ and $f^{\prime}: \tilde M%
{\to} \tilde M$ are supersmooth, and $(i) {\quad}{{\pi}_E}{\circ}f=f^{\prime}%
{\circ}{{\pi}_B};$ $(ii) {\quad} f:{\pi}^{-1}_B {(x)} {\to} {\pi}^{-1}_E {%
\circ} f^{\prime}(x)$ is a vs--space homomorphism.
We denote by
$$
u=(x,y)=({\hat x},{\theta },{\hat y},{\zeta })=\{u^\alpha =(x^I,y^A)=({{\hat
x}^i},{\theta }^{\hat i},{{\hat y}^a},{\zeta }^{\hat a})=({{\hat x}^i}%
,x^{\hat i},{{\hat y}^a},y^{\hat a})\}
$$
the local coordinates in ${\tilde {{\cal E}}}$ and write their
transformations as
$$
x^{I^{\prime }}=x^{I^{\prime }}({x^I}),{\quad }srank({\frac{{\partial }%
x^{I^{\prime }}}{{\partial }x^I}})=(n,k),\eqno(1)
$$
$y^{A^{\prime }}=Y_A^{A^{\prime }}(x){y}^A,$ where $Y_A^{A^{\prime }}(x){\in
}G(m,l,\Lambda ).$

For local coordinates and geometric objects on ts-bundle $T\tilde M$ we
shall not distinguish indices of coordinates on the base and in the fibre
and write, for instance,
$$
u=(x,y)=({\hat x},{\theta },{\hat y},{\zeta })=\{u^\alpha =({x^I},{y^I})=({{%
\hat x}^i},{\theta }^{\hat i},{{\hat y}^i},{\zeta }^{\hat i})=({{\hat x}^i}%
,x^{\hat i},{{\hat y}^i},y^{\hat i})\}.
$$
We shall use Greek indices for marking local coordinates on both s--vector
and usual vector bundles.

\subsection{Distinguished vector superbundles}

Some recent considerations in mathematical physics are based on the
so--called k--jet spaces (see, for instance, \cite{sau,sar,asa89}). In order
to formulate a systematic theory of connections and of geometric structures
on k--jet bundles,
 in a manner following the approaches \cite{yan} and \cite
{ma87,ma94} R. Miron and Gh. Atanasiu \cite{mirata} introduced the concept
of k--osculator bundle for which a fiber of k-jets is changed into a
k--osculator fiber representing an element of k--order curve. Such
considerations are connected with geometric constructions on tangent bundles
of higher order. On the other hand for developments in modern supersymmetric
Kaluza--Klein theories (see, for instance, \cite{sal}) a substantial
interest would present a variant of ''osculator'' space for which the higher
order tangent s--space distributions are of different dimensions. The second
part of this section is devoted to the definition of such type distinguished
vector superbundle spaces.

A vector superspace ${\cal F}^{<z>}$ of dimension $(m,l)$ is a distinguished
vector superspace ( dvs--space ) if it is decomposed into an invariant 
oriented
direct sum ${\cal F}^{<z>}={\cal F}_{(1)}\oplus {\cal F}_{(2)}\oplus
...\oplus {\cal F}_{(z)}$ of vs--spaces ${\cal F}_{(p)},\dim {\cal F}%
_{(p)}=(m_{(p)},l_{(p)}),$ where $(p)=(1),(2),...,(z),%
\sum_{p=1}^{p=z}m_{(p)}=m,\sum_{p=1}^{p=z}l_{(p)}=l.$

Coordinates on ${\cal F}^{<p>}$ will be parametrized as
$$
{y}^{<p>}=(y_{(1)},y_{(2)},...,y_{(p)})=({\hat y}_{(1)},{\zeta }_{(1)},{\hat
y}_{(2)},{\zeta }_{(2)},...,{\hat y}_{(p)},{\zeta }_{(p)})=
$$
$$
\{y^{<A>}=({{\hat y}^{<a>}},{\zeta }^{<\hat a>})=({{\hat y}^{<a>}},y^{<\hat
a>})\},
$$
where bracketed indices are correspondingly split on ${\cal F}_{(p)}$%
--components:
$$
<A>=\left( A_{(1)},A_{(2)},...,A_{(p)}\right)
,<a>=(a_{(1)},a_{(2)},...,a_{(p)})
$$
$$
\mbox{ and }<\widehat{a}>=(\widehat{a}_{(1)}\widehat{a}_{(2)},...,\widehat{a}%
_{(p)}),\eqno(2)
$$
For simplicity, we shall also write (2) as $<A>=\left(
A_1,A_2,...,A_p\right) ,<a>=(a_1,a_2,...,a_p)$ and $<\widehat{a}>=(\widehat{a%
}_1\widehat{a}_2,...,\widehat{a}_p)$ if this will give not rise to
ambiguities.

A distinguished vector superbundle ( dvs--bundle )\\ $\widetilde{{\cal E}}%
^{<z>}=(\tilde E^{<z>},\pi ^{<d>},{\cal F}^{<d>},\tilde M),$ with surjective
projection $\pi ^{<z>}:\tilde E^{<z>}\rightarrow \tilde M,$ where $\tilde M$
and $\tilde E^{<z>}$ are respectively base and total s--spaces and the
dvs--space ${\cal F}^{<z>}$ is the standard fibre.

A dvs--bundle $\widetilde{{\cal E}}^{<z>}$ is constructed as an oriented set
of vs--bundles $\pi ^{<p>}:\tilde E^{<p>}\rightarrow \tilde E^{<p-1>}$ (with
typical fiber ${\cal F}^{<p>},p=1,2,...,z);$ $\tilde E^{<0>}=\tilde M.$ We
shall use index $z~(p)$ as to denote the total (intermediate) numbers of
consequent vs--bundle coverings of $\tilde M.$

Local coordinates on $\widetilde{{\cal E}}^{<p>}$ are denoted as%
$$
u_{(p)}=(x,y_{<p>})=(x,y_{(1)},y_{(2)},...,y_{(p)})=
$$
$$
({\hat x},{\theta },{\hat y}_{<p>},{\zeta }_{<p>})=({\hat x},{\theta },{\hat
y}_{(1)},{\zeta }_{(1)},{\hat y}_{(2)},{\zeta }_{(2)},...,{\hat y}_{(p)},{%
\zeta }_{(p)})=
$$
$$
\{u^{<\alpha >}=(x^I,y^{<A>})=({{\hat x}^i},{\theta }^{\hat i},{{\hat y}%
^{<a>}},{\zeta }^{<\hat a>})=({{\hat x}^i},x^{\hat i},{{\hat y}^{<a>}}%
,y^{<\hat a>}) =$$
$$(x^I = y^{A_{0}}, y^{A_{1}},..., y^{A_{p}},...,y^{A_{z}}) \}$$
(in our further considerations we shall consider different variants of
splitting of indices of geometric objects).

Instead of (1) the coordinate transforms for dvs--bundles\\ $\{u^{<\alpha
>}=(x^I,y^{<A>})\}\rightarrow \{u^{<\alpha ^{\prime }>}=(x^{I^{\prime
}},y^{<A^{\prime }>})\}$ are given by recurrent maps:%
$$
x^{I^{\prime }}=x^{I^{\prime }}({x^I}),{\quad }srank({\frac{{\partial }%
x^{I^{\prime }}}{{\partial }x^I}})=(n,k),\eqno(3)
$$
$$
y_{(1)}^{A_1^{\prime }}=K_{A_1}^{A_1^{\prime }}(x){y}%
_{(1)}^{A_1},K_{A_1}^{A_1^{\prime }}(x){\in }G(m_{(1)},l_{(1)},\Lambda ),
$$
$$
..................................................
$$
$$
y_{(p)}^{A_p^{\prime }}=K_{A_p}^{A_p^{\prime }}(u_{(p-1)}){y}%
_{(p)}^{A_p},K_{A_p}^{A_p^{\prime }}(u_{(p-1)}){\in }G(m_{(p)},l_{(p)},%
\Lambda ),
$$
$$
.................................................
$$
$$
y_{(z)}^{A_z^{\prime }}=K_{A_z}^{A_z^{\prime }}(u_{(z-1)}){y}%
_{(z)}^{A_z},K_{A_z}^{A_z^{\prime }}(u_{(z-1)}){\in }G(m_{(z)},l_{(z)},%
\Lambda ).
$$
In brief we write transforms (3) as
$$
x^{I^{\prime }}=x^{I^{\prime }}(x^I),~y^{<A^{\prime }>}=K_{<A>}^{<A^{\prime
}>}y^{<A>}.
$$
More generally, we shall consider matrices $K_{<\alpha >}^{<\alpha ^{\prime
}>}=(K_I^{I^{\prime }},K_{<A>}^{<A^{\prime }>}),$ where $K_I^{I^{\prime
}}\doteq \frac{\partial x^{I^{\prime }}}{\partial x^I}.$

In consequence the local coordinate bases of the module of ds--vector fields 
$\Xi (\widetilde{{\cal E}}^{<z>}),$
$$
\partial _{<\alpha >}=(\partial _I,\partial _{<A>})=(\partial _I,\partial
_{(A_1)},\partial _{(A_2)},...,\partial _{(A_z)})=
$$
$$
\frac \partial {\partial u^{<\alpha >}}=(\frac \partial {\partial x^I},\frac
\partial {\partial y_{(1)}^{A_1}},\frac \partial {\partial
y_{(2)}^{A_2}},...,\frac \partial {\partial y_{(z)}^{A_z}})\eqno(4)
$$
(the dual coordinate bases are denoted as%
$$
d^{<\alpha >}=(d^I,d^{<A>})=(d^I,d^{(A_1)},d^{(A_2)},...,d^{(A_z)})=
$$
$$
du^{<\alpha >}=(dx^I,dy^{(A_1)},dy^{(A_2)},...,dy^{(A_z)})\quad )\eqno(5)
$$
are transformed as%
$$
\partial _{<\alpha >}=(\partial _I,\partial _{<A>})=(\partial _I,\partial
_{(A_1)},\partial _{(A_2)},...,\partial _{(A_z)})\rightarrow \partial
_{<\alpha >}=
$$
$$
(\partial _I,\partial _{<A>})=(\partial _I,\partial _{(A_1)},\partial
_{(A_2)},...,\partial _{(A_z)})
$$
$$
\frac \partial {\partial x^I}=K_I^{I^{\prime }}\frac \partial {\partial
x^{I^{\prime }}}+Y_{(1,0)I}^{A_1^{\prime }}\frac \partial {\partial
y_{(1)}^{A_1^{\prime }}}+Y_{(2,0)I}^{A_2^{\prime }}\frac \partial {\partial
y_{(2)}^{A_2^{\prime }}}+...+Y_{(z,0)I}^{A_z^{\prime }}\frac \partial
{\partial y_{(z)}^{A_z^{\prime }}},\eqno(6)
$$
$$
\frac \partial {\partial y_{(1)}^{A_1}}=K_{A_1}^{A_1^{\prime }}\frac
\partial {\partial y_{(1)}^{A_1^{\prime }}}+Y_{(2,1)A_1}^{A_2^{\prime
}}\frac \partial {\partial y_{(2)}^{A_2^{\prime
}}}+...+Y_{(z,1)A_1}^{A_z^{\prime }}\frac \partial {\partial
y_{(z)}^{A_z^{\prime }}},
$$
$$
\frac \partial {\partial y_{(2)}^{A_2}}=K_{A_2}^{A_2^{\prime }}\frac
\partial {\partial y_{(2)}^{A_2^{\prime }}}+Y_{(3,2)A_2}^{A_3^{\prime
}}\frac \partial {\partial y_{(3)}^{A_3^{\prime
}}}+...+Y_{(z,2)A_2}^{A_z^{\prime }}\frac \partial {\partial
y_{(z)}^{A_z^{\prime }}},
$$
$$
........................................................
$$
$$
\frac \partial {\partial y_{(z-1)}^{A_{z-1}}}=K_{A_{z-1}}^{A_{z-1}^{\prime
}}\frac \partial {\partial y_{(z-1)}^{A_{z-1}^{\prime
}}}+Y_{(z,z-1)A_{s-1}}^{A_z^{\prime }}\frac \partial {\partial
y_{(z)}^{A_z^{\prime }}},
$$
$$
\frac \partial {\partial y_{(z)}^{A_z}}=K_{A_z}^{A_z^{\prime }}\frac
\partial {\partial y_{(z)}^{A_z^{\prime }}}.
$$

$Y$--matrices from (6) are partial derivations of corresponding
combinations of $K$--coefficients from coordinate transforms (3),
$$
Y_{A_f}^{A_p^{\prime }}=\frac{\partial (K_{A_p}^{A_p^{\prime }}~y^{A_p})}{%
\partial y^{A_f}},~f<p.
$$

In brief we denote respectively ds--coordinate transforms of coordinate bases
(4)\ and (5)\ as%
$$
\partial _{<\alpha >}=(K_{<\alpha >}^{<\alpha ^{\prime }>}+Y_{<\alpha
>}^{<\alpha ^{\prime }>})~\partial _{<\alpha ^{\prime }>}\mbox{ and }%
~d^{<\alpha >}=(K_{<\alpha ^{\prime }>}^{<\alpha >}+Y_{<\alpha ^{\prime
}>}^{<\alpha >})d^{<\alpha ^{\prime }>},
$$
where matrix $K_{<\alpha >}^{<\alpha ^{\prime }>}$, its s-inverse $%
K_{<\alpha ^{\prime }>}^{<\alpha >}$, as well $Y_{<\alpha >}^{<\alpha
^{\prime }>}$ and $Y_{<\alpha ^{\prime }>}^{<\alpha >}$ are paramet\-riz\-ed
according to (6). In order to illustrate geometric properties of some of
our transforms it is useful to introduce matrix operators and to consider
in explicit form the parametrizations of matrices under consideration. For
instance, in operator form the transforms (6)
$$
{\bf \partial =}\widehat{{\bf Y}}{\bf \partial }^{\prime },
$$
are characterized by matrices of type
$$
{\bf \partial =}\partial _{<\alpha >}=\left(
\begin{array}{c}
\partial _I \\
\partial _{A_1} \\
\partial _{A_2} \\
... \\
\partial _{A_z}
\end{array}
\right) =\left(
\begin{array}{c}
\frac \partial {\partial x^I} \\
\frac \partial {\partial y_{(1)}^{A_1}} \\
\frac \partial {\partial y_{(2)}^{A_2}} \\
... \\
\frac \partial {\partial y_{(z)}^{A_z}}
\end{array}
\right) ,{\bf \partial }^{\prime }{\bf =}\partial _{<\alpha ^{\prime
}>}=\left(
\begin{array}{c}
\partial _{I^{\prime }} \\
\partial _{A_1^{\prime }} \\
\partial _{A_2^{\prime }} \\
... \\
\partial _{A_z^{\prime }}
\end{array}
\right) =\left(
\begin{array}{c}
\frac \partial {\partial x^{I^{\prime }}} \\
\frac \partial {\partial y_{(1)}^{A_1^{\prime }}} \\
\frac \partial {\partial y_{(2)}^{A_2^{\prime }}} \\
... \\
\frac \partial {\partial y_{(z)}^{A_z^{\prime }}}
\end{array}
\right)
$$
and
$$
\widehat{{\bf Y}}{\bf =}\widehat{Y}_{<\alpha >}^{<\alpha ^{\prime }>}=\left(
\begin{array}{ccccc}
K_I^{I^{\prime }} & Y_{(1,0)I}^{A_1^{\prime }} & Y_{(2,0)I}^{A_2^{\prime }}
& ... & Y_{(z,0)I}^{A_z^{\prime }} \\
0 & K_{A_1}^{A_1^{\prime }} & Y_{(2,1)A_1}^{A_2^{\prime }} & ... &
Y_{(z,1)A_1}^{A_z^{\prime }} \\
0 & 0 & K_{A_2}^{A_2^{\prime }} & ... & Y_{(z,2)A_2}^{A_z^{\prime }} \\
... & ... & ... & ... & ... \\
0 & 0 & 0 & ... & K_{A_z}^{A_z^{\prime }}
\end{array}
\right) .
$$

We note that we obtain a supersimmetric generalization of the
Miron--Atanasiu \cite{mirata} osculator bundle
$\left( Osc^z\tilde M,\pi
,\tilde M\right) $ if the fiber space  is taken to be a
direct sum of $z$ vector s-spaces of the same dimension $\dim {\cal F}=\dim
\widetilde{M},$ i.e. ${\cal F}^{<d>}={\cal F}\oplus {\cal F}\oplus ...\oplus
{\cal F}.$ In this case the $K$ and $Y$ matrices from (3) and (6)
satisfy identities:%
$$
K_{A_1}^{A_1^{\prime }}=K_{A_2}^{A_2^{\prime }}=...=K_{A_z}^{A_z^{\prime }},
$$
$$
Y_{(1,0)A}^{A^{\prime }}=Y_{(2,1)A}^{A^{\prime
}}=...=Y_{(z,z-1)A}^{A^{\prime }},
$$
$$
.............................
$$
$$
Y_{(p,0)A}^{A^{\prime }}=Y_{(p+1,1)A}^{A^{\prime
}}=...=Y_{(z,z-1)A}^{A^{\prime }},\quad (p=2,...,z-1).
$$
For $s=1$ the $Osc^1\widetilde{M}$ is the ts--bundle $T\widetilde{M}.$

Introducing projection $\pi _0^z\doteq \pi ^{<z>}:\widetilde{{\cal E}}%
^{<z>}\rightarrow \widetilde{M}$ we can also consider projections $\pi
_{p_2}^{p_1}:\widetilde{{\cal E}}^{<p_1>}\rightarrow \widetilde{{\cal E}}%
^{<p_2>}~\quad (p_2<p_1)$ defined as
$$
\pi _{s_2}^{s_1}(x,y^{(1)},...,y^{(p_1)})=(x,y^{(1)},...,y^{(p_2)}).
$$
The s-differentials $d\pi _{p_2}^{p_1}:T(\widetilde{{\cal E}}%
^{<p_1>})\rightarrow T(\widetilde{{\cal E}}^{<p_2>})$ of maps $\pi
_{p_2}^{p_1}$ in turn define vertical dvs-subbundles $V_{h+1}=Kerd\pi
_h^{p_1}~(h=0,1,..,p_1-1)$ of the tangent dvs-bundle $T(\widetilde{{\cal E}}%
^{<z>})~$( the dvs-space $V_1=V$ is the vertical dvs-subbundle on $%
\widetilde{{\cal E}}^{<z>}.$ The local fibres of dvs-subbundles $V_h$
determines this regular s-distribution $V_{h+1}:u\in \widetilde{{\cal E}}%
^{<z>}\rightarrow V_{h+1}(u)\subset T(\widetilde{{\cal E}}^{<z>})$ for which
one holds inclusions $V_z\subset V_{z-1}\subset ...\subset V_1.\,$ The
enumerated properties of vertical dvs--subbundles are explicitly illustrated
by transformation laws (6) for distinguished local bases.

\section{Nonlinear Connections in DVS--Bundles}

The purpose of this section is to present an introduction into geometry of
the nonlinear connection structures in dvs--bundles. The concept of nonlinear
connection ( N--connection ) was introduced in the fra\-me\-work of Finsler
geometry \cite{car,car35,kaw} (the global definition of N--connec\-ti\-on is
given in \cite{barth}). It should be noted here that the N--connection
( splitting ) field could play an important rule in modeling various variants
of dynamical reduction from higher dimensional to lower dimensional
s--spaces with (or not) different types of local anisotropy. In monographs
\cite{ma87,ma94} there are contained detailed investigations of geometrical
properties of N-connection structures in v--bundles and different
generalizations of Finsler geometry and some proposals (see Chapter XII in
\cite{ma87}, written by S. Ikeda) on physical interpretation of
N--connection in the framework of ''unified'' field theory with interactions
nonlocalized by y--dependencies are discussed. We emphasize that N--connection
is a different geometrical object from that introduced by using nonlinear
realizations of gauge groups and supergroups (see, for instance, the
collection of works on supergravity \cite{sal} and approaches to gauge
gravity \cite{ts,pon}).To make the presentation to aid rapid assimilation we
shall have realized our geometric constructions firstly for vs--bundles then
we shall extend them for higher order extensions, i.e. for general
dvs--bundles.

\subsection{N-connections in vs--bundles}

Let consider the definitions of N--connection structure \cite{vlasg} in a
vs-bundle $\tilde {{\cal E}}=(\tilde E,{\pi }_E,\tilde M)$ whose type fibre
is ${\hat {{\cal F}}}$ and ${{\pi }^T}:T\tilde {{\cal E}}{\to }T\tilde M$ is
the superdifferential of the map ${{\pi }_E}$ (${{\pi }^T}$ is a
fibre-preserving morphism of the ts-bundle $(T\tilde {{\cal E}},{{\tau }_E}%
,\tilde M)$ to $\tilde E$ and of ts-bundle $(T\tilde M,{\tau },\tilde M)$ to
$\tilde M$). The kernel of this vs-bundle morphism being a subbundle of $%
(T\tilde E,{{\tau }_E},\tilde E)$ is called the vertical subbundle over $%
\tilde {{\cal E}}$ and denoted by $V\tilde {{\cal E}}=(V\tilde E,{{\tau }_V}%
,\tilde E)$. Its total space is $V\tilde {{\cal E}}={{\bigcup }_{u\in \tilde
{{\cal E}}}}{\quad }{V_u},{\quad }$ where ${V_u}={ker}{{\pi }^T},{\quad }{u{%
\in }\tilde {{\cal E}}}.$ A vector
$$
Y={Y^\alpha }{\frac{{\partial }}{{{\partial }{u^\alpha }}}}={Y^I}{\frac{{%
\partial }}{{{\partial }{x^I}}}}+{Y^A}{\frac{{\partial }}{{{\partial }{y^A}}}%
}={Y^i}{\frac{{\partial }}{{{\partial }{x^i}}}}+{Y^{\hat i}}{\frac{{\partial
}}{{{\partial }{{\theta }^{\hat i}}}}}+{Y^a}{\frac{{\partial }}{{{\partial }{%
y^a}}}}+{Y^{\hat a}}{\frac \partial {\partial {{\zeta }^{\hat a}}}}
$$
tangent to $\tilde {{\cal E}}$ in the point $u\in \tilde {{\cal E}}$ is
locally represented as
$$
(u,Y)=({u^\alpha },{Y^\alpha })=({x^I},{y^A},{Y^I},{Y^A})=({{\hat x}^i},{{%
\theta }^{\hat i}},{{\hat y}^a},{{\zeta }^{\hat a}},{{\hat Y}^i},{Y^{\hat i}}%
,{{\hat Y}^a},{Y^{\hat a}}).
$$

A nonlinear connection, N--connection, in vs--bundle $\tilde {{\cal E}}$ is a
splitting on the left of the exact sequence
$$
0{\longmapsto }{V\tilde {{\cal E}}}\stackrel{i}{\longmapsto }{T\tilde {{\cal %
E}}}{\longmapsto }{{T\tilde {{\cal E}}{/}V\tilde {{\cal E}}}}{\longmapsto }0,%
\eqno(7)
$$
i.e. a morphism of vs-bundles $N:T\tilde {{\cal E}}\in {V\tilde {{\cal E}}}$
such that $N{\circ }i$ is the identity on $V\tilde {{\cal E}}$.

The ker\-nel of the mor\-phism $N$ is called the hor\-i\-zon\-tal
sub\-bun\-dle and denoted by
$$
(H\tilde E,{{\tau }_E},\tilde E).
$$
From the exact sequence (7) one follows that N--connection structure can be
equivalently defined as a distribution $\{{{{\tilde E}_u}\to {{H_u}\tilde E}}%
,{{T_u}\tilde E}={{H_u}\tilde E}{\oplus }{{V_u}\tilde E}\}$ on $\tilde E$
defining a global decomposition, as a Whitney sum,
$$
T{\tilde {{\cal E}}}=H{\tilde {{\cal E}}}+V{\tilde {{\cal E}}}.\eqno(8)
$$

To a given N-connection we can associate a covariant s-derivation on ${%
\tilde M}:$

$$
{\bigtriangledown }_X{Y}=X^I{\{{\frac{{\partial Y^A}}{{\partial x^I}}}+{N_I^A%
}(x,Y)\}}s_A,\eqno(9)
$$
where $s_A$ are local independent sections of $\tilde {{\cal E}},{\quad }Y={%
Y^A}s_A$ and $X={X^I}s_I$.

S--differentiable func\-tions $N^{A}_{I}$ from (3) writ\-ten as func\-tions
on $x^I$ and $y^{A},\\ N^{A}_{I}(x,y),$ are called the coefficients of the
N--connection and satisfy these transformation laws under coordinate
transforms (1) in ${\cal E}$:
$$
N^{A^{\prime}}_{I^{\prime}}{\frac{{\partial x^{I^{\prime}}}}{{\partial x^{I}}%
}}=M^{A^{\prime}}_{A} N^{A}_{I}- {\frac{\partial {M^{A^{\prime}}_{A}{(x)}}}{%
\partial x^I}} {y^A}.%
$$

If coefficients of a given N-connection are s--differentiable with respect to
coordinates $y^A$ we can introduce (additionally to covariant nonlinear
s-derivation (9)) a linear covariant s-derivation $\hat D$ (which is a
generalization for vs--bundles of the Berwald connection \cite{berw}) given
as follows:
$$
{{\hat D}_{({\frac{{\partial }}{{\partial x^I}}})}}({\frac{{\partial }}{{%
\partial y^A}}})={{{\hat N}^B}_{AI}}({\frac{{\partial }}{{\partial y^B}}}),{%
\quad }{{\hat D}_{({\frac{{\partial }}{{\partial y^A}}})}}({\frac{{\partial }%
}{{\partial y^B}}})=0,
$$
where

$$
{{\hat N}^A}_{BI}(x,y)={\frac{{{\partial }{{N^A}_I}{(x,y)}}}{{\partial y^B}}}%
\eqno(10)
$$
and
$$
{{{\hat N}^A}_{BC}}{(x,y)}=0.
$$
For a vector field on ${\tilde {{\cal E}}}{\quad }Z={Z^I}{\frac \partial {{%
\partial x^I}}}+{Y^A}{\frac \partial {{\partial y^A}}}$ and $B={B^A}{(y)}{%
\frac \partial {{\partial y^A}}}$ being a section in the vertical s--bundle $%
(V\tilde E,{{\tau }_V},\tilde E)$ the linear connection (10) defines
s--derivation (compare with (9)):
$$
{{\hat D}_Z}B=[{Z^I}({\frac{\partial B^A}{\partial x^I}}+{\hat N}%
_{BI}^AB^B)+Y^B{\frac{\partial B^A}{\partial y^B}}]{\frac \partial {\partial
y^A}}.
$$

Another important characteristic of a N--connection is its curvature
 ( N--connection curvature ):
$$
\Omega ={\frac 12}{\Omega }_{IJ}^A{dx^I}\land {dx^J}\otimes {\frac \partial
{\partial y^A}}
$$
with local coefficients
$$
{\Omega }_{IJ}^A={\frac{\partial N_I^A}{\partial x^J}}-{(-)}^{|IJ|}{\frac{%
\partial N_J^A}{\partial x^I}}+N_I^B{\hat N}_{BJ}^A-{(-)}^{|IJ|}N_J^B{\hat N}%
_{BI}^A,\eqno(11)
$$
where for simplicity we have written ${(-)}^{{\mid K\mid }{\mid J\mid }}={(-)%
}^{\mid {KJ}\mid }.$

We note that lin\-ear con\-nec\-tions are par\-tic\-u\-lar cases of
N--connections locally pa\-ra\-met\-rized as $N_I^A{(x,y)}=N_{BI}^A{(x)}%
x^Iy^B,$ where functions $N_{BI}^A{(x)}$, defined on $\tilde M,$ are called
the Christoffel coefficients.

\subsection{N--connections in dvs--bundles}

In order to define a N--connection into dvs--bundle $\widetilde{{\cal E}}%
^{<z>}$ we consider a s--sub\-bund\-le $N\left( \widetilde{{\cal E}}%
^{<z>}\right) $ of the ts-bundle $T\left( \widetilde{{\cal E}}^{<z>}\right) $%
for which one holds (see \cite{sau} and \cite{mirata} respectively for jet
and osculator bundles) the Whitney sum (compare with (8))%
$$
T\left( \widetilde{{\cal E}}^{<z>}\right) =N\left( \widetilde{{\cal E}}%
^{<z>}\right) \oplus V\left( \widetilde{{\cal E}}^{<z>}\right) .
$$
$N\left( \widetilde{{\cal E}}^{<z>}\right) $ can be also interpreted as a
regular s--distribution (horizontal distribution being supplementary to the
vertical s-distribution $V\left( \widetilde{{\cal E}}^{<z>}\right) )$
determined by maps $N:u\in \widetilde{{\cal E}}^{<z>}\rightarrow N(u)\subset
T_u\left( \widetilde{{\cal E}}^{<z>}\right) .$

The condition of existence of a N--connection in a dvs--bundle $\widetilde{%
{\cal E}}^{<z>}$ can be proved as in \cite{ma87,ma94,mirata}: It is required
that $\widetilde{{\cal E}}^{<z>}$ is a paracompact s--differentiable (in our
case) manifold.

Locally a N--connection in $\widetilde{{\cal E}}^{<z>}$ is given by its
coefficients
$$
N_{(01)I}^{A_1}(u),(N_{(02)I}^{A_2}(u),
N_{(12)A_1}^{A_2}(u)),...(N_{(0p)I}^{A_p}(u),N_{(1p)A_1}^{A_p}(u),...
N_{(p-1p)A_{p-1}}^{A_p}(u)),...,
$$
$$
(N_{(0z)I}^{A_z}(u),N_{(1z)A_1}^{A_z}(u),...,N_{(pz)A_p}^{A_z}(u),...,
N_{(z-1z)A_{z-1}}^{A_z}(u)),
$$
where, for instance, $%
(N_{(0p)I}^{A_p}(u),N_{(1p)A_1}^{A_p}(u),...,N_{(p-1p)A_{p-1}}^{A_p}(u))$
are components of N--connection in vs-bundle $\pi ^{<p>}:\tilde
E^{<p>}\rightarrow \tilde E^{<p-1>}.$ Here we note that if a
N-con\-nect\-i\-on structure is defined we must correlate to it the local
partial derivatives on $\widetilde{{\cal E}}^{<z>}$ by considering instead of
local coordinate bases (4) and (5) the so--called locally adapted bases
( la--bases )
$$
\delta _{<\alpha >}=(\delta _I,\delta _{<A>})=(\delta _I,\delta
_{(A_1)},\delta _{(A_2)},...,\delta _{(A_s)})=
$$
$$
\frac \delta {\partial u^{<\alpha >}}=(\frac \delta {\partial x^I},\frac
\delta {\partial y_{(1)}^{A_1}},\frac \delta {\partial
y_{(2)}^{A_2}},...,\frac \delta {\partial y_{(z)}^{A_z}})\eqno(12)
$$
(the dual la--bases are denoted as%
$$
\delta ^{<\alpha >}=(\delta ^I,\delta ^{<A>})=(\delta ^I,\delta
^{(A_1)},\delta ^{(A_2)},...,\delta ^{(A_z)})=
$$
$$
\delta u^{<\alpha >}=(\delta x^I,\delta y^{(A_1)},\delta
y^{(A_2)},...,\delta y^{(A_z)})\quad )\eqno(13)
$$
with components parametrized as
$$
\delta _I=\partial _I-N_I^{A_1}\partial _{A_1}-N_I^{A_2}\partial
_{A_2}-...-N_I^{A_{z-1}}\partial _{A_{z-1}}-N_I^{A_z}\partial _{A_z},%
\eqno(14)
$$
$$
\delta _{A_1}=\partial _{A_1}-N_{A_1}^{A_2}\partial
_{A_2}-N_{A_1}^{A_3}\partial _{A_3}-...-N_{A_1}^{A_{z-1}}\partial
_{A_{z-1}}-N_{A_1}^{A_z}\partial _{A_z},
$$
$$
\delta _{A_2}=\partial _{A_2}-N_{A_2}^{A_3}\partial
_{A_3}-N_{A_2}^{A_4}\partial _{A_4}-...-N_{A_2}^{A_{z-1}}\partial
_{A_{z-1}}-N_{A_2}^{A_z}\partial _{A_z},
$$
$$
..............................................................
$$
$$
\delta _{A_{z-1}}=\partial _{A_{z-1}}-N_{A_{z-1}}^{A_z}\partial _{A_z},
$$
$$
\delta _{A_z}=\partial _{A_z},
$$
or, in matrix form, as%
$$
{\bf \delta }_{\bullet }{\bf =}\widehat{{\bf N}}(u)\times {\bf \partial }%
_{\bullet },$$
where%
$$
{\bf \delta }_{\bullet }{\bf =}\delta _{<\alpha >}=\left(
\begin{array}{c}
\delta _I \\
\delta _{A_1} \\
\delta _{A_2} \\
... \\
\delta _{A_z}
\end{array}
\right) =\left(
\begin{array}{c}
\frac \delta {\partial x^I} \\
\frac \delta {\partial y_{(1)}^{A_1}} \\
\frac \delta {\partial y_{(2)}^{A_2}} \\
... \\
\frac \delta {\partial y_{(z)}^{A_z}}
\end{array}
\right) ,{\bf \partial }_{\bullet }{\bf =}\partial _{<\alpha >}=\left(
\begin{array}{c}
\partial _I \\
\partial _{A_1} \\
\partial _{A_2} \\
... \\
\partial _{A_z}
\end{array}
\right) =\left(
\begin{array}{c}
\frac \partial {\partial x^I} \\
\frac \partial {\partial y_{(1)}^{A_1}} \\
\frac \partial {\partial y_{(2)}^{A_2}} \\
... \\
\frac \partial {\partial y_{(z)}^{A_z}}
\end{array}
\right) .
$$
and%
$$
\widehat{{\bf N}}{\bf =}\left(
\begin{array}{ccccc}
1 & -N_I^{A_1} & -N_I^{A_2} & ... & -N_I^{A_z} \\
0 & 1 & -N_{A_1}^{A_2} & ... & -N_{A_1}^{A_z} \\
0 & 0 & 1 & ... & -N_{A_2}^{A_z} \\
... & ... & ... & ... & ... \\
0 & 0 & 0 & ... & 1
\end{array}
\right) $$
In generalized index form we write the matrix (6) as $\widehat{N}_{<\beta
>}^{<\alpha >},$ where, for instance, $\widehat{N}_J^I=\delta _J^I,\widehat{N%
}_{B_1}^{A_1}=\delta _{B_1}^{A_1},...,\widehat{N}_I^{A_1}=-N_I^{A_1},...,%
\widehat{N}_{A_1}^{A_z}=-N_{A_1}^{A_z},\widehat{N}%
_{A_2}^{A_z}=-N_{A_2}^{A_z},...~.$

So in every point $u\in \widetilde{{\cal E}}^{<z>}$ we have this invariant
decomposition:%
$$
T_u\left( \widetilde{{\cal E}}^{<d>}\right) =N_0(u)\oplus N_1(u)\oplus
...\oplus N_{z-1}(u)\oplus V_z(u),\eqno(15)
$$
where $\delta _I\in N_0,\delta _{A_1}\in N_1,...,\delta _{A_{z-1}}\in
N_{z-1},\partial _{A_z}\in V_z.$

We note that for the osculator s--bundle $\left( Osc^z\tilde M,\pi ,\tilde
M\right) $ there is an additional (we consider the N--adapted variant)
s--tangent structure
$$
J:\chi \left( Osc^z\tilde M\right) \rightarrow \chi \left( Osc^z\tilde
M\right)
$$
defined as
$$
\frac \delta {\partial y_{(1)}^I}=J\left( \frac \delta {\partial x^I}\right)
,...,\frac \delta {\partial y_{(z-1)}^I}=J\left( \frac \delta {\partial
y_{(z-2)}^I}\right) ,\frac \partial {\partial y_{(z)}^I}=J\left( \frac
\delta {\partial y_{(z-1)}^I}\right) \eqno(16)
$$
(in this case $I$- and $A$--indices take the same values and we can not
distinguish them), by considering vertical $J$--distributions
$$
N_0=N,N_1=J\left( N_0\right) ,...,N_{z-1}=J\left( N_{z-2}\right) .
$$
In consequence, for the la-adapted bases on $\left( Osc^z\tilde M,\pi
,\tilde M\right) $ there is written this N--connection matrix:%
$$
{\bf N=}N_{<I>}^{<J>}=\left(
\begin{array}{ccccc}
1 & -N_{(1)I}^J & -N_{(2)I}^J & ... & -N_{(z)I}^J \\
0 & 1 & -N_{(1)I}^J & ... & -N_{(z-1)I}^J \\
0 & 0 & 1 & ... & -N_{(z-2)I}^J \\
... & ... & ... & ... & ... \\
0 & 0 & 0 & ... & 1
\end{array}
\right) .\eqno(17)
$$

There is a unique distinguished local decomposition of every s--vector $X\in
\chi \left( \widetilde{{\cal E}}^{<z>}\right) $ on la--base (12):%
$$
X=X^{(H)}+X^{(V_1)}+...+X^{(V_z)},\eqno(18)
$$
by using the horizontal, $h,$ and verticals, $v_1,v_2,...,v_z$ projections:%
$$
X^{(H)}=hX=X^I\delta _I,~X^{(V_1)}=v_1X=X^{(A_1)}\delta
_{A_1},...,X^{(V_z)}=v_zX=X^{(A_z)}\delta _{A_z}.
$$

With respect to coordinate transforms (4 ) the la--bases (12) and
ds--vector components (18) are correspondingly transformed as
$$
\frac \delta {\partial x^I}=\frac{{\partial }x^{I^{\prime }}}{{\partial }x^I}%
\frac \delta {\partial x^{I^{\prime }}},~\frac \delta {\partial
y_{(p)}^{A_p}}=K_{A_p}^{A_p^{\prime }}\frac \delta {\partial
y_{(p)}^{A_p^{\prime }}},\eqno(19)
$$
and%
$$
X^{I^{\prime }}=\frac{{\partial }x^{I^{\prime }}}{{\partial }x^I}%
X^I,~X^{(A_p^{\prime })}=K_{A_p}^{A_p^{\prime }}X^{(A_p^{\prime })},\forall
p=1,2,...z.
$$

Under changing of coordinates (3) the local coefficients of a nonlinear
connection transform as follows:%
$$
Y_{<\alpha >}^{<\alpha ^{\prime }>}\widehat{N}_{<\alpha ^{\prime }>}^{<\beta
^{\prime }>}=\widehat{N}_{<\alpha >}^{<\beta >}(K_{<\beta >}^{<\beta
^{\prime }>}+Y_{<\beta >}^{<\beta ^{\prime }>})
$$
(we can obtain these relations by putting (19) and (6) into (14) where
$\widehat{N}_{<\alpha ^{\prime }>}^{<\beta ^{\prime }>}$ satisfy $\delta
_{<\alpha ^{\prime }>}=\widehat{N}_{<\alpha ^{\prime }>}^{<\beta ^{\prime
}>}\partial _{<\beta ^{\prime }>}).$

For dual la-bases (13) we have these N--connection ''prolongations of
differentials'':%
$$
\delta x^I=dx^I,
$$
$$
\delta y^{A_1}=dy_{(1)}^{A_1}+M_{(1)I}^{A_1}dx^I,\eqno(20)
$$
$$
\delta
y^{A_2}=dy_{(2)}^{A_2}+M_{(2)A_1}^{A_2}dy_{(1)}^{A_1}+M_{(2)I}^{A_2}dx^I,
$$
$$
................................
$$
$$
\delta
y^{A_s}=dy_{(s)}^{A_s}+M_{(s)A_1}^{As}dy_{(1)}^{A_1}+
M_{(s)A_2}^{As}dy_{(2)}^{A_2}+...+M_{(z)I}^{A_z}dx^I,
$$
where $M_{(\bullet )\bullet }^{\bullet }$ are the dual coefficients of the
N-connection which can be expressed explicitly by recurrent formulas through
the components of N--connection $N_{<A>}^{<I>}.$ To do this we shall rewrite
formulas (20) in matrix form:%
$$
{\bf \delta }^{\bullet }={\bf d}^{\bullet }\times {\bf M}(u),
$$
where
$$
{\bf \delta }^{\bullet }=\left(
\begin{array}{ccccc}
\delta x^I & \delta y^{A_1} & \delta y^{A_2} & ... & \delta y^{A_s}
\end{array}
\right) ,~{\bf d}^{\bullet }=\left(
\begin{array}{ccccc}
dx^I & dy_{(1)}^{A_1} & \delta y_{(2)}^{A_2} & ... & \delta y_{(s)}^{A_s}
\end{array}
\right)
$$
and%
$$
{\bf M=}\left(
\begin{array}{ccccc}
1 & M_{(1)I}^{A_1} & M_{(2)I}^{A_2} & ... & M_{(z)I}^{A_z} \\
0 & 1 & M_{(2)A_1}^{A_2} & ... & M_{(z)A_1}^{A_z} \\
0 & 0 & 1 & ... & M_{(z)A_2}^{A_z} \\
... & ... & ... & ... & ... \\
0 & 0 & 0 & ... & 1
\end{array}
\right) ,
$$
and then, taking into consideration that bases ${\bf \partial }_{\bullet }(%
{\bf \delta }_{\bullet })$ and ${\bf d}^{\bullet }({\bf \delta }^{\bullet })$
are mutually dual, to compute the components of matrix ${\bf M}$ being
s--inverse to matrix ${\bf N}$ (see (17 )). We omit these simple but
tedious calculus for general dvs-bundles and, for simplicity, we present the
basic formulas for osculator s--bundle $\left( Osc^z\tilde M,\pi ,\tilde
M\right) $ when $J$--distribution properties (16) and (17) alleviates
the problem. For common type of indices on $\tilde M$ and higher order
extensions on $Osc^z\tilde M$ the dual la--base is expressed as
$$
\delta x^I=dx^I,
$$
$$
\delta y_{(1)}^I=dy_{(1)}^I+M_{(1)J}^Idx^J,
$$
$$
\delta y_{(2)}^I=dy_{(2)}^I+M_{(1)J}^Idy_{(1)}^J+M_{(2)J}^Idx^J,
$$
$$
................................
$$
$$
\delta
y_{(z)}^I=dy_{(z)}^I+M_{(1)J}^Idy_{(s-1)}^J+M_{(2)J}^Idy_{(z-2)}^J+...+
M_{(z)J}^Idx^J,
$$
with $M$--coefficients computed by recurrent formulas:%
$$
M_{(1)J}^I=N_{(1)J}^I,\eqno(21)
$$
$$
M_{(2)J}^I=N_{(2)J}^I+N_{(1)K}^IM_{(1)J}^K,
$$
$$
..............
$$
$$
M_{(s)J}^I=N_{(s)J}^I+N_{(s-1)K}^IM_{(1)J}^K+...+N_{(2)K}^IM_{(z-2)J}^K+
N_{(1)K}^IM_{(z-1)J}^K.
$$
One holds these transformation law for dual coefficients (21) with respect
to coordinate transforms (3) :
$$
M_{(1)J}^KY_{(0,0)K}^{I^{\prime }}=M_{(1)K^{\prime }}^{I^{\prime
}}Y_{(0,0)J}^{K^{\prime }}+Y_{(1,0)J}^{I^{\prime }},
$$
$$
M_{(2)J}^KY_{(0,0)K}^{I^{\prime }}=M_{(2)K^{\prime }}^{I^{\prime
}}Y_{(0,0)J}^{K^{\prime }}+M_{(1)K^{\prime }}^{I^{\prime
}}Y_{(1,0)J}^{K^{\prime }}+Y_{(2,0)J}^{I^{\prime }},
$$
$$
................................
$$
$$
M_{(z)J}^KY_{(0,0)K}^{I^{\prime }}=M_{(z)K^{\prime }}^{I^{\prime
}}Y_{(0,0)J}^{K^{\prime }}+M_{(z-1)K^{\prime }}^{I^{\prime
}}Y_{(1,0)J}^{K^{\prime }}+...+M_{(1)K^{\prime }}^{I^{\prime
}}Y_{(z-1,0)J}^{K^{\prime }}+Y_{(z,0)J}^{I^{\prime }}.
$$
(the proof is a straightforward regroupation of terms after we have put
(3) into (21)).

Finally, we note that curvatures of a N-connection in a dvs-bundle $%
\widetilde{{\cal E}}^{<z>}$ can be introduced in a manner similar to that
for usual vs-bundles (see (11) by a consequent step by step inclusion of
higher dimension anisotropies :
$$
\Omega _{(p)}={\frac 12}{\Omega }_{(p)\alpha _{p-1}\beta _{p-1}}^{A_p}{%
\delta u}^{\alpha _{p-1}}\land {\delta u}^{\beta _{p-1}}\otimes \frac \delta
{\partial y_{(p)}^{A_p}},~p=1,2,...,z,
$$
with local coefficients
$$
{\Omega }_{(p)\beta _{p-1}\gamma _{p-1}}^{A_p}=\frac{\delta N_{\beta
_{p-1}}^{A_p}}{\partial u_{(p-1)}^{\gamma _{p-1}}}-{(-)}^{|\beta
_{p-1}\gamma _{p-1}|}\frac{\delta N_{\gamma _{p-1}}^{A_p}}{\partial
u_{(p-1)}^{\beta _{p-1}}}+
$$
$$
N_{\beta _{p-1}}^{D_p}{\hat N}_{D_p\gamma _{p-1}}^{A_p}-{(-)}^{|\beta
_{p-1}\gamma _{p-1}|}N_{\gamma _{p-1}}^{D_p}{\hat N}_{D_p\beta _{p-1}}^{A_p},%
$$
where ${\hat N}_{D_p\gamma _{p-1}}^{A_p}=\frac{\delta N_{\gamma
_{p-1}}^{A_p}}{\partial y_{(p)}^{D_p}}$ (we consider $y^{A_0}\simeq x^I).$

\section{Geometric Objects in DVS--Bundles}

The geometry of the dvs--bundles is very rich and could have various
applications in theoretical and mathematical physics. In this section we
shall present the main results from the geometry of total spaces of 
dvs--bundles.

\subsection{D--tensors and d--connections in dvs--bundles}

By using adapted bases (12) and (13) one introduces algebra $DT({\tilde {%
{\cal E}}}^{<z>})$ of distinguished tensor s--fields (ds--fields, ds--tensors,
ds--objects) on $\tilde {{\cal E}}^{<z>},{\quad }{\cal T}={\cal T}%
_{qq_1q_2...q_z}^{pp_1p_{2....}p_z},$ which is equivalent to the tensor
algebra of vs-bundle ${\pi }_{hv_1v_2...v_z}:H\tilde {{\cal E}}^{<z>}{\oplus
}V_1\tilde {{\cal E}}^{<z>}{\oplus }V_2\tilde {{\cal E}}^{<z>}\oplus ...{%
\oplus }V_s\tilde {{\cal E}}^{<z>}{\to }\tilde {{\cal E}}^{<z>},$ hereafter
briefly denoted as ${{\tilde {{\cal E}}}_{dz}}.$ An element $Q{\in {\cal T}}%
_{qq_1q_2...q_z}^{pp_1p_{2....}p_z},$ , ds-field of type $\left(
\begin{array}{ccccc}
p & p_1 & p_2 & ... & p_z \\
q & q_1 & q_2 & ... & q_z
\end{array}
\right) ,$ can be written in local form as
$$
Q={Q}_{{J_1}{\dots }{J_q}{B_1}{\dots }{B}_{q_1}{C}%
_1...C_{q_2}...F_1...F_{q_s}}^{{I_1}{\dots }{I_p}{A_1}{\dots }{A}%
_{p_1}E_1...E_{p_2}...D_1....D_{p_s}}{(u)}{{\delta }_{I_1}}\otimes {\dots }%
\otimes {{\delta }_{I_p}}\otimes {d^{J_1}}\otimes {\dots }\otimes {d^{J_q}}%
\otimes
$$
$$
{{\partial }_{A_1}}\otimes {\dots }\otimes {\partial }_{A_{p_1}}\otimes {{%
\delta }^{B_1}{\delta }^{B_1}}\otimes {\dots }\otimes {\delta }%
^{B_{q_1}}\otimes {{\partial }_{E_1}}\otimes {\dots }\otimes {\partial }%
_{E_{p_2}}\otimes {{\delta }^{C_1}}\otimes {\dots }\otimes {\delta }%
^{C_{q_2}}\otimes ...
$$
$$
\otimes {{\partial }_{D_1}}\otimes {\dots }\otimes {\partial }%
_{D_{p_z}}\otimes {{\delta }^{F_1}}\otimes {\dots }\otimes {\delta }%
^{F_{qz}}.\eqno(22)
$$

In addition to ds--tensors we can introduce ds--objects with various s--group
and coordinate transforms adapted to global splitting (15).

A linear distinguished connection, d--connection, in dvs--bundle $\tilde {%
{\cal E}}^{<z>}$ is a linear connection $D$ on $\tilde {{\cal E}}^{<z>}$
which preserves by parallelism the horizontal and vertical distributions in $%
\tilde {{\cal E}}^{<z>}$.

By a linear connection of a s--manifold we understand a linear connection in
its tangent bundle.

Let denote by $\Xi (\tilde M)$ and $\Xi (\tilde {{\cal E}}^{<p>}),$
respectively, the modules of vector fields on s-manifold $\tilde M$ and
vs-bundle $\tilde {{\cal E}}^{<p>}$ and by ${\cal F}{(\tilde M)}$ and ${\cal %
F}{(\tilde {{\cal E}}}^{<p>}{)},$ respectively, the s-modules of functions
on $\tilde M$ and on $\tilde {{\cal E}}^{<p>}.$

It is clear that for a given global splitting into horizontal and verticals
s--subbund\-les (15) we can associate operators of horizontal and vertical
covariant derivations (h- and v--derivations, denoted respectively as $%
D^{(h)} $ and $D^{(v_1v_2...v_z)}$) with properties:
$$
{D_X}Y=(XD)Y={D_{hX}}Y+{D}_{v_1X}Y+{D}_{v_2X}Y+...+{D}_{v_zX}Y,
$$
where
$$
D_X^{(h)}{Y}=D_{hX}{Y},{\quad }D_X^{(h)}f=(hX)f
$$
and
$$
D_X^{(v_p)}{Y}=D_{v_pX}{Y},{\quad }D_X^{(v_p)}f=(v_pX)f,~(p=1,...,z)
$$
for every $f\in {\cal F}(\tilde M)$ with decomposition of vectors $X,Y\in {%
\Xi }(\tilde {{\cal E}}^{<z>})$ into horizontal and vertical parts, $%
X=hX+v_1X+....+v_zX{\quad }$ and ${\quad }Y=hY+v_1Y+...+v_zY.$

The local coefficients of a d--connection $D$ in $\tilde {{\cal E}}^{<z>}$
with respect to the local adapted frame (5) separate into corresponding
distinguished groups. We introduce horizontal local coefficients\\ $%
(L_{JK}^I,L_{<B>K}^{<A>})=({L^I}_{JK}(u),{L}_{B_1K}^{A_1}{(}u),{L}%
_{B_2K}^{A_2}{(}u),...,{L}_{B_zK}^{A_z}{(}u))$ of $D^{(h)}$ such that
$$
D_{({\frac \delta {\delta x^K}})}^{(h)}{\frac \delta {\delta x^J}}={L^I}%
_{JK}(u){\frac \delta {\delta x^I}},D_{({\frac \delta {\delta x^K}}%
)}^{(h)}\frac \delta {\delta y_{(p)}^{B_p}}={L}_{B_pK}^{A_p}{(u)}\frac
\delta {\delta y_{(p)}^{A_p}},(p=1,...,z),
$$
$$
D_{({\frac \delta {\delta x^k}})}^{(h)}q={\frac{{\delta q}}{\delta x^K}},
$$
and $p$--vertical local coefficients\\ $(C_{J<C>}^I,C_{<B><C>}^{<A>})=({C}%
_{JC_p}^I(u),{C}_{B_1C_p}^{A_1}(u),{C}_{B_2C_p}^{A_2}(u),...,{C}%
_{B_zC_p}^{A_z}(u))$ $(p=1,...,z)$ such that
$$
D_{(\frac \delta {\delta y^{C_p}})}^{(v_p)}{\frac \delta {\delta x^J}}={C}%
_{JC_p}^I(u){\frac \delta {\delta x^I}},D_{(\frac \delta {\delta
y^{C_p}})}^{(v_p)}\frac \delta {\delta y_{(f)}^{B_f}}={C}_{B_fC_p}^{A_f}%
\frac \delta {\delta y_{(f)}^{A_f}},D_{(\frac \delta {\delta
y^{C_p}})}^{(v_p)}q=\frac{\delta q}{\partial y^{C_p}},
$$
where $q\in {\cal F}(\tilde {{\cal E}}^{<z>}),$ $f=1,...,z.$

The covariant ds--de\-ri\-va\-ti\-on along vector 
$X=X^I{\frac \delta {\delta x^I}}%
+Y^{A_1}\frac \delta {\delta y^{A_1}}+...+Y^{A_z}\frac \delta {\delta
y^{A_z}}$ of a ds--tensor field $Q,$ for instance, of type $\left(
\begin{array}{cc}
p & p_r \\
q & q_r
\end{array}
\right) ,1\leq r\leq z,{\quad }$ see (22), can be written as
$$
{D_X}Q=D_X^{(h)}Q+D_X^{(v_1)}Q+...+D_X^{(v_z)}Q,
$$
where h-covariant derivative is defined as
$$
D_X^{(h)}Q=X^KQ_{JB_r{\mid }K}^{IA_r}{\delta }_I{\otimes \delta }_{A_r}{%
\otimes }d^I{\otimes }{\delta }^{B_r},
$$
with components

$$
Q_{JB_r{\mid }K}^{IA_r}={\frac{\delta Q_{JB_r}^{IA_r}}{\partial x^K}}+{L^I}%
_{HK}Q_{JB_R}^{HA_r}+{L}_{C_iK}^{A_r}Q_{JB_i}^{IC_r}-{L^H}%
_{JK}Q_{HB_r}^{IA_r}-{L}_{B_rK}^{C_r}Q_{JC_r}^{IA_r},
$$
and v$_p$-covariant derivatives defined as
$$
{D}_X^{(v_p)}Q={X}^{C_p}{Q}_{JB_r\perp C_p}^{IA_r}{\delta }_I{\otimes }{%
\partial }_{A_r}{\otimes }d^I{\otimes }{\delta }^{B_r},
$$
with components
$$
{Q}_{JB_r\perp C_p}^{IA_r}=\frac{\delta Q_{JB_R}^{IA_r}}{\partial y^{C_p}}+{%
C^I}_{HC_p}Q_{JB_R}^{HA_r}+{C}_{F_rC_p}^{A_r}Q_{JB_R}^{IF_r}-{C}%
_{JC_p}^HQ_{HF_R}^{IA_r}-{C}_{B_rC_p}^{F_r}Q_{JF_R}^{IA_r}..
$$

The above presented formulas show that $D{\Gamma }=(L,...,{L}_{(p)},...,{C}%
,...,C_{(p)},...)$ are the local coefficients of the d-connection $D$ with
respect to the local frame $({\frac \delta {\delta x^I}},{\frac \partial
{\partial y^a}}).$ If a change (3) of local coordinates on $\tilde {{\cal E%
}}^{<z>}$ is performed, by using the law of transformation of local frames
(19),we obtain the following transformation laws of the local coefficients
of a d--connect\-i\-on:
$$
{L^{I^{\prime }}}_{J^{\prime }M^{\prime }}={\frac{\partial x^{I^{\prime }}}{%
\partial x^I}}{\frac{\partial x^J}{\partial x^{J^{\prime }}}}{\frac{\partial
x^M}{\partial x^{M^{\prime }}}}{L^I}_{JM}+{\frac{\partial x^{I^{\prime }}}{%
\partial x^M}}{\frac{{\partial }^2x^M}{{{\partial x^{J^{\prime }}}{\partial
x^{M^{\prime }}}}}},\eqno(23)
$$
$$
{L}_{(f)B_f^{\prime }M^{\prime }}^{A_f^{\prime }}=K_{A_f}^{A_f^{\prime
}}K_{B_f^{\prime }}^{B_f}{\frac{\partial x^M}{\partial x^{M^{\prime }}}L}%
_{(f)B_fM}^{A_f}+K_{C_f}^{A_f^{\prime }}\frac{\partial K_{B_f^{\prime
}}^{C_f}}{\partial x^{M^{\prime }}},
$$
$$
................
$$
$$
{C_{(p)J^{\prime }C_p^{\prime }}^{I^{\prime }}}={\frac{\partial x^{I^{\prime
}}}{\partial x^I}}{\frac{\partial x^J}{\partial x^{J^{\prime }}}}%
K_{C_p}^{C_p^{\prime }}{C_{(p)JC_p}^I},...,{C_{B_f^{\prime }C_p^{\prime
}}^{A_f^{\prime }}}=K_{A_f}^{A_f^{\prime }}{K}_{B_f^{\prime
}}^{B_f}K_{C_p^{\prime }}^{C_p}{C_{B_fC_p}^{A_f},...}.
$$
As in the usual case of tensor calculus on locally isotropic spaces the
transformation laws (23) for d--connections differ from those for
ds-tensors, which are written (for instance, we consider transformation laws
for ds--tensor (22)) as
$$
{Q}_{{J}^{\prime }{_1}{\dots }{J}^{\prime }{_q}{B}^{\prime }{_1}{\dots }{B}%
_{q_1}^{\prime }{C}_1^{\prime }...C_{q_2}^{\prime }...F_1^{\prime
}...F_{q_s}^{\prime }}^{{I}^{\prime }{_1}{\dots }{I}^{\prime }{_p}{A}%
^{\prime }{_1}{\dots }{A}_{p_1}^{\prime }E_1^{\prime }...E_{p_2}^{\prime
}...D_1^{\prime }....D_{p_s}^{\prime }}=
$$
$$
{\frac{\partial x^{I_1^{\prime }}}{\partial x^{I_1}}{\frac{\partial x^{J_1}}{%
\partial x^{J_1^{\prime }}}}\dots }K_{A_1}^{A_1^{\prime }}K_{B_1^{\prime
}}^{B_1}{\dots K}_{D_{p_s}}^{D_{p_s}^{\prime }}{K}_{F_{p_s}^{\prime
}}^{F_{p_s}}{Q}_{{J_1}{\dots }{J_q}{B_1}{\dots }{B}_{q_1}{C}%
_1...C_{q_2}...F_1...F_{q_s}}^{{I_1}{\dots }{I_p}{A_1}{\dots }{A}%
_{p_1}E_1...E_{p_2}...D_1....D_{p_s}}.
$$

To obtain local formulas on usual higher order anisotropic spaces we have to
restrict us with even components of geometric objects by changing, formally,
capital indices $(I,J,K,...)$ into $(i,j,k,a,..)$ and s--derivation and
s--commutation rules into those for real number fields on usual manifolds.

\subsection{ Torsions and curvatures of d--connections}

Let $\tilde {{\cal E}}^{<z>}$ be a dvs--bundle endowed with N-connection and
d-connec\-ti\-on structures. The torsion of a d--connection is introduced as
$$
T(X,Y)=[X,DY\}-[X,Y\},{\quad }X,Y{\subset }{\Xi }{(\tilde M)}.
$$
One holds the following invariant decomposition (by using h-- and
v--projections associated to N):
$$
T(X,Y)=T(hX,hY)+T(hX,v_1Y)+T(v_1X,hX)+T(v_1X,v_1Y)+...
$$
$$
+T(v_{p-1}X,v_{p-1}Y)+T(v_{p-1}X,v_pY)+T(v_pX,v_{p-1}X)+T(v_pX,v_pY)+...
$$
$$
+T(v_{z-1}X,v_{z-1}Y)+T(v_{z-1}X,v_zY)+T(v_zX,v_{z-1}X)+T(v_zX,v_zY).
$$
Taking into ac\-count the skew\-su\-per\-sym\-me\-try of $T$ and the
equa\-tions
$$
h[v_pX,v_pY\}=0,...,v_f[v_pX,v_pY\}=0,f\neq p,
$$
we can verify that the torsion of a d--connection is completely determined by
the following ds-tensor fields:
$$
hT(hX,hY)=[X(D^{(h)}{h})Y\}-h[hX,hY\},...,
$$
$$
v_pT(hX,hY)=-v_p[hX,hY\},...,
$$
$$
hT(hX,v_pY)=-D_Y^{(v_p)}{hX}-h[hX,v_pY\},...,
$$
$$
v_pT(hX,v_pY)=D_X^{(h)}{v}_p{Y}-v_p[hX,v_pY\},...,
$$
$$
v_fT(v_fX,v_fY)=[X(D^{(v_f)}{v}_f)Y\}-v_f[v_fX,v_fY\},...,
$$
$$
v_pT(v_fX,v_fY)=-v_p[v_fX,v_fY\},...,
$$
$$
v_fT(v_fX,v_pY)=-D_Y^{(v_p)}{v}_f{X}-v_f[v_fX,v_pY\},...,
$$
$$
v_pT(v_fX,v_pY)=D_X^{(v_f)}{v}_p{Y}-v_p[v_fX,v_pY\},...,f<p,
$$
$$
v_{z-1}T(v_{z-1}X,v_{x-1}Y)=[X(D^{(v_{z-1})}{v}_{z-1})Y%
\}-v_{z-1}[v_{z-1}X,v_{z-1}Y\},...,
$$
$$
v_zT(v_{z-1}X,v_{z-1}Y)=-v_z[v_{z-1}X,v_{z-1}Y\},
$$
$$
v_{z-1}T(v_{z-1}X,v_zY)=-D_Y^{(v_z)}{v}_{z-1}{X}-v_{z-1}[v_{z-1}X,v_zY%
\},...,
$$
$$
v_zT(v_{z-1}X,v_zY)=D_X^{(v_{z-1})}{v}_z{Y}-v_z[v_{z-1}X,v_zY\}.
$$
where $X,Y\in {{\Xi }(\tilde {{\cal E}}^{<z>})}.$ In order to get the local
form of the ds--tensor fields which determine the torsion of d--connection 
$D{\Gamma }$ (the torsions of $D{\Gamma }$) we use equations
$$
[{\frac \delta {\partial x^J}},{\frac \delta {\partial x^K}}\}={R}%
_{JK}^{<A>}\frac \delta {\partial y^{<A>}},~[{\frac \delta {\partial x^J}},{%
\frac \delta {\partial y^{<B>}}}\}={\frac{\delta {N_J^{<A>}}}{\partial
y^{<B>}}}{\frac \delta {\partial y^{<A>}}},
$$
where
$$
{R}_{JK}^{<A>}=\frac{\delta N_J^{<A>}}{\partial x^K}-{(-)}^{\mid {KJ}\mid }%
\frac{\delta N_K^{<A>}}{\partial x^J},
$$
and introduce notations
$$
hT({\frac \delta {\delta x^K}},{\frac \delta {\delta x^J}})={T^I}_{JK}{\frac
\delta {\delta x^I}},\eqno(24)
$$
$$
v_1T{({\frac \delta {\delta x^K}},{\frac \delta {\delta x^J}})}={\tilde T}%
_{KJ}^{A_1}\frac \delta {\partial y^{A_1}},
$$
$$
hT({\frac \delta {\partial y^{<A>}}},{\frac \delta {\partial x^J}})={\tilde P%
}_{J<A>}^I{\frac \delta {\delta x^I}},{...,}
$$
$$
v_pT(\frac \delta {\partial y^{B_p}},{\frac \delta {\delta x^J}})={P}%
_{JB_p}^{<A>}{\frac \delta {\partial y^{<A>}}},...,
$$
$$
v_pT(\frac \delta {\partial y^{C_p}},\frac \delta {\partial y^{B_f}})={S}%
_{B_fC_p}^{<A>}{\frac \delta {\partial y^{<A>}}}.
$$

Now we can compute the local components of the torsions, introduced in
(24), with respect to a la--frame of a d--connection\\ $D{\Gamma }=(L,...,{L}%
_{(p)},...,{C},...,C_{(p)},...)$ $:$
$$
{T^I}_{JK}={L^I}_{JK}-{(-)}^{\mid {JK}\mid }{L^I}_{KJ},\eqno(25)
$$
$$
{{\tilde T}^{<A>}}_{JK}={R^{<A>}}_{JK},{{\tilde P}^I}_{J<B>}={C^I}_{J<B>},
$$
$$
{P^{<A>}}_{J<B>}={\frac{\delta {N_J^{<A>}}}{\partial y^{<B>}}}-{L^{<A>}}%
_{<B>J},
$$
$$
{S^{<A>}}_{<B><C>}={C^{<A>}}_{<B><C>}-{(-)}^{\mid <B><C>\mid }{C^{<A>}}%
_{<C><B>}.
$$
The even and odd components of torsions (25) can be specified in explicit
form by using decompositions of indices into even and odd parts $(I=(i,{\hat
i}),J=(j,{\hat j}),..)$, for instance,
$$
{T^i}_{jk}={L^i}_{jk}-{L^i}_{kj},{\quad }{T^i}_{j{\hat k}}={L^i}_{j{\hat k}}+%
{L^i}_{{\hat k}j},
$$
$$
{T^{\hat i}}_{jk}={L^{\hat i}}_{jk}-{L^{\hat i}}_{kj},{\dots },
$$
and so on (see \cite{ma87,ma94}).

Another important characteristic of a d--connection $D\Gamma $ is its
curvature:
$$
R(X,Y)Z=D_{[X}D_{Y\}}-D_{[X,Y\}}Z,
$$
where $X,Y,Z\in {\Xi }(\tilde E^{<z>}).$ From the properties of h- and
v-projections it follows that

$$
v_pR(X,Y)hZ=0,...,hR(X,Y)v_pZ=0,v_fR(X,Y)v_pZ=0,f\neq p,\eqno(26)
$$
and
$$
R(X,Y)Z=hR(X,Y)hZ+v_1R(X,Y)v_1Z+...+v_zR(X,Y)v_zZ,
$$
where $X,Y,Z\in {\Xi }(\tilde E^{<z>}).$ Tak\-ing into ac\-count
prop\-er\-ties (26) and the equa\-tions
$$
R(X,Y)=-{(-)}^{\mid XY\mid }R(Y,X)
$$
we prove that the curvature of a d-connection $D$ in the total space of a
vs-bundle $\tilde E^{<z>}$ is completely determined by the following
ds-tensor fields:
$$
R(hX,hY)hZ=({D_{[X}^{(h)}}D_{Y\}}^{(h)}-D_{[hX,hY\}}^{(h)}\eqno(27)
$$
$$
-D_{[hX,hY\}}^{(v_1)}-...D_{[hX,hY\}}^{(v_{z-1})}-D_{[hX,hY\}}^{(v_z)})hZ,
$$

$$
R(hX,hY)v_pZ=(D_{[X}^{(h)}D_{Y\}}^{(h)}-D_{[hX,hY\}}^{(h)}-
$$
$$
D_{[hX,hY\}}^{(v_1)})v_pZ-...-D_{[hX,hY\}}^{(v_{p-1})})v_pZ-D_{[hX,hY%
\}}^{(v_p)})v_pZ,
$$
$$
R(v_pX,hY)hZ=(D_{[X}^{(v_p)}D_{Y\}}^{(h)}-D_{[v_pX,hY\}}^{(h)}-
$$
$$
D_{[v_pX,hY\}}^{(v_{_1})})hZ-...-D_{[v_pX,hY\}}^{(v_{p-_1})}-D_{[v_pX,hY%
\}}^{(v_p)})hZ,
$$
$$
R(v_fX,hY)v_pZ=(D_{[X}^{(v_f)}D_{Y\}}^{(h)}-D_{[v_fX,hY\}}^{(h)}-
$$
$$
D_{[v_fX,hY\}}^{(v_1)})v_1Z-...-D_{[v_fX,hY%
\}}^{(v_{p-1})})v_{p-1}Z-D_{[v_fX,hY\}}^{(v_p)})v_pZ,
$$
$$
R(v_fX,v_pY)hZ=(D_{[X}^{(v_f)}D_{Y\}}^{(v_p)}-D_{[v_fX,v_pY%
\}}^{(v_1)})hZ-...
$$
$$
-D_{[v_fX,v_pY\}}^{(v_{z-1})}-D_{[v_fX,v_pY\}}^{(v_z)})hZ,
$$
$$
R(v_fX,v_qY)v_pZ=(D_{[X}^{(v_f)}D_{Y\}}^{(v_q)}-D_{[v_fX,v_qY%
\}}^{(v_1)})v_1Z-...
$$
$$
-D_{[v_fX,v_qY\}}^{(v_{p-1})})v_{p-1}Z-D_{[v_fX,v_qY\}}^{(v_p)})v_pZ,
$$
where
$$
{D_{[X}^{(h)}}{D_{Y\}}^{(h)}}={D_X^{(h)}}{D_Y^{(h)}}-{{(-)}^{\mid XY\mid }}{%
D_Y^{(h)}}{D_X^{(h)}},~
$$
$$
{D_{[X}^{(h)}}{D}_{Y\}}^{(v_p)}={D_X^{(h)}}{D}_Y^{(v_p)}-{(-)}^{|Xv_pY|}{D}%
_Y^{(v_p)}{D_X^{(h)},}
$$
$$
{D}_{[X}^{(v_p)}{D_{Y\}}^{(h)}}={D}_X^{(v_p)}{D_Y^{(h)}}-{(-)}^{\mid
v_pXY\mid }{D_Y^{(h)}}{D}_X^{(v_p)},~
$$
$$
~{D}_{[X}^{(v_f)}{D}_{Y\}}^{(v_p)}={D}_X^{(v_f)}{D}_Y^{(v_p)}-{(-)}%
^{|v_fXv_pY|}{D}_Y^{(v_p)}{D}_X^{(v_f)}.
$$
The local components of the ds--tensor fields (27) are introduced as
follows:
$$
R({{\delta }_K},{{\delta }_J}){{\delta }_H}={{{R_H}^I}_{JK}}{{\delta }_I},{~}%
R({{\delta }_K},{{\delta }_J})\delta {_{<B>}}={R}_{<B>.JK}^{.<A>}\delta
_{<A>},\eqno(28)
$$
$$
R({\delta _{<C>}},{{\delta }_K}){{\delta }_J}=P_{JK<C>}^I{{{\delta }_I},~}%
R\left( \delta _{<C>},\delta _K\right) \delta
_{<B>}=P_{<B>.K<C>}^{.<A>}\delta _{<A>},
$$
$$
R({\delta _{<C>}},{{\delta }_{<B>}}){{\delta }_J}=S_{J.<B><C>}^{.I}{{{\delta
}_I}},R({\delta _{<D>}},{{\delta }_{<C>}}){{\delta }_{<B>}}%
=S_{<B>.<C><D>}^{.<A>}{{{\delta }_{<A>}}}.
$$
Putting the components of a d--connection 
$D{\Gamma }=(L,...,{L}_{(p)},...,{C}%
,...,C_{(p)},...)$ in (28), by a direct computation, we obtain these
locally adapted components of the curvature (curvatures):
$$
{{R_H}^I}_{JK}={{\delta }_K}{L^I}_{HJ}-{(-)}^{\mid KJ\mid }{{\delta }_J}{L^I}%
_{HK}+
$$
$$
{L^M}_{HJ}{L^I}_{MK}-{(-)}^{\mid KJ\mid }{L^M}_{HK}{L^I}_{MJ}+{C^I}_{H<A>}{%
R^{<A>}}_{JK},
$$
$$
{R_{<B>\cdot JK}^{\cdot <A>}}={{\delta }_K}{L^{<A>}}_{<B>J}-{(-)}^{\mid
KJ\mid }{{\delta }_J}{L^{<A>}}_{<B>K}+
$$
$$
{L^{<C>}}_{<B>J}{L^{<A>}}_{<C>K}-{(-)}^{\mid KJ\mid }{L^{<C>}}_{<B>K}+{%
C^{<A>}}_{<B><C>}{R^{<C>}}_{JK},
$$
$$
{P_{J\cdot K<A>}^{\cdot I}}={\delta _{<A>}}{L^I}_{JK}-{C^I}_{J<A>{\mid }K}+{%
C^I}_{J<B>}{P^{<B>}}_{K<A>},\eqno(29)
$$
$$
{{P_{<B>}}^{<A>}}_{K<C>}={\delta _{<C>}}{L^{<A>}}_{<B>K}-{C^{<A>}}_{<B><C>{%
\mid }K}+{C^{<A>}}_{<B><D>}{P^{<D>}}_{K<C>},
$$
$$
{S_{J\cdot <B><C>}^{\cdot I}}={\delta _{<C>}}{C^I}_{J<B>}-{(-)}^{\mid
<B><C>\mid }{\delta _{<B>}}{C^I}_{J<C>}+
$$
$$
{C^{<H>}}_{J<B>}{C^I}_{<H><C>}-{(-)}^{\mid <B><C>\mid }{C^{<H>}}_{J<C>}{C^I}%
_{<H><B>},
$$
$$
{{S_{<B>}}^{<A>}}_{<C><D>}={\delta _{<D>}}{C^{<A>}}_{<B><C>}-{(-)}^{\mid
<C><D>\mid }{\delta _{<C>}}{C^{<A>}}_{<B><D>}+
$$
$$
{C^{<E>}}_{<B><C>}{C^{<A>}}_{<E><D>}-{(-)}^{\mid <C><D>\mid }{C^{<E>}}%
_{<B><D>}{C^{<A>}}_{<E><C>}.
$$
Even and odd components of curvatures (29) can be written out by splitting
indices into even and odd parts, for instance,
$$
{{R_h}^i}_{jk}={{\delta }_k}{L^i}_{hj}-{{\delta }_j}{{L^i}_{hk}+{{L^m}_{hj}}{%
{L^i}_{mk}}-{{L^m}_{hk}}{{L^i}_{mj}}}+{{C^i}_{h<a>}}{{R^{<a>}}_{jk}},
$$
$$
{{R_h}^i}_{j{\hat k}}={{\delta }_{\hat k}}{L^i}_{hj}+{{\delta }_j}{L^i}_{h{%
\hat k}}+{L^m}_{hj}{L^i}_{m{\hat k}}+{L^m}_{h{\hat k}}{L^i}_{mj}+{C^i}_{h<a>}%
{R^{<a>}}_{j{\hat k}}{\quad },{\dots }.
$$
(we omit formulas for the even--odd components of
curvatures because we shall not use them in this work).

\subsection{Bianchi and Ricci identities}
The torsions and curvatures of every linear connection $D$ on a vs--bundle $%
\tilde {{\cal E}}^{<z>}$ \\satisfy the following generalized Bianchi
identities:
$$
\sum_{SC}{[(D_X{T})(Y,Z)-R(X,Y)Z+T(T(X,Y),Z)]}=0,
$$

$$
\sum_{SC}{[(D_X{R})(U,Y,Z)+R(T(X,Y)Z)U]}=0,\eqno(30)
$$
where $\sum_{SC}$ means the respective supersymmretric cyclic sum over $%
X,Y,Z $ and $U.$ If $D$ is a d--connection, then by using (26) and
$$
v_p(D_X{R})(U,Y,hZ)=0,{~}h({D_X}R(U,Y,v_pZ)=0,v_f(D_X{R})(U,Y,v_pZ)=0,
$$
the identities (30) become
$$
\sum_{SC}[h({D_X}T)(Y,Z)-hR(X,Y)Z+hT(hT(X,Y),Z)+
$$
$$
hT(v_1T(X,Y),Z)+...+hT(v_zT(X,Y),Z)]=0,
$$
$$
\sum_{SC}{[v}_f{({D_X}T)(Y,Z)}-{v}_f{R(X,Y)Z+}
$$
$$
{v}_f{T(hT(X,Y),Z)+}\sum\limits_{p\geq f}{v}_f{T(v}_p{T(X,Y),Z)]}=0,
$$
$$
\sum_{SC}{[h({D_X}R)(U,Y,Z)+hR(hT(X,Y),Z)U+}
$$
$$
{hR(v}_1{T(X,Y),Z)U+...+{hR(v}_z{T(X,Y),Z)U}]}=0,
$$
$$
\sum_{SC}{[v}_f{({D_X}R)(U,Y,Z)+v}_f{R(hT(X,Y),Z)U+}
$$
$$
\sum\limits_{p\geq f}{v}_f{R(v}_p{T(X,Y),Z)U]}=0.\eqno(31)
$$

In order to get the component form of these identities we insert
correspondingly in (31) these values of triples $(X,Y,Z)$,\ ($=({{\delta }%
_J},{{\delta }_K},{{\delta }_L}),$ or\\ $({\delta _{<D>}},{\delta _{<C>}},{%
\delta _{<B>}})$), and put successively $U={\delta }_H$ and $U={\delta }%
_{<A>}.$ Taking into account (24),(25) and (27),(28) we obtain:
$$
\sum_{SC[L,K,J\}}[{T^I}_{JK{\mid }H}+{T^M}_{JK}{T^J}_{HM}+{R^{<A>}}_{JK}{C^I}%
_{H<A>}-{{R_J}^I}_{KH}]=0,
$$
$$
\sum_{SC[L,K,J\}}[{{R^{<A>}}_{JK{\mid }H}}+{T^M}_{JK}{R^{<A>}}_{HM}+{R^{<B>}}%
_{JK}{P^{<A>}}_{H<B>}]=0,
$$
{\cal
$$
{C^I}_{J<B>{\mid }K}-{(-)}^{\mid JK\mid }{C^I}_{K<B>{\mid }J}-{T^I}_{JK{\mid
<}B>}+{C^M}_{J<B>}{T^I}_{KM}-
$$
}%
$$
{(-)}^{\mid JK\mid }{C^M}_{K<B>}{T^I}_{JM}+{T^M}_{JK}{C^I}_{M<B>}+{P^{<D>}}%
_{J<B>}{C^I}_{K<D>}
$$
{\cal
$$
-{(-)}^{\mid KJ\mid }{P^{<D>}}_{K<B>}{C^I}_{J<D>}+{{P_J}^I}_{K<B>}-{(-)}%
^{\mid KJ\mid }{{P_K}^I}_{J<B>}=0,
$$
$$
{P^{<A>}}_{J<B>{\mid }K}-{(-)}^{\mid KJ\mid }{P^{<A>}}_{K<B>{\mid }J}-{%
R^{<A>}}_{JK\perp <B>}+{C^M}_{J<B>}{R^{<A>}}_{KM}-
$$
}%
$$
{(-)}^{\mid KJ\mid }{C^M}_{K<B>}{R^{<A>}}_{JM}+{T^M}_{JK}{P^{<A>}}_{M<B>}+{%
P^{<D>}}_{J<B>}{P^{<A>}}_{K<D>}-
$$
{\cal
$$
{(-)}^{\mid KJ\mid }{P^{<D>}}_{K<B>}{P^{<A>}}_{J<D>}-{{R^{<D>}}_{JK}}{{%
S^{<A>}}_{B<D>}}+{R_{<B>\cdot JK}^{\cdot <A>}}=0,
$$
$$
{C^I}_{J<B>\perp <C>}-{(-)}^{\mid <B><C>\mid }{C^I}_{J<C>\perp <B>}+{C^M}%
_{J<C>}{C^I}_{M<B>}-
$$
}%
$$
{(-)}^{\mid <B><C>\mid }{C^M}_{J<B>}{C^I}_{M<C>}+{S^{<D>}}_{<B><C>}{C^I}%
_{J<D>}-{S_{J\cdot <B><C>}^{\cdot I}}=0,
$$
$$
{P^{<A>}}_{J<B>\perp <C>}-{(-)}^{\mid <B><C>\mid }{P^{<A>}}_{J<C>\perp <B>}+%
$$
$$
{S^{<A>}}_{<B><C>\mid J}+ {C^M}_{J<C>}{P^{<A>}}_{M<B>}-%
$$
$$
{(-)}^{\mid <B><C>\mid }{C^M}_{J<B>}{P^{<A>}}_{M<C>}+{P^{<D>}}_{J<B>}{S^{<A>}%
}_{<C><D>}-
$$
{\cal
$$
{(-)}^{\mid <C><B>\mid }{P^{<D>}}_{J<C>}{S^{<A>}}_{<B><D>}+{S^{<D>}}_{<B><C>}%
{P^{<A>}}_{J<D>}+
$$
$$
{{P_{<B>}}^{<A>}}_{J<C>}-{(-)}^{\mid <C><B>\mid }{{P_{<C>}}^{<A>}}_{J<B>}=0,
$$
$$
\sum_{SC[<B>,<C>,<D>\}}[{S^{<A>}}_{<B><C>\perp <D>}+
$$
}%
$$
{S^{<F>}}_{<B><C>}{S^{<A>}}_{<D><F>}-{{S_{<B>}}^{<A>}}_{<C><D>}]=0,
$$
{\cal
$$
\sum_{SC[H,J,L\}}[{{R_K}^I}_{HJ\mid L}-{T^M}_{HJ}{{R_K}^I}_{LM}-{{R^{<A>}}%
_{HJ}P}{_{K\cdot L<A>}^{\cdot I}}]=0,
$$
$$
\sum_{SC[H,J,L\}}[{R_{<D>\cdot HJ\mid L}^{\cdot <A>}}-{{T^M}_{HJ}R}{%
_{<D>\cdot LM}^{\cdot <A>}}-{{R^{<C>}}_{HJ}}{{{P_{<D>}}^{<A>}}_{L<C>}}]=0,
$$
$$
{P_{K\cdot J<D>\mid L}^{\cdot I}}-{(-)}^{\mid LJ\mid }{P_{K\cdot L<D>\mid
J}^{\cdot I}}+{{R_K}^I}_{LJ\perp <D>}+{C^M}_{L<D>}{{R_K}^I}_{JM}-
$$
}%
$$
{(-)}^{\mid LJ\mid }{C^M}_{J<D>}{{R_K}^I}_{LM}-{T^M}_{JL}{P_{K\cdot
M<D>}^{\cdot I}}+
$$
{\cal
$$
{P^{<A>}}_{L<D>}{P_{K\cdot J<A>}^{\cdot I}}-{(-)}^{\mid LJ\mid }{{P^{<A>}}%
_{J<D>}P}{_{K\cdot L<A>}^{\cdot I}}-{{R^{<A>}}_{JL}S}{_{K\cdot
<A><D>}^{\cdot I}}=0,
$$
$$
{{P_{<C>}}^{<A>}}_{J<D>\mid L}-{(-)}^{\mid LJ\mid }{{P_{<C>}}^{<A>}}%
_{L<D>\mid J}+{R_{<C>\cdot LJ\mid <D>}^{\cdot <A>}}+
$$
$$
{C^M}_{L<D>}{{R_{<C>}}^{<A>}}_{JM}-{(-)}^{\mid LJ\mid }{C^M}_{J<D>}{{R_{<C>}}%
^{<A>}}_{LM}-
$$
}%
$$
{T^M}_{JL}{{P_{<C>}}^{<A>}}_{M<D>}+{P^{<F>}}_{L<D>}{{P_{<C>}}^{<A>}}_{J<F>}-
$$
{\cal
$$
{(-)}^{\mid LJ\mid }{P^{<F>}}_{J<D>}{{P_{<C>}}^{<A>}}_{L<F>}-{R^{<F>}}_{JL}{{%
S_{<C>}}^{<A>}}_{F<D>}=0,
$$
}%
$$
{P_{K\cdot J<D>\perp <C>}^{\cdot I}-(-)}^{\mid <C><D>\mid }{P_{K\cdot
J<C>\perp <D>}^{\cdot I}+{S_K}_{\ <D><C>|J}^I+}
$$
$$
{C^M}_{J<D>}{P_{K\cdot M<C>}^{\cdot I}}-{(-)}^{\mid <C><D>\mid }{C^M}_{J<C>}{%
P_{K\cdot M<D>}^{\cdot I}}+
$$
{\cal
$$
{P^{<A>}}_{J<C>}{S_{K\cdot <D><A>}^{\cdot I}}-{(-)}^{\mid <C><D>\mid }{%
P^{<A>}}_{J<D>}{S_{K\cdot <C><A>}^{\cdot I}}+
$$
$$
{S^{<A>}}_{<C><D>}{P_{K\cdot J<A>}^{\cdot I}}=0,
$$
$$
{{P_{<B>}}^{<A>}}_{J<D>\perp <C>}-{(-)}^{\mid <C><D>\mid }{{P_{<B>}}^{<A>}}%
_{J<C>\perp <D>}+{{S_{<B>}}^{<A>}}_{<C><D>\mid J}+
$$
}%
$$
{C^M}_{J<D>}{{P_{<B>}}^{<A>}}_{M<C>}-{(-)}^{\mid <C><D>\mid }{C^M}_{J<C>}{{%
P_{<B>}}^{<A>}}_{M<D>}+
$$
{\cal
$$
{P^{<F>}}_{J<C>}{{S_{<B>}}^{<A>}}_{<D><F>}-{(-)}^{\mid <C><D>\mid }{P^{<F>}}%
_{J<D>}{{S_{<B>}}^{<A>}}_{<C><F>}+
$$
$$
{S^{<F>}}_{<C><D>}{{P_{<B>}}^{<A>}}_{J<F>}=0,
$$
}%
$$
\sum\limits_{SC[<B>,<C>,<D>\}}[S_{K.<B><C>\perp
<D>}^{.I}-S_{.<B><C>}^{<A>}S_{K.<D><A>}^{.I}]=0,
$$
{\cal
$$
\sum\limits_{SC[<B>,<C>,<D>\}}[S_{<F>.<B><C>\perp
<D>}^{.<A>}-S_{.<B><C>}^{<E>}S_{<F>.<E><A>}^{.<A>}]=0,
$$
}where $\sum\limits_{SC[<B>,<C>,<D>\}}i$s the supersymmetric cyclic sum over
indices $<B>,$ $<C>,<D>.$

As a consequence of a corresponding arrangement of (27) we obtain the
Ricci identities (for simplicity we establish them only for ds--vector
fi\-elds, although they may be written for every ds-tensor field):
$$
{D_{[X}^{(h)}}{D_{Y\}}^{(h)}}hZ=R(hX,hY)hZ+{D_{[hX,hY\}}^{(h)}}%
hZ+\sum\limits_{f=1}^z{D}_{[hX,hY\}}^{(v_f)}hZ,\eqno(32)
$$
$$
{D}_{[X}^{(v_p)}{D_{Y\}}^{(h)}}hZ=R(v_pX,hY)hZ+{D}_{[v_pX,hY\}}^{(h)}hZ+\sum%
\limits_{f=1}^z{D}_{[v_pX,hY\}}^{(v_f)}hZ,
$$
$$
{D}_{[X}^{(v_p)}D_{Y\}}^{(v_P)}=R(v_pX,v_pY)hZ+\sum\limits_{f=1}^z{D}%
_{[v_pX,v_pY\}}^{(v_f)}hZ
$$
and
$$
{D_{[X}^{(h)}}{D_{Y\}}^{(h)}}v_pZ=R(hX,hY)v_pZ+{D_{[hX,hY\}}^{(h)}}%
v_pZ+\sum\limits_{f=1}^z{D}_{[hX,hY\}}^{(v_f)}v_pZ,\eqno(33)
$$
$$
D_{[X}^{(v_f)}D_{Y\}}^{(h)}v_pZ=R(v_fX,hY)v_pZ+\sum%
\limits_{q=1}^zD_{[v_fX,hY\}}^{(v_q)}v_pZ+\sum\limits_{q=1}^zD_{[v_fX,hY%
\}}^{(v_q)}v_pZ,
$$
$$
D_{[X}^{(v_q)}D_{Y\}}^{(v_f)}v_pZ=R(v_qX,v_fY)v_pZ+\sum%
\limits_{s=1}^zD_{[v_fX,v_fY\}}^{(v_s)}v_pZ.
$$
Putting $X={X^I}(u){\frac \delta {\delta x^I}}+{X^{<A>}}(u){\frac \delta
{\partial y^{<A>}}}$ and taking into account the local form of the h- and
v-covariant s-derivatives and (24),(25),(27),(28) we can express
respectively identities (32) and (33) in this form:
$$
{X^{<A>}}_{\mid K\mid L}-{(-)}^{\mid KL\mid }{X^{<A>}}_{\mid L\mid K}=
$$
$$
{{{R_{<B>}}^{<A>}}_{KL}}{X^{<B>}}-{T^H}_{KL}{X^{<A>}}_{\mid H}-{R^{<B>}}_{KL}%
{X^{<A>}}_{\perp <B>},
$$
$$
{X^I}_{\mid K\perp <D>}-{(-)}^{\mid K<D>\mid }{X^I}_{\perp <D>\mid K}=
$$
$$
{P_{H\cdot K<D>}^{\cdot I}}{X^H}-{C^H}_{K<D>}{X^I}_{\mid H}-{P^{<A>}}_{K<D>}{%
X^I}_{\perp <A>},
$$
$$
{X^I}_{\perp <B>\perp <C>}-{(-)}^{\mid <B><C>\mid }{X^I}_{\perp <C>\perp
<B>}=
$$
$$
{S_{H\cdot <B><C>}^{\cdot I}}{X^H}-{S^{<A>}}_{<B><C>}{X^I}_{\perp <A>}
$$
and
$$
{X^{<A>}}_{\mid K\mid L}-{(-)}^{\mid KL\mid }{X^{<A>}}_{\mid L\mid K}=
$$
$$
{{R_{<B>}}^{<A>}}_{KL}{X^{<B>}}-{T^H}_{KL}{X^{<A>}}_{\mid H}-{R^{<B>}}_{KL}{%
X^{<A>}}_{\perp <B>},
$$
$$
{X^{<A>}}_{\mid K\perp <B>}-{(-)}^{\mid <B>K\mid }{X^{<A>}}_{\perp <B>\mid
K}=
$$
$$
{{P_{<B>}}^{<A>}}_{KC}{X^C}-{C^H}_{K<B>}{X^{<A>}}_{\mid H}-{P^{<D>}}_{K<B>}{%
X^{<A>}}_{\perp <D>},
$$
$$
{X^{<A>}}_{\perp <B>\perp <C>}-{(-)}^{\mid <C><B>\mid }{X^{<A>}}_{\perp
<C>\perp <B>}=
$$
$$
{{S_{<D>}}^{<A>}}_{<B><C>}{X^{<D>}}-{S^{<D>}}_{<B><C>}{X^{<A>}}_{\perp <D>}.
$$
We note that the above presented formulas \cite{v96jpa1} generalize for
higher order anisotropy the similar ones for locally anisotropic superspaces
\cite{vlasg}.

\subsection{Cartan structure equations in dvs--bundles}
Let consider a ds--tensor field on $\tilde {{\cal E}}^{<z>}$:
$$
t={t_{<A>}^I}{\delta }_I{\otimes }{\delta ^{<A>}}.
$$
The d-connection 1-forms ${\omega }_J^I$ and ${{\tilde \omega }_{<B>}^{<A>}}$
are introduced as%
$$
Dt=(D{t_{<A>}^I}){\delta }_I{\otimes }{\delta }^{<A>}
$$
with
$$
Dt_{<A>}^I=dt_{<A>}^I+{\omega }_J^I{t_{<A>}^J}-{{\tilde \omega }_{<A>}^{<B>}}%
{t_{<B>}^I}=t_{<A>\mid J}^I{dx^J}+t_{<A>\perp <B>}^I{\delta }y^{<B>}.
$$
For the d-connection 1-forms of a d-connection $D$ on $\tilde {{\cal E}%
}^{<z>}$ defined by ${{\omega }_J^I}$ and ${{\tilde \omega }_{<B>}^{<A>}}$
one holds the following structure equations:
$$
d({d^I})-{d^H}\wedge {\omega }_H^I=-{\Omega },~d{({{\delta }^{<A>}})}-{{%
\delta }^{<B>}}\wedge {{\tilde \omega }_{<B>}^{<A>}}=-{{\tilde \Omega }^{<A>}%
},
$$
$$
d{{\omega }_J^I}-{{\omega }_J^H}\wedge {{\omega }_H^I}=-{{\Omega }_J^I},~d{{%
\tilde \omega }_{<B>}^{<A>}}-{{\tilde \omega }_{<B>}^{<C>}}\wedge {{\tilde
\omega }_{<C>}^{<A>}}=-{{\tilde \Omega }_{<B>}^{<A>}},
$$
in which the torsion 2-forms ${\Omega }^I$ and ${{\tilde \Omega }^{<A>}}$
are given respectively by formulas:
$$
{{\Omega }^I}={\frac 12}{T^I}_{JK}{d^J}\wedge {d^K}+{\frac 12}{C^I}_{J<C>}{%
d^J}\wedge {{\delta }^{<C>}},
$$
$$
{{\tilde \Omega }^{<A>}}={\frac 12}{R^{<A>}}_{JK}{d^J}\wedge {d^K}+{\frac 12}%
{P^{<A>}}_{J<C>}{d^J}\wedge {{\delta }^{<C>}}+{\frac 12}{S^{<A>}}_{<B><C>}{{%
\delta }^{<B>}}\wedge {{\delta }^{<C>}},
$$
and
$$
{{\Omega }_J^I}={\frac 12}{{R_J}^I}_{KH}{d^K}\wedge {d^H}+{\frac 12P}{%
_{J\cdot K<C>}^{\cdot I}}{d^K}\wedge {{\delta }^{<C>}}+{\frac 12S}{_{J\cdot
K<C>}^{\cdot I}}{{\delta }^{<B>}}\wedge {{\delta }^{<C>}},
$$
$$
{{\tilde \Omega }_{<B>}^{<A>}}={\frac 12R}{_{<B>\cdot KH}^{\cdot <A>}}{d^K}%
\wedge {d^H}+%
$$
$$
{\frac 12}{{P_{<B>}}^{<A>}}_{K<C>}{d^K}\wedge {{\delta }^{<C>}}+{\frac 12}{{%
S_{<B>}}^{<A>}}_{<C><D>}{{\delta }^{<C>}}\wedge {{\delta }^{<D>}}.
$$
We have defined the exterior product on s--space to satisfy the property
$$
{{\delta }^{<\alpha >}}\wedge {{\delta }^{<\beta >}}=-{(-)}^{\mid <\alpha
><\beta >\mid }{{\delta }^{<\beta >}}\wedge {{\delta }^{<\alpha >}}.%
$$

\subsection{Metrics in dvs--bundles}

The base $\tilde M$ of dvs--bundle $\tilde {{\cal E}}^{<z>}$is considered to
be a connected and paracompact s--manifold.

A metric structure on the total space $\tilde E^{<z>}$ of a dvs-bundle $%
\tilde {{\cal E}}^{<z>}$is a supersymmetric, second order, covariant
s--tensor field
$$
G=G_{<\alpha ><\beta >}\partial ^{<\alpha >}\otimes \partial ^{<\beta >}
$$
which in every point $u\in \tilde {{\cal E}}^{<z>}$ is given by
nondegenerate supersymmetric matrix $G_{<\alpha ><\beta >}=G({{\partial }%
_{<\alpha >}},{{\partial }_{<\beta >}}){\quad }$ (with nonvanishing
superdeterminant, $sdetG\not =0).$

The metric and N--connection structures on $\tilde {{\cal E}}^{<z>}$ are
compatible if there are satisfied conditions:
$$
G({{\delta }_I},{{\partial }_{<A>}})=0,G(\delta _{A_f},{\partial }%
_{A_p})=0,~z\geq p>f\geq 1,
$$
or, in consequence,

$$
{G_{I<A>}}-{N_I^{<B>}}{h_{<A><B>}}=0,{G}_{A_fA_p}-{N}_{A_f}^{B_p}{h}%
_{A_pB_p}=0,\eqno(34)
$$
where
$$
{G_{I<A>}}=G({{\partial }_I},{{\partial }_{<A>}}),{G}_{A_fA_p}=G({\partial }%
_{A_f},{\partial }_{A_p}).
$$
From (34) one follows
$$
{N_I^{<B>}}={h^{<B><A>}}{G_{I<A>}},~{N}_{A_f}^{A_p}={h}^{A_pB_p}{G}%
_{A_fB_p},...,
$$
where matrices $h^{<A><B>},{h}^{A_pB_p},...$ are respectively s--inverse to
matrices
$$
h_{<A><B>}=G({{\partial }_{<A>}},{{\partial }_{<B>}}),h_{A_pB_p}=G({\partial
}_{A_p},{\partial }_{B_p}).
$$
So, in this case, the coefficients of N-connection are uniquely determined
by the components of the metric on $\tilde {{\cal E}}^{<z>}.$

A compatible with N--connection metric on $\tilde {{\cal E}}^{<z>}$ is
written in irreducible form as
$$
G(X,Y)=G(hX,hY)+G(v_1X,v_1Y)+...+G(v_zX,v_zY),{\quad }X,Y\in {\Xi (\tilde {%
{\cal E}}^{<z>})},
$$
and looks locally as
$$
G=g_{{\alpha }{\beta }}{(u)}{{\delta }^\alpha }\otimes {{\delta }^\beta }%
=g_{IJ}{d^I}\otimes {d^J}+h_{<A><B>}{{\delta }^{<A>}}\otimes {{\delta }^{<B>}%
}=
$$

$$
g_{IJ}{d^I}\otimes {d^J}+h_{A_1B_1}{\delta }^{A_1}\otimes {\delta }%
^{B_1}+h_{A_2B_2}{\delta }^{A_2}\otimes {\delta }^{B_2}+...+h_{A_zB_z}{%
\delta }^{A_z}\otimes {\delta }^{B_z}.\eqno(35)
$$

A d--connection $D$ on $\tilde {{\cal E}}^{<z>}$ is metric, or compatible
with metric $G$, if conditions
$$
{D_{<\alpha >}}{G_{<{\beta ><}{\gamma >}}}=0
$$
are satisfied.

A d--connection $D$ on $\tilde {{\cal E}}^{<z>}$ provided with a metric $G$
is a metric d--con\-nec\-ti\-on if and only if

$$
{D_X^{(h)}}{(hG)}=0,{D_X^{(h)}}{(v}_p{G)}=0,{D}_X^{(v_p)}{(hG)}=0,{D}%
_X^{(v_f)}{(v}_p{G)}=0\eqno(36)
$$
for every $,f,p=1,2,...,z,$ and $X\in {\Xi (\tilde {{\cal E}}^{<z>})}.$
Conditions (36) are written in locally adapted form as
$$
g_{IJ\mid K}=0,g_{IJ\perp <A>}=0,h_{<A><B>\mid K}=0,h_{<A><B>\perp <C>}=0.
$$

In every dvs--bundle provided with compatible N--connection and metric
structures one exists a metric d-connection (called the canonical
d--connection associated to $G)$ depending only on components of G-metric and
N--connection. Its local coefficients $C{\Gamma }=({{\grave L}^I}_{JK},{{%
\grave L}^{<A>}}_{<B>K},{{\grave C}^I}_{J<C>},{{\grave C}^{<A>}}_{<B><C>})$
are as follows:
$$
{{\grave L}^I}_{JK}={\frac 12}{g^{IH}}({{{\delta }_K}g_{HJ}+{{\delta }_J}%
g_{HK}-{{\delta }_H}g_{JK}}),\eqno(37)
$$
$$
{{\grave L}^{<A>}}_{<B>K}={\delta _{<B>}}{N_K^{<A>}}+
$$

$$
{\frac 12}{h^{<A><C>}}[{{{\delta }_{<K>}}{h_{<B><C>}}-(\delta }%
_{<B>}N_K^{<D>})h_{<\dot D><C>}{-{(\delta }_{<C>}N_K^{<D>})h_{<\dot D><B>}}%
],
$$
 $$
{{\grave C}^I}_{J<C>}={\frac 12}{g^{IK}\delta }{_{<C>}}{g_{JK}},
$$
$$
{{\grave C}^{<A>}}_{<B><C>}={\frac 12}{h^{<A><D>}(\delta }_{<C>}h_{<D><B>}+{%
\delta }_{<B>}h_{<D><C>}-{\delta }_{<D>}h_{<B><C>}.
$$
We emphasize that, in general, the torsion of $C\Gamma $--connection (37)
does not vanish.

It should be noted here that on dvs-bundles provided with N-connection and
d-connection and metric really it is defined a multiconnec\-ti\-on 
ds--struc\-tu\-re,
i.e. we can use in an equivalent geometric manner different types of d-
connections with various properties. For example, for modeling of some
physical processes we can use an extension of the Berwald d--connection
$$
B{\Gamma }=({{L^I}_{JK}},\delta _{<B>}N_K^{<A>},0,{C}_{<B><C>}^{<A>}),%
\eqno(38)
$$
where ${L^I}_{JK}={{\grave L}^I}_{JK}$ and ${C^{<A>}}_{<B><C>}={{\grave C}%
^{<A>}}_{<B><C>},$ which is hv-metric, i.e. satisfies conditions:
$$
{D_X^{(h)}}hG=0,...,{D}_X^{(v_p)}v_pG=0,...,{D}_X^{(v_z)}v_zG=0
$$
for every $X\in \Xi {(\tilde {{\cal E}}^{<z>})},$ or in locally adapted
coordinates,
$$
g_{IJ\mid K}=0,h_{<A><B>\perp <C>}=0.
$$

As well we can introduce the Levi--Civita connection%
$$
\{{\frac{<{\alpha >}}{<{{\beta ><}{\gamma >}}}}\}={\frac 12}{G^{<{\alpha ><}{%
\beta >}}({{\partial }_{<\beta >}}{G_{<\tau ><\gamma >}}+{{\partial }%
_{<\gamma >}}{G_{<\tau ><\beta >}}-{{\partial }_{<\tau >}}{G_{<\beta
><\gamma >}})},
$$
constructed as in the Riemann geometry from components of metric $G_{<{%
\alpha ><}{\beta >}}$ by using partial derivations ${{\partial }_{<\alpha >}}%
={\frac \partial {\partial u^{<\alpha >}}}=({\frac \partial {\partial x^I}},{%
\frac \partial {\partial y^{<A>}}}),$ which is metric but not a d-connection.

In our further considerations we shall largely use the Christoffel
d--symbols defined similarly as components of Levi--Civita connection but by
using la--partial de\-ri\-va\-ti\-ons,
$$
{{{\tilde \Gamma }^{<\alpha >}}_{<\beta ><\gamma >}}={\frac 12}{G^{<\alpha
><\tau >}}({{\delta }_{<\beta >}}{G_{<\tau ><\gamma >}}+{{\delta }_{<\gamma
>}}{G_{<\tau ><\beta >}}-{{\delta }_{<\tau >}}{G_{<\beta ><\gamma >}}),%
\eqno(39)
$$
having components
$$
C{\tilde \Gamma }=({L^I}_{JK},0,0,{C^{<A>}}_{<B><C>}),
$$
where coefficients ${L^I}_{JK}$ and ${C^{<A>}}_{<B><C>}$ must be com\-puted
as in formu\-las (38).

We can express arbitrary d--connection as a deformation of the background
d--connection (38):
$$
{{{\Gamma }^{<\alpha >}}_{<\beta ><\gamma >}}={{\tilde \Gamma }_{\cdot
<\beta ><\gamma >}^{<\alpha >}}+{{P^{<\alpha >}}_{<\beta ><\gamma >}},%
\eqno(40)
$$
where ${{P^{<\alpha >}}_{<\beta ><\gamma >}}$ is called the deformation
ds-tensor. Putting splitting (40) into (25) and (29) we can express
torsion ${T^{<\alpha >}}_{<\beta ><\gamma >}$ and curvature\\ ${{R_{<\beta >}%
}^{<\alpha >}}_{<\gamma ><\delta >}$ of a d-connection ${{\Gamma }^{<\alpha
>}}_{<\beta ><\gamma >}$ as respective deformations of torsion ${{\tilde T}%
^{<\alpha >}}_{<\beta ><\gamma >}$ and torsion ${\tilde R}_{<\beta >\cdot
<\gamma ><\delta >}^{\cdot <\alpha >}$ for connection ${{\tilde \Gamma }%
^{<\alpha >}}_{<\beta ><\gamma >}{\quad }:$

$$
{{T^{<\alpha >}}_{<\beta ><\gamma >}}={{\tilde T}_{\cdot <\beta ><\gamma
>}^{<\alpha >}}+{{\ddot T}_{\cdot <\beta ><\gamma >}^{<\alpha >}}
$$
and
$$
{{{R_{<\beta >}}^{<\alpha >}}_{<\gamma ><\delta >}}={{\tilde R}_{<\beta
>\cdot <\gamma ><\delta >}^{\cdot <\alpha >}}+{{\ddot R}_{<\beta >\cdot
<\gamma ><\delta >}^{\cdot <\alpha >}},
$$
where
$$
{{\tilde T}^{<\alpha >}}_{<\beta ><\gamma >}={{\tilde \Gamma }^{<\alpha >}}%
_{<\beta ><\gamma >}-{(-)}^{\mid <\beta ><\gamma >\mid }{{\tilde \Gamma }%
^{<\alpha >}}_{<\gamma ><\beta >}+{w^{<\alpha >}}_{<\gamma ><\delta >},
$$
$$
~{{\ddot T}^{<\alpha >}}_{<\beta ><\gamma >}={{\ddot \Gamma }^{<\alpha >}}%
_{<\beta ><\gamma >}-{(-)}^{\mid <\beta ><\gamma >\mid }{{\ddot \Gamma }%
^{<\alpha >}}_{<\gamma ><\beta >},
$$
and%
$$
{{\tilde R}_{<\beta >\cdot <\gamma ><\delta >}^{\cdot <\alpha >}}={{\delta }%
_{<\delta >}}{{\tilde \Gamma }^{<\alpha >}}_{<\beta ><\gamma >}-{(-)}^{\mid
<\gamma ><\delta >\mid }{{\delta }_{<\gamma >}}{{\tilde \Gamma }^{<\alpha >}}%
_{<\beta ><\delta >}+
$$
$$
{{{\tilde \Gamma }^{<\varphi >}}_{<\beta ><\gamma >}}{{{\tilde \Gamma }%
^{<\alpha >}}_{<\varphi ><\delta >}}-{(-)}^{\mid <\gamma ><\delta >\mid }{{{%
\tilde \Gamma }^{<\varphi >}}_{<\beta ><\delta >}}{{{\tilde \Gamma }%
^{<\alpha >}}_{<\varphi ><\gamma >}}+{{\tilde \Gamma }^{<\alpha >}}_{<\beta
><\varphi >}{w^{<\varphi >}}_{<\gamma ><\delta >},
$$
$$
{{\ddot R}_{<\beta >\cdot <\gamma ><\delta >}^{\cdot <\alpha >}}={{\tilde D}%
_{<\delta >}}{{P^{<\alpha >}}_{<\beta ><\gamma >}}-{(-)}^{\mid <\gamma
><\delta >\mid }{{\tilde D}_{<\gamma >}}{{P^{<\alpha >}}_{<\beta ><\delta >}}%
+
$$
$$
{{P^{<\varphi >}}_{<\beta ><\gamma >}}{{P^{<\alpha >}}_{<\varphi ><\delta >}}%
-{(-)}^{\mid <\gamma ><\delta >\mid }{{P^{<\varphi >}}_{<\beta ><\delta >}}{{%
P^{<\alpha >}}_{<\varphi ><\gamma >}}
$$
$$
+{{P^{<\alpha >}}_{<\beta ><\varphi >}}{{w^{<\varphi >}}_{<\gamma ><\delta >}%
},
$$
the nonholonomy coefficients ${w^{<\alpha >}}_{<\beta ><\gamma >}$ are
defined as
$$
[{\delta }_{<\alpha >},{\delta }_{<\beta >}\}={{\delta }_{<\alpha >}}{{%
\delta }_{<\beta >}}-{(-)}^{|<\alpha ><\beta >|}{{\delta }_{<\beta >}}{{%
\delta }_{<\alpha >}}={w^{<\tau >}}_{<\alpha ><\beta >}{{\delta }_{<\tau >}}.%
$$

We emphasize that if from geometric point of view all considered
d--con\-nect\-i\-ons are ''equ\-al in rights'', the con\-struct\-ion of
physical models on la--spaces requires an explicit fixing of the type of
d--con\-nect\-ion and metric structures.

\section{Higher Order Tangent S--bun\-dles}

The aim of this section is to present a study of supersymmetric extensions
from $\widetilde{M}$ to $T\tilde M$ and $Osc^{(z)}\widetilde{M}$ and to
consider corresponding prolongations of Riemann and ge\-ne\-ra\-li\-zed
Fins\-ler
structures (on classical and new approaches to Finsler geometry, its
generalizations and applications in physics se, for example, \cite
{fin,car,run,ma87,ma94,asa,asa88,mat,mk,am,az94,bog,bej}).

The presented in the previous section basic results on dvs-bundles ${\tilde {%
{\cal E}}^{<z>}}$ pro\-vid\-ed with N-connection, d-connection and metric
structures can be correspondingly adapted to the osculator s--bundle $\left(
Osc^z\tilde M,\pi ,\tilde M\right) .$ In this case the dimension of the base
space and typical higher orders fibre coincides and we shall not distinguish
indices of geometrical objects.

Coefficients of a d--connection $D\Gamma
(N)=(L_{JM}^I,C_{(1)JM}^I,...,C_{(z)JM}^I)$ in $Osc^z\tilde M,$ with respect
to a la--base are introduced as to satisfy equations%
$$
D_{\frac \delta {\delta x^I}}\frac \delta {\delta y_{(f)}^I}=L_{IJ}^M\frac
\delta {\delta y_{(f)}^M},~D_{\frac \delta {\delta y_{(p)}^J}}\frac \delta
{\delta y_{(f)}^I}=C_{(p)IJ}^M\frac \delta {\delta y_{(f)}^M}, \eqno(41)%
$$
$$
(f=0,1,...,z;p=1,...,z,\mbox{ and }y_{(0)}^I=x^I).
$$

A metric structure on $Osc^z\tilde M$ is ds--tensor s--symmetric field $%
g_{IJ}(u_{(z)})=g_{IJ}(x,y_{(1)},y_{(2)},...,y_{(z)})$ of type $%
(0,2),srank|g_{ij}|=(n,m).$ The N--lift of Sasaki type of $g_{IJ}$ is given
by (see (35)) defines a global Riemannian s--structure (if $\widetilde{M}$
is a s--differentiable, paracompact s-manifold):%
$$
G=g_{IJ}(u_{(z)})dx^I\otimes dx^J+g_{IJ}(u_{(z)})dy_{(1)}^I\otimes
dy_{(1)}^J+...+g_{IJ}(u_{(z)})dy_{(z)}^I\otimes dy_{(z)}^J.\eqno(42)
$$
The condition of compatibility of a d--connection (41) with metric (42)
is expressed as
$$
D_XG=0,\forall X\in \Xi (Osc^z\tilde M),
$$
or, by using d--covariant partial derivations $|_{(p)}$ defined by
coefficients\\ $(L_{JM}^I,C_{(1)JM}^I,...,C_{(z)JM}^I),$
$$
g_{IJ|M=0,~}g_{IJ|_{(p)M}}=0,(p=1,...,z).
$$
An example of compatible with metric d--connection is given by Christof\-fel
d--sym\-bols (see (39)):%
$$
L_{IJ}^M=\frac 12g^{MK}\left( \frac{\delta g_{KJ}}{\partial x^I}+\frac{%
\delta g_{IK}}{\partial x^J}-\frac{\delta g_{IJ}}{\partial x^K}\right) ,%
$$
$$
C_{(p)IJ}^M=\frac 12g^{MK}\left( \frac{\delta g_{KJ}}{\partial y_{(p)}^I}+%
\frac{\delta g_{IK}}{\partial y_{(p)}^J}-\frac{\delta g_{IJ}}{\partial
y_{(p)}^K}\right) ;p=1,2,...,z.
$$

\subsection{Supersymmetric ex\-ten\-si\-ons of Fins\-ler spaces}

We start our considerations with the ts-bundle $T\tilde M.$ An s-vector $%
X\in \Xi (T\tilde M)$ is decomposed with respect to la--bundles as%
$$
X=X(u)^I\delta _I+Y(u)^I\partial _I,
$$
where $u=u^\alpha =(x^I,y^J)$ local coordinates. The s--tangent structures
(16) are transformed into a global map

$$
J:\Xi (T\tilde M)\to \Xi (T\tilde M)
$$
which does not depend on N-connection structure:
$$
J({\frac \delta {\delta x^I}})={\frac \partial {\partial y^I}}
$$
and%
$$
J({\frac \partial {\partial y^I}})=0.
$$
This endomorphism is called the natural ( or canonical ) almost tangent
structure on $T\tilde M;$ it has the properties:
$$
1)J^2=0,{\quad }2)ImJ=KerJ=VT\tilde M
$$
and 3) the Nigenhuis s--ten\-sor,
$$
{N_J}(X,Y)=[JX,JY\}-J[JX,Y\}-J[X,JY]
$$
$$
(X,Y\in \Xi (TN))
$$
identically vanishes, i.e. the natural almost tangent structure $J$ on $%
T\tilde M$ is integrable.

A generalized Lagrange superspace, GLS--space, is a pair\\ ${GL}%
^{n,m}=(\tilde M,g_{IJ}(x,y))$, \quad where \quad $g_{IJ}(x,y)$ \quad is a
ds--tensor field on \\ \quad ${{\tilde {T{\tilde M}}}=T\tilde M-\{0\}},$%
\quad s--symmetric of superrank $(n,m).$

We call $g_{IJ}$ as the fundamental ds--tensor, or metric ds--tensor, of
GLS-space.

There exists an unique d-connection $C\Gamma (N)$ which is compatible with $%
g_{IJ}{(u)}$ and has vanishing torsions ${T^I}_{JK}$ and ${S^I}_{JK}$ (see
formulas (25) rewritten for ts-bundles). This connection, depending only
on $g_{IJ}{(u)}$ and ${N_J^I}{(u)}$ is called the canonical metric
d-connection of GLS-space. It has coefficients
$$
{L^I}_{JK}={\frac 12}{g^{IH}}({\delta }_J{g_{HK}}+{\delta }_H{g_{JK}}-{%
\delta }_H{g_{JK}}),
$$
$$
{C^I}_{JK}={\frac 12}{g^{IH}}({\partial }_J{g_{HK}}+{\partial }_H{g_{JK}}-{%
\partial }_H{g_{JK}}).
$$
There is a unique normal d-connection $D\Gamma (N)=({\bar L}_{\cdot JK}^I,{%
\bar C}_{\cdot JK}^I)$ which is metric and has a priori given torsions ${T^I}%
_{JK}$ and ${S^I}_{JK}.$ The coefficients of $D\Gamma (N)$ are the following
ones:
$$
{\bar L}_{\cdot JK}^I={L^I}_{JK}-\frac 12g^{IH}(g_{JR}{T^R}_{HK}+g_{KR}{T^R}%
_{HJ}-g_{HR}{T^R}_{KJ}),
$$
$$
{\bar C}_{\cdot JK}^I={C^I}_{JK}-\frac 12g^{IH}(g_{JR}{S^R}_{HK}+g_{KR}{S^R}%
_{HJ}-g_{HR}{S^R}_{KJ}),
$$
where ${L^I}_{JK}$ and ${C^I}_{JK}$ are the same as for the $C\Gamma (N)$%
--connection (41).

The Lagrange spaces were introduced \cite{ker} in order to geometrize the
concept of Lagrangian in mechanics (the Lagrange geometry is studied in
details in \cite{ma87,ma94}). For s-spaces we present this generalization:

A Lagrange s--space, LS--space, $L^{n,m}=(\tilde M,g_{IJ}),$ is defined as a
particular case of GLS-space when the ds--metric on $\tilde M$ can be
expressed as
$$
g_{IJ}{(u)}={\frac 12}{\frac{{\partial }^2L}{{{\partial y^I}{\partial y^J}}}}%
,\eqno(43)
$$
where $L:T\tilde M\to \Lambda ,$ is a s-differentiable function called a
s-Lagrangian on $\tilde M.$

Now we consider the supersymmetric extension of Fins\-ler space:
A Finsler s--metric on $\tilde M$ is a function $F_S:T\tilde M\to \Lambda $
having the properties:

1. The restriction of $F_S$ to ${\tilde {T\tilde M}}=T\tilde M\setminus
\{0\} $ is of the class $G^\infty $ and F is only supersmooth on the image
of the null cross--section in the ts-bundle to $\tilde M.$

2. The restriction of F to ${\tilde {T\tilde M}}$ is positively homogeneous
of degree 1 with respect to ${(y^I)}$, i.e. $F(x,{\lambda }y)={\lambda }%
F(x,y),$ where ${\lambda }$ is a real positive number.

3. The restriction of F to the even subspace of $\tilde {T\tilde M}$ is a
positive function.

4. The quadratic form on ${\Lambda }^{n,m}$ with the coefficients
$$
g_{IJ}{(u)}={\frac 12}{\frac{{\partial }^2F^2}{{{\partial y^I}{\partial y^J}}%
}}
$$
defined on $\tilde {T\tilde M}$ is nondegenerate.

A pair $F^{n,m}=(\tilde M,F)$ which consists from a supersmooth
s-ma\-ni\-fold $\tilde M$ and a Finsler s--metric is called a Finsler
superspace, FS--space.

It's obvious that FS--spaces form a particular class of LS--spaces with
s-Lagran\-gi\-an $L={F^2}$ and a particular class of GLS--spaces with metrics
of type (58).

For a FS--space we can introduce the supersymmetric variant of nonlinear
Car\-tan con\-nec\-ti\-on \cite{car,run} :
 $$
N_J^I{(x,y)}={\frac \partial {\partial y^J}}G^{*I},
$$
where
$$
G^{*I}={\frac 14}g^{*IJ}({\frac{{\partial }^2{\varepsilon }}{{\partial y^I}{%
\partial x^K}}}{y^K}-{\frac{\partial {\varepsilon }}{\partial x^J}}),{\quad }%
{\varepsilon }{(u)}=g_{IJ}{(u)}y^Iy^J,
$$
and $g^{*IJ}$ is inverse to $g_{IJ}^{*}{(u)}={\frac 12}{\frac{{\partial }%
^2\varepsilon }{{{\partial y^I}{\partial y^J}}}}.$ In this case the
coefficients of canonical metric d-connection (25) gives the
supersymmetric variants of coefficients of the Cartan connection of Finsler
spaces. A similar remark applies to the Lagrange superspaces.

\subsection{Higher order prolongations of
Riemann,Finsler and Lagrange s--spaces}

The geometric constructions on $T\widetilde{M}$ from the previus subsection
have corresponding generalizations to the $Osc^{(z)}\widetilde{M}$
s--bundle. The basic idea is similar to that used for prolongations of
geometric structures (see \cite{morim} for prolongations on tangent bundle).
Having defined a metric structure $g_{IJ}(x)$ on a s-manifold $\widetilde{M}$
we can extend it to the $Osc^z\tilde M$ s--bundle by considering $%
g_{IJ}(u_{(z)})=g_{IJ}(x)$ in (42). R. Miron and Gh. Atanasiu \cite{mirata}
solved the problem of prolongations of Finsler and Lagrange structures on
osculator bundle. In this subsection we shall analyze supersymmetric
extensions of Finsler and Lagrange structures as well present a brief
introduction into geometry of higher order Lagrange s-spaces.

Let $F^{n,m}=(\tilde M,F)$ be a FS--space with the fundamental function $%
F_S:T\tilde M\to \Lambda $ on $\widetilde{M}.$ A prolongation of $F$ on $%
Osc^z\tilde M$ is given by a map%
$$
(F\circ \pi _1^z)(u_{(z)})=F(u_{(1)})
$$
and corresponding fundamental tensor
$$
g_{IJ}(u_{(1)})=\frac 12\frac{\partial ^2F^2}{\partial y_{(1)}^I\partial
y_{(1)}^J},
$$
for which
$$
(g_{IJ}\circ \pi _1^z)(u_{(z)})=g_{IJ}(u_{(1)}).
$$
So, $g_{IJ}(u_{(1)})$ is a ds--tensor on\\ $\widetilde{Osk^z\widetilde{M}}%
=Osc^z\tilde M/\{0\}=\{(u_{(z)})\in Osc^z\tilde M,srank|y_{(1)}^I|=1\}.$

The Christoffel d--symbols
$$
\gamma _{IJ}^M(u^{(1)})=\frac 12g^{MK}(u_{(1)})(\frac{\partial
g_{KI}(u_{(1)})}{\partial x^J}+\frac{\partial g_{JK}(u_{(1)})}{\partial x^I}-%
\frac{\partial g_{IJ}(u_{(1)})}{\partial x^K})
$$
define the Cartan nonlinear connection \cite{car35}:%
$$
G_{(N)J}^I=\frac 12\frac \partial {\partial y_{(1)}^J}(\gamma
_{KM}^Iy_{(1)}^Ky_{(1)}^M).\eqno(44)
$$
The dual coefficients for the N-connection (21) are recurrently computed
by using (44) and operator
$$
\Gamma =y_{(1)}^I\frac \partial {\partial x^I}+2~y_{(2)}^I\frac \partial
{\partial y_{(1)}^I}+...+z~y_{(z)}^I\frac \partial {\partial y_{(z-1)}^I},
$$
$$
M_{(1)J}^I=G_{(N)J}^I,
$$
$$
M_{(2)J}^I=\frac 12[\Gamma G_{(N)J}^I+G_{(N)K}^IM_{(1)J}^K],
$$
$$
..............
$$
$$
M_{(z)J}^I=\frac 1z[\Gamma M_{(z-1)J}^I+G_{(N)K}^IM_{(z-1)J}^K].
$$

The prolongations of FS--spaces can be generalized for Lagrange s--spaces
(on Lagrange spaces and theirs higher order extensions see \cite
{ma87,ma94,mirata} and on supersymmetric extensions of Finsler geometry see
\cite{vlasg}). Let $L^{n,m}=(\tilde M,g_{IJ})$ be a Lagrange s--space. The
Lagrangian $L:T\widetilde{M}\rightarrow \Lambda \,$ can be extended on $%
Osc^z\tilde M$ by using maps of the Lagrangian, $(L\circ \pi
_1^z)(u_{(z)})=L\left( u_{(1)}\right) ,$ and, as a consequence, of the
fundamental tensor (43), $(g_{IJ}\circ \pi _1^z)(u_{(z)})=g_{IJ}\left(
u_{(1)}\right) .$

\subsection{Higher order Lagrange s--spaces}

We introduce the notion of Lagrangian of z--order on a differentiable
s--manifold $\widetilde{M}$ as a map $L^z:Osc^z\tilde M$ $\rightarrow
\Lambda .$ In order to have concordance with the definitions proposed by
\cite{mirata} we require the even part of the fundamental ds--tensor to be
of constant signature. Here we also note that questions to considered in
this subsection, being an supersymmetric approach, are connected with the
problem of elaboration of the so--called higher order analytic mechanics
(see, for instance, \cite{cram,lib,leo,sau}).

A Lagrangian s--differentiable of order $z$ ($z=1,2,3,...)$ on
s-differentiable s--manifold $\widetilde{M}$ is an application $L^{(z)}:Osc^z%
\widetilde{M}\rightarrow \Lambda ,$ s--differentiable on $\widetilde{Osk^z%
\widetilde{M}}$ and smooth in the points of $Osc^z\widetilde{M}$ where $%
y_{(1)}^I=0.$

It is obvious that
$$
g_{IJ}(x,y_{(1)},...,y_{(z)})=\frac 12\frac{\partial ^2L^{(z)}}{\partial
y_{(z)}^I\partial y_{(z)}^J}
$$
is a ds--tensor field because with respect to coordinate transforms (3)
one holds transforms%
$$
K_I^{I^{\prime }}K_J^{J^{\prime }}g_{I^{\prime }J^{\prime }}=g_{IJ.}
$$

A Lagrangian $L$ is regular if $srank|g_{IJ}|=(n,m).$

A Lagrange s--space of $z$--order is a pair $L^{(z,n,m)}=(\widetilde{M}%
,L^{(z)}),$ where $L^{(z)}$ is a s--differentiable regular Lagrangian of $z$%
--order, and with ds--tensor $g_{IJ}$ being of constant signature on the
even part of the basic s--manifold.

For details on nonsupersymmetric osculator bundles we cite \cite{mirata}

\section{Superstrings in Higher Order Anisotropic S--Spaces}

This section considers the basic formalism for superstrings in dvs--bundles.
We shall begin our study with nonsupersymmetric two dimensional higher order
anisotropic sigma models. Then we shall analyze supersimmetric extensions
and locally anisotropic generalizations of the Green--Schvarz action.

\subsection{Two dimensional higher order an\-isotropic sigma s--models}

Let $\widehat{{\cal E}}^{<z>}$ be a higher order anisotropic space (not
superspace) with coordinates $\widehat{u}^{<\alpha >}=\widehat{u}(z)=(%
\widehat{x}^i=\widehat{x}(z),\widehat{y}^{<a>}=\widehat{y}%
(z))=(x^i,y^{a_1},....,y^{a_p},...,y^{a_z)},$ d--metric $\widehat{g}%
_{<\alpha ><\beta >};$ we use denotation $\left( N_2,\gamma _{\ddot a\ddot
e}\right) $ for a two dimensional world sheet with metric $\gamma _{\ddot
a\ddot e}(z^{\ddot u})$ of signature (+,-) and local coordinates $z=z^{\ddot
u},$ where $\ddot a,\ddot e,\ddot u,...=1,2.$

The action of a bosonic string in a dv--bundle $\widehat{{\cal E}}^{<z>}$ is
postulated as
$$
I_\sigma =\frac 1{\lambda ^2}\int d^2z\{\frac 12\sqrt{\gamma }\gamma ^{\ddot
a\ddot e}\partial _{\ddot a}\widehat{u}^{<\alpha >}(z)\partial _{\ddot e}%
\widehat{u}^{<\beta >}(z)\},\eqno(45)
$$
which defines the so called two dimensional sigma model ($\sigma $--model)
with d--metric $\widehat{g}_{<\alpha ><\beta >}$ in higher order
anisotropic spaces (dv--bundles with N--connec\-ti\-on) and $\lambda $ being
constant. A detailed study of different modifications of the
model (45) is given in  also \cite{vlags,vlagsr}. Here we shall
consider a supersymmetric generalization of the string action by applying
the techniques of two dimensional (1,1)--supersymmetry by changing of scalar
fields $u(z)$ into real {\sf N=1} s--fields (without constraints; for
locally isotropic constructions see \cite{bag,brad,cz,gat,how,oliv}) $%
\widehat{u}(z,\theta )$ which are polynoms with respect to Maiorana
anticommuting spinor coordinate $\theta :$%
$$
\widehat{u}^{<\alpha >}(z,\theta )=\widehat{u}^{<\alpha >}(z)+\overline{%
\theta }\lambda ^{<\alpha >}(z)+\frac 12\overline{\theta }\theta F^{<\alpha
>}(z).\eqno(46)
$$

We adopt next conventions and denotations with respect to two--dimension
Dirac matrices $\gamma _{\ddot a}$ and matrix of charge conjugation $C:$%
$$
\{\gamma _{\ddot a},\gamma _{\ddot e}\}=2\eta _{\ddot a\ddot e}{\bf 1,~}%
tr(\gamma _{\ddot e}\gamma _{\ddot a})=2\eta _{\ddot e\ddot a},~(\gamma
_5)^2=1;
$$
$$
\widehat{\partial }=\gamma ^{\ddot e}\partial _{\ddot e};~C\gamma _{\ddot
a}^TC^{-1}=-\gamma _{\ddot a},~C=-C^T=C^{-1}.
$$
For Maiorana spinors $\theta _{\tilde a},\chi _{\tilde n},...$ one holds
relations
$$
\overline{\theta }=\theta ^{+}\gamma ^0,~\overline{\theta ^{\tilde a}}%
=C^{\tilde a\tilde n}\theta _{\tilde n},
$$
$$
\overline{\theta }\chi =\overline{\chi }\theta ,\overline{\theta }\gamma
_{\ddot a}\chi =-\overline{\chi }\gamma _{\ddot a}\theta ,\overline{\theta }%
\gamma _5\chi =-\overline{\chi }\gamma _5\theta .
$$

Let introduce in the two dimensional (1,1)--superspace the covariant
derivations%
$$
D_{\tilde n}=\frac \partial {\partial \overline{\theta }^{\tilde n}}-i(%
\widehat{\partial }\theta )_{\tilde n},
$$
satisfying algebra%
$$
\{D_{\tilde n},D_{\tilde o}\}=2i(\widehat{\partial }C)_{\tilde n\tilde
o}\equiv 2i\partial _{\tilde n\tilde o}
$$
and the integration measure on anticommuting variables with properties%
$$
\int d\theta _{\tilde n}=0,\int d\theta _{\tilde n}\theta ^{\tilde a}=\delta
_{\tilde n}^{\tilde a},\frac 1{2i}\int d^2\theta (\overline{\theta }\theta
)=1.
$$

The (1,1)--supersymmetric generalization of (45) in terms of s--fields
(46) is written as%
$$
I_{\sigma S}=\frac 1{8i\pi \alpha ^{\prime }}\int d^2z\int d^2\theta \{%
\widehat{g}_{<\alpha ><\beta >}(\widehat{u})-\widehat{b}_{<\alpha ><\beta >}(%
\widehat{u})\}\overline{D}\widehat{u}^{<\alpha >}(1+\gamma _5)D\widehat{u}%
^{<\beta >},
$$
where $\lambda ^2=2\pi \alpha ^{\prime }$ which is a higher order
anisotropic generalization of the Curtright--Zachos \cite{cz} nonlinear
sigma model. Integrating on $\theta \,$ and excluding auxiliary fields $%
F^{<\alpha >}$ according to theirs algebraic equations we obtain from the
last expression:%
$$
I_{\sigma S}=\frac 12\int d^2z\{\widehat{g}_{<\alpha ><\beta >}\partial
^{\ddot e}\widehat{u}^{<\alpha >}\partial _{\ddot e}\widehat{u}^{<\beta >}+%
\eqno(47)
$$
$$
\varepsilon ^{\ddot e\ddot \imath }\widehat{b}_{<\alpha ><\beta >}\partial
_{\ddot e}\widehat{u}^{<\alpha >}\partial _{\ddot \imath }\widehat{u}%
^{<\beta >}+i\widehat{g}_{<\alpha ><\beta >}\overline{\lambda }^{<\alpha
>}\gamma ^{\ddot e}\widehat{D}_{\ddot e}^{(-)}\lambda ^{<\beta >}+
$$
$$
\frac 18\widehat{R}_{<\beta ><\alpha ><\gamma ><\delta >}^{(-)}\overline{%
\lambda }^{<\alpha >}(1+\gamma _5)\lambda ^{<\gamma >}\overline{\lambda }%
^{<\beta >}(1+\gamma _5)\lambda ^{<\delta >}\},
$$
where
$$
\widehat{D}_{\ddot e}^{(\pm )}\lambda ^{<\beta >}=[\delta _{<\alpha
>}^{<\beta >}\partial _{\ddot e}+\widehat{\widetilde{\Gamma }}_{<\alpha
><\gamma >}^{<\beta >}\partial _{\ddot e}\widehat{u}^{<\gamma >}\pm \widehat{%
B}_{<\alpha ><\gamma >}^{<\beta >}\varepsilon _{\ddot e\ddot o}\partial
^{\ddot o}\widehat{u}^{<\gamma >}]\lambda ^{<\gamma >},
$$
$\widehat{\widetilde{\Gamma }}_{<\alpha ><\gamma >}^{<\beta >}$ are
Christoffel d--symbols\ (39) on dv--bundle. In order to have compatible
with the N--connection structure motions of la--strings we consider
\cite{vlags,vlagsr} these relations between ds--tensor $b_{<\alpha ><\beta >},$
strength\\ $\widehat{B}_{<\alpha ><\beta ><\gamma >}=\delta _{[<\alpha >}%
\widehat{b}_{<\beta ><\gamma >\}}$ and torsion $T_{<\alpha ><\gamma
>}^{<\beta >}$ (see (29)):%
$$
\delta _{<\alpha >}\widehat{b}_{<\beta ><\gamma >}=\widehat{g}_{<\alpha
><\delta >}\widehat{T}_{<\beta ><\gamma >}^{<\delta >},\eqno(48)
$$
with s--integrability conditions
$$
\Omega _{a_pa_s}^{a_f}\delta _{a_f}\widehat{b}_{<\beta ><\gamma >}=\delta
_{[a_h}\widehat{T}_{a_s\}<\beta ><\gamma >},~(f<p,s;p,s=0,1,...,z),\eqno(49)
$$
where $\Omega _{a_pa_s}^{a_f}$ are the coefficients of the N--connection
curvature (11). In this case we can express $\widehat{B}_{<\alpha ><\beta
><\gamma >}=\widehat{T}_{[<\alpha ><\beta ><\gamma >\}}.$ Conditions (48)
and (49) define a model of higher order anisotropic superstrings when the $%
\sigma $--modes s--antisymmetric strength is introduced from the higher
order anisotropic background torsion. More general constructions are
possible by using normal coordinates locally adapted to both N--connection
and torsion structures on background s--spaces. For simplicity, we omit such
considerations in this work.

Ds--tensor $\widehat{R}_{<\alpha ><\beta ><\gamma ><\delta >}$ from (47)
denotes the curvature with torsion $B:$%
$$
\widehat{R}_{<\beta ><\alpha ><\gamma ><\delta >}^{(\pm )}[\widehat{\Gamma }%
^{(\pm )}]=\widetilde{R}_{<\beta ><\alpha ><\gamma ><\delta >}\mp
$$
$$
D_{<\gamma >}\widehat{B}_{<\alpha ><\beta ><\delta >}\pm D_{<\delta >}%
\widehat{B}_{<\alpha ><\beta ><\gamma >}+
$$
$$
\widehat{B}_{<\tau ><\alpha ><\gamma >}\widehat{B}_{<\delta ><\beta
>}^{<\tau >}-\widehat{B}_{<\tau ><\alpha ><\delta >}\widehat{B}_{<\gamma
><\beta >}^{<\tau >},
$$
where $\widetilde{R}_{<\beta ><\alpha ><\gamma ><\delta >}$ is the curvature
of the torsionless Christoffel d--symbols (39),%
$$
\widehat{\Gamma }_{<\beta ><\gamma >}^{<\alpha >(\pm )}=\widehat{g}^{<\alpha
><\tau >}\widehat{\Gamma }_{<\tau ><\beta ><\gamma >}^{(\pm )},\ \widehat{%
\Gamma }_{<\tau ><\beta ><\gamma >}^{(\pm )}=\widehat{\widetilde{\Gamma }}%
_{<\tau ><\beta ><\gamma >}\pm B_{<\tau ><\beta ><\gamma >}^{(\pm )},
$$
$$
\widetilde{\Gamma }_{<\tau ><\beta ><\gamma >}=\frac 12(\delta _{<\beta >}%
\widetilde{g}_{<\tau ><\gamma >}+\delta _{<\gamma >}\widetilde{g}_{<\beta
><\tau >}-\delta _{<\tau >}\widetilde{g}_{<\beta ><\gamma >}).
$$

In order to define a locally supersymmetric generalization of the model
(47) we consider a supersymmetric calculus of the set of (1,1)--multiplets
of higher order anisotropic matter ($\widehat{\varphi }^{<\alpha >},\widehat{%
\lambda }^{<\alpha >}=\lambda ^{<\alpha >})$ with the multiplet of
(1,1)--supergravity $\left( e_{\ddot e}^{\underline{\ddot e}},\psi _{\ddot
e}\right) $ in two dimensions (see $\cite{ketnp,ket,ketn}$ for locally
isotropic constructions).

The global supersymmetric variant of action (47), for $2\pi \alpha ^{\prime
}=1,$ is written as
$$
I_0[\varphi ,\lambda ]=\frac 12\int d^2z[\widehat{g}_{<\alpha ><\beta
>}\partial ^{\underline{\ddot e}}\widehat{u}^{<\alpha >}\partial _{%
\underline{\ddot e}}\widehat{u}^{<\beta >}+\widehat{b}_{<\alpha ><\beta
>}\varepsilon ^{\underline{\ddot e}\ \underline{\ddot a}}\partial _{%
\underline{\ddot e}}\widehat{u}^{<\alpha >}\partial _{\underline{\ddot a}}%
\widehat{u}^{<\beta >}+\eqno(50)
$$
$$
i\widehat{g}_{<\alpha ><\beta >}\overline{\lambda }^{<\alpha >}\gamma ^{%
\underline{\ddot e}}(D_{\underline{\ddot e}}\lambda )^{<\beta >}+i\widehat{B}%
_{<\alpha ><\beta ><\gamma >}\overline{\lambda }^{<\alpha >}\gamma _5\gamma
^{\underline{\ddot e}}(\partial _{\underline{\ddot e}}\widehat{u}^{<\beta
>})\lambda ^{<\gamma >}+
$$
$$
\frac 16\widehat{\widetilde{R}}_{<\beta ><\alpha ><\gamma ><\delta >}(%
\overline{\lambda }^{<\alpha >}\lambda ^{<\gamma >})(\overline{\lambda }%
^{<\beta >}\lambda ^{<\delta >})-
$$
$$
\frac 14\widehat{\widetilde{D}}_{<\varepsilon >}B_{<\alpha ><\beta ><\tau >}(%
\overline{\lambda }^{<\alpha >}\gamma _5\lambda ^{<\beta >})(\overline{%
\lambda }^{<\tau >}\lambda ^{<\varepsilon >})-
$$
$$
\frac 14\widetilde{B}_{<\alpha ><\beta ><\tau >}\widetilde{B}_{<\gamma
><\delta >}^{<\tau >}(\overline{\lambda }^{<\alpha >}\gamma _5\lambda
^{<\beta >})(\overline{\lambda }^{<\gamma >}\gamma _5\lambda ^{<\delta >})],
$$
where covariant derivation $\widehat{\widetilde{D}}_{<\varepsilon >}$ is
defined by torsionless Christoffel d--symbols.

The action (50) is invariant under global s--transforms with Maiorana
spinor parameter $\varepsilon :$%
$$
\bigtriangleup \widehat{\varphi }^{<\alpha >}=\overline{\varepsilon }\
\lambda ^{<\alpha >},
$$
$$
\bigtriangleup \lambda ^{<\alpha >}=-i(\widehat{\partial }\widehat{\varphi }%
^{<\alpha >})\varepsilon +\frac \varepsilon 2(\widehat{\widetilde{\Gamma }}%
_{<\beta ><\gamma >}^{<\alpha >}\overline{\lambda }^{<\beta >}\lambda
^{<\gamma >}-\widetilde{B}_{<\beta ><\gamma >}^{<\alpha >}\overline{\lambda }%
^{<\beta >}\gamma _5\lambda ^{<\gamma >}).
$$

Defining Maiorana--Weyl spinors $\lambda _{\pm }^{<\alpha >}$ (MW--spinors)\
instead of $\lambda ^{<\alpha >}$ we can rewrite the action (50) in a more
convenient form:%
$$
I_0[\varphi ,\lambda _{\pm }]=\frac 12\int d^2z[\widetilde{g}_{<\alpha
><\beta >}\partial ^{\underline{\ddot e}}\widetilde{u}^{<\alpha >}\partial _{%
\underline{\ddot e}}\widetilde{u}^{<\beta >}+\widehat{b}_{<\alpha ><\beta
>}\varepsilon ^{\underline{\ddot e}\ \underline{\ddot a}}\partial _{%
\underline{\ddot e}}\widehat{u}^{<\alpha >}\partial _{\underline{\ddot a}}%
\widehat{u}^{<\beta >}+\eqno(50a)
$$
$$
i\widehat{g}_{<\alpha ><\beta >}\overline{\lambda }_{+}^{<\alpha >}(\widehat{%
D}^{+}\lambda _{+})^{<\beta >}+i\widehat{g}_{<\alpha ><\beta >}\overline{%
\lambda }_{-}^{<\alpha >}(\widehat{D}^{-}\lambda _{-})^{<\beta >}+
$$
$$
\frac 14\widehat{\widetilde{R}}_{<\beta ><\alpha ><\gamma ><\delta >}^{+}(%
\overline{\lambda }^{<\alpha >}\gamma _5\lambda ^{<\beta >})(\overline{%
\lambda }^{<\gamma >}\gamma _5\lambda ^{<\delta >})],
$$
with s--symmetric transformation law%
$$
\bigtriangleup \varphi ^{<\alpha >}=\bigtriangleup _{+}\varphi ^{<\alpha
>}+\bigtriangleup _{-}\varphi ^{<\alpha >}=\overline{\varepsilon }_{+}\
\lambda _{-}^{<\alpha >}+\overline{\varepsilon }_{-}\ \lambda _{+}^{<\alpha
>},
$$
$$
\bigtriangleup \lambda _{\pm }^{<\alpha >}=-i(\widehat{\partial }u^{<\alpha
>})\varepsilon _{\mp }-\widetilde{\Gamma }_{<\beta ><\gamma >}^{<\alpha
>(\pm )}\lambda _{\pm }^{<\beta >}\bigtriangleup _{\pm }\varphi ^{<\gamma
>}.
$$
For simplicity,  we shall omit ''hats'' on
geometrical objects if ambiguities connected with indices for manifolds and
supermanifolds will not arise.

In string theories one considers variations of actions of type (50) with
respect to s--symmetric transformation laws and decompositions with respect
to powers of $\lambda .$ Coefficients proportional to $\lambda ^5$ vanishes
because they do not contain derivations of $\varepsilon (z)$--parameters. In
order to compensate the therms proportional to $\lambda $ and $\lambda ^3$
one adds the so--called Nether term
$$
I^{(N)}=\frac 12\int d^2z[2g_{<\alpha ><\beta >}(\partial _{\underline{\ddot
e}}\varphi ^{<\alpha >})(\overline{\lambda }^{<\beta >}\gamma ^{\underline{%
\ddot o}}\gamma ^{\underline{\ddot e}}\psi _{\underline{\ddot o}})-
$$
$$
\frac i3B_{<\alpha ><\beta ><\gamma >}(\overline{\lambda }^{<\alpha >}\gamma
_5\gamma ^{\underline{\ddot e}}\lambda ^{<\beta >})(\overline{\lambda }%
^{<\gamma >}\psi _{\underline{\ddot e}})-%
$$
$$
\frac i3B_{<\alpha ><\beta ><\gamma >}(\overline{\lambda }^{<\alpha >}\gamma
^{\underline{\ddot e}}\lambda ^{<\beta >})(\overline{\lambda }^{<\gamma
>}\gamma _5\psi _{\underline{\ddot e}}),
$$
where $\psi _{\underline{\ddot e}}$ is the higher order an\-isot\-rop\-ic
generalization of Maiorana gravitino with s--symmetric transformation law%
$$
\bigtriangleup \psi _{\underline{\ddot e}}=-\partial _{\underline{\ddot e}%
}\varepsilon +...
$$

From the standard variation, but locally adapted to the N--connection, of
the $I^{(N)}$ with a next covariantization (with respect to $\left( e_{\ddot
e}^{\underline{\ddot e}},\psi _{\ddot e}\right) )$ of the theory. In result
(it's convenient to use MW--spinors) we introduce this action:%
$$
I=\frac 12\int d^2z\ e\ [\gamma ^{\ddot a\ddot u}g_{<\alpha ><\beta
>}\partial _{\ddot a}u^{<\alpha >}\partial _{\ddot u}u^{<\beta
>}+e^{-1}\varepsilon ^{\ddot a\ddot u}b_{<\alpha ><\beta >}\partial _{\ddot
a}u^{<\alpha >}\partial _{\ddot u}u^{<\beta >}+\eqno(51)
$$
$$
ig_{<\alpha ><\beta >}\overline{\lambda }_{+}^{<\alpha >}(\widehat{D}%
^{+}\lambda _{+})^{<\beta >}+ig_{<\alpha ><\beta >}\overline{\lambda }%
_{-}^{<\alpha >}(\widehat{D}^{-}\lambda _{-})^{<\beta >}+
$$
$$
\frac 14\widetilde{R}_{<\beta ><\alpha ><\gamma ><\delta >}^{+}(\overline{%
\lambda }^{<\alpha >}\gamma _5\lambda ^{<\beta >})(\overline{\lambda }%
^{<\gamma >}\gamma _5\lambda ^{<\delta >})]+
$$
$$
g_{<\alpha ><\beta >}(2\partial _{\ddot a}u^{<\alpha >}+\overline{\psi }%
_{\ddot a}\lambda _{+}^{<\alpha >}+\overline{\psi }_{\ddot a}\lambda
_{-}^{<\alpha >})\times
$$
$$
(\overline{\lambda }_{+}^{<\beta >}\{\gamma ^{\ddot a\ddot
u}+e^{-1}\varepsilon ^{\ddot a\ddot u}\}\psi _{\ddot u}+(\overline{\lambda }%
_{-}^{<\beta >}\{\gamma ^{\ddot a\ddot u}-e^{-1}\varepsilon ^{\ddot a\ddot
u}\}\psi _{\ddot u})+
$$
$$
\frac{2i}3B_{<\alpha ><\beta ><\gamma >}\{(\overline{\lambda }_{+}^{<\alpha
>}\gamma ^{\ddot a}\overline{\lambda }_{+}^{<\beta >})(\overline{\lambda }%
_{+}^{<\gamma >}\psi _{\ddot a})-(\overline{\lambda }_{-}^{<\alpha >}\gamma
^{\ddot a}\overline{\lambda }_{-}^{<\beta >})(\overline{\lambda }%
_{-}^{<\gamma >}\psi _{\ddot a})\}],
$$
where $e=\det |e_{\ddot e}^{\underline{\ddot e}}|,$ for which the higher
order anisotropic laws of supersymmetric transforms holds:%
$$
\bigtriangleup e_{\ddot e}^{\underline{\ddot e}}=2i\overline{\varepsilon }%
\gamma ^{\underline{\ddot e}}\psi _{\ddot e},\bigtriangleup \psi _{\ddot
e}=-D_{\ddot e}\varepsilon ,\eqno(52)
$$
$$
\bigtriangleup \varphi ^{<\alpha >}=\bigtriangleup _{+}\varphi ^{<\alpha
>}+\bigtriangleup _{-}\varphi ^{<\alpha >}=\overline{\varepsilon }_{+}\
\lambda _{-}^{<\alpha >}+\overline{\varepsilon }_{-}\ \lambda _{+}^{<\alpha
>},
$$
$$
\bigtriangleup \lambda _{\pm }^{<\alpha >}=-i(\widehat{\partial }u^{<\alpha
>}+\{\overline{\lambda }_{+}^{<\alpha >}\psi _{\ddot a}+\overline{\lambda }%
_{-}^{<\alpha >}\psi _{\ddot a}\}\gamma ^{\ddot a})\varepsilon _{\mp }-%
\widetilde{\Gamma }_{<\beta ><\gamma >}^{<\alpha >(\pm )}\lambda _{\pm
}^{<\beta >}\bigtriangleup _{\pm }\varphi ^{<\gamma >}.
$$

Restricting our considerations in (51) and (52) only with $\lambda
_{+}^{<\alpha >}$--spinors and $\varepsilon _{-}$--parameters, when $%
\varepsilon _{+},\lambda _{-}^{<\alpha >}=0,$ we obtain the action for the
heterotic higher order anisotropic string on background $\left( g_{<\alpha
><\beta >},b_{<\alpha ><\beta >}\right) $ with (1,0)--local supersymmetry $%
\left( \psi _{\ddot a}\rightarrow \psi _{\ddot a(-)}\right) .$ This action
can be interpreted as the ''minimal'' interaction of the higher order
anisotropic (1,0)--matter $\left( \varphi ^{<\alpha >},\lambda _{+}^{<\alpha
>}\right) $ with (1,0)--supergravity $\left( e_{\ddot e}^{\underline{\ddot e}%
},\psi _{\ddot e}\right) .$

\subsection{Locally anisotropic heterotic strings}

As an illustration of application of s--field methods in locally anisotropic
s--spaces we shall construct the action for a model of higher order
anisotropic s--string.

The (1,0)--superspaces can be parametrized by two Bose coordinates $%
(z^{\ddagger },z^{=})$ and one Fermy coordinate $\theta ^{+};\ddot
u=(z^{\ddagger },z^{=},\theta ^{+}).$ One represents vector indices as $%
(++,--)\equiv (\ddagger ,=)$ taking into account that by $(+,-)$ there are
denoted spirality $\pm 1/2.$

The standard derivations
$$
D_{\ddot A}=\{D_{+},\partial _{\ddagger },\partial _{=}\};D_{+}=\frac
\partial {\partial \theta ^{+}}+i\theta ^{+}\partial _{+},
$$
in the flat (1,0)--superspace \cite{sak} satisfy algebra%
$$
\{D_{+},D_{+}\}=2i\partial _{\ddagger },~\partial _{\ddagger }\partial
_{=}=\Box ,~[\partial _{\underline{a}},D_{+}]=[\partial _{\underline{a}%
},\partial _{\underline{b}}]=0,
$$
and s--space integration measure%
$$
\int d\theta _{+}=\frac \partial {\partial \theta ^{+}},d^3\ddot
u^{-}=d^2zd\theta _{+}.
$$

In the flat (1,0)--superspace one defines scalar and spinor s--fields%
$$
\varphi (z,\theta )=A(z)+\theta ^{+}\lambda _{+}(z),\psi _{-}(z,\theta
)=\eta _{-}(z)+\theta ^{+}F(z)
$$
and action
$$
I=\int d^3\ddot uL=\int d^2z(D_{+}L)_{|\theta =0}
$$
with a charged Lagrangian, $L=L_{-},$ in order to have the Lorentz
invariance.

The (1,0)--multiplet of supergravity is described by a set of covariant
derivations%
$$
\bigtriangledown _{\ddot A}=E_{\ddot A}^{\underline{\ddot U}}D_{\underline{%
\ddot U}}+\omega _{\ddot A}^{(M)}\equiv E_{\ddot A}+\Omega _{\ddot A},%
\eqno(53)
$$
where $E_{\ddot A}^{\underline{\ddot U}}$ is a s--vielbein and $\omega
_{\ddot A}^{(M)}$ is the Lorentz connection with L--generator $M:$%
$$
[M,\lambda _{\pm }]=\pm \frac 12\lambda _{\pm }.
$$

The covariant constraints in s--space (for (1,0)--supergravity \cite
{bro,gates}) are given by relations:%
$$
\{\nabla _{+},\nabla _{+}\}=2i\nabla _{\ddagger },~[\nabla _{+},\nabla
_{=}]=-2i\Sigma ^{+}M,\eqno(54)
$$
$$
[\nabla _{+},\nabla _{\ddagger }]=0,~[\nabla _{\ddagger },\nabla
_{=}]=-\Sigma ^{+}\nabla _{+}-R^{(2)}M,
$$
where$R^{(2)}=2\nabla _{+}\Sigma ^{+}$ and $\Sigma ^{+}$ defines the
covariant strength of (1,0)--supergravity in s--space.

As a consequence of (54) only $E_{+}^{=},E_{=}^{\ddagger },E_{+}^{\ddagger
},E_{+}^{+}$ and $E_{=}^{=}$ are independent s--fields; the rest of
components of vielbein and connection can be expressed through them. The
conditions of covariance of derivations (54) lead to these transformation
laws with respect to d--coordinate and local Lorentz transforms with
corresponding parameters $K^{\underline{\ddot U}}$ and $\Lambda _l:$%
$$
\bigtriangledown _{\ddot A}^{\prime }=e^K\bigtriangledown _{\ddot
A}e^{-K},~K=K^{\underline{\ddot U}}D_{\underline{\ddot U}}+\Lambda _lM;
$$
for vielbeins we have%
$$
\bigtriangleup E_{\ddot A}^{\underline{\ddot U}}=-\nabla _{\ddot A}K^{%
\underline{\ddot U}}+K^{\underline{\ddot I}}D_{\underline{\ddot I}}E_{\ddot
A}^{\underline{\ddot U}}+E_{\ddot A}^{\underline{\ddot O}}K^{\underline{%
\ddot I}}[D_{\underline{\ddot I}},D_{\underline{\ddot O}}\}^{\underline{%
\ddot U}}+\Lambda _l[M,E_{\ddot A}^{\underline{\ddot U}}],
$$
from which one follows that%
$$
\bigtriangleup E_{+}^{\ddagger }=K^{\underline{\ddot U}}D_{\underline{\ddot U%
}}E_{+}^{\ddagger }+2iE_{+}^{+}K^{+}-E_{+}^{\underline{\ddot U}}D_{%
\underline{\ddot U}}K^{\ddagger }+\frac 12\Lambda _lE_{+}^{\ddagger }.
$$

It is convenient to use the s--symmetric gauge (when $E_{+}^{\ddagger }=0),$%
$$
K^{+}=-\frac i2\left( E_{+}^{+}\right) ^{-1}\nabla _{+}K^{\ddagger },
$$
and to introduce the Lorentz--invariant scalar s--field $S$ and Lorentz
compensator $L$ satisfying correspondingly conditions%
$$
E_{+}^{+}(E_{=}^{=})^{1/2}=e^{-S}~\mbox{ and }%
~E_{+}^{+}(E_{=}^{=})^{-1/2}=e^L.
$$
In a Lorentz invariant theory we can always choose the gauge $L=0;$ in this
gauge the s--conformal transforms are accompanied by a corresponding
compensating Lorentz transform with parameter%
$$
\Lambda _l=(E_{+}^{+})^{-1}\nabla _{+}K^{+}-\frac 12(E_{=}^{=})^{-1}\nabla
_{=}K^{=}=\frac 12(\nabla _{\ddagger }K^{\ddagger }-\nabla _{=}K^{=})+...
$$

The solution of constraints (54) in the s--symmetric gauge $%
E_{+}^{\ddagger }=L=0$ and in the linear approximation is \cite{gates}%
$$
\nabla _{+}=(1-\frac S2)D_{+}+H_{+}^{=}\partial _{=}-(D_{+}S+\partial
_{=}H_{+}^{=})M,
$$
$$
\nabla _{\ddagger }=(1-S)\partial _{\ddagger }+i[\frac 12(\partial
_{=}H_{+}^{=})+(D_{+}S)]D_{+}-
$$
$$
i[D_{+}H_{+}^{=}]\partial _{=}-(\partial _{\ddagger }S-iD_{+}\partial
_{=}H_{+}^{=})M,
$$
$$
\nabla _{=}=(1-S)\partial _{=}-\frac i2[(D_{+}H_{=}^{\ddagger
})]D_{+}+H_{=}^{\ddagger }\partial _{\ddagger }+(\partial _{=}S+\partial
_{\ddagger }H_{=}^{\ddagger })M,
$$
$$
\Sigma ^{+}=\frac i2[D_{+}(\partial _{\ddagger }H_{=}^{\ddagger }+2\partial
_{=}S)+\partial _{=}^2H_{+}^{=}]+...,
$$
where s--fields $\left( H_{+}^{=},H_{=}^{\ddagger }\right) $ and $S$ are
prepotentials of the system. These s--potentials have to be used in the
quantum field theory.

The linearized expression for s--field density $E^{-1}=Ber(E_{\ddot A}^{%
\underline{\ddot U}})$ is computed as
$$
E^{-1}=s\det (E_{\ddot A}^{\underline{\ddot U}})=e^{3S/2}[1+iH_{=}^{\ddagger
}(D_{+}H_{+}^{=}+H_{+}^{=}\partial _{=}H_{+}^{=})]^{-1}.
$$

The action for heterotic string in higher order an\-isot\-rop\-ic
(1,0)--superspace accounting for the background of massless modes of locally
anisotropic graviton, antisymmetric d--tensor, dilaton and gauge bosons is
introduced in a manner similar to locally isotropic models \cite
{henty,hull,ket,ketn} but with corresponding extension to distinguished and
locally adapted to N--connection geometric objects:%
$$
I_{HS}=\frac 1{4\pi \alpha ^{\prime }}\int d^3\ddot u^{-}E^{-1}\{i\nabla
_{+}u^{<\alpha >}\nabla _{=}u^{<\beta >}[g_{<\alpha ><\beta >}(u)+b_{<\alpha
><\beta >}(u)]+\eqno(55)
$$
$$
\Psi _{(-)}^{|I|}[\delta _{|I||J|}\nabla _{+}+A_{|I||J|}^{+}(u)]\Psi
_{(-)}^{|J|}+\alpha ^{\prime }\Phi (u)\Sigma ^{+}\},
$$
where $A_{|I||J|}^{+}(u)=A_{|I||J|<\alpha >}\nabla _{+}u^{<\alpha >}$ is the
gauge boson background, $\Phi (u)$ is the dilaton field and $\Psi
_{(-)}^{|I|}$ are hetrotic fermions, $|I|,|J|,...=1,2,...,,{\sf N.}$

\section{Background D--Field Methods for $\sigma$--Mo\-dels}

The background--quantum decomposition of superfields of (1,0) higher order
anisotropic supergravity considered in previous section can be performed in
a standard manner \cite{gates} by taking into account the distinguished
character of geometrical objects on locally anisotropic s--spaces;
constraints (54) should be solved in terms of background--covariant
derivations and quantum s--fields $\left( H_{+}^{=},H_{=}^{\ddagger
},S\right) ,$ quantum s--fields $L$ and $E_{+}^{\ddagger }$ are gauged in an
algebraic manner (not introducing into considerations ghosts) and note that
by using quantum scale transforms we can impose gauge $S=0$ (also without
Faddeev--Popov ghosts). Finally, after a background--quantum decomposition
of superfields of (1.0) higher order anisotropic supergravity in the just
pointed out manner we can fix the quantum gauge invariance, putting zero
values for quantum fields (in absences of supergravitational and conformal
anomalies and for topological trivial background configurations). In this
case all (1,0) supergravity fields can be considered as background ones.

The fixing of gauge symmetry as a vanishing of quantum s--fields induces the
ghost action%
$$
I_{FP}=\int d^3z^{-}E\{b_{=}^{\ddagger }\nabla _{+}c^{=}+b_{+}^{=}\nabla
_{=}c^{\ddagger }\}.\eqno(56)
$$

The aim of this section is to compute the renormalized effective action,
more exactly, it anomaly part on the background of s--fields (1,0) higher
order anisotropic supergravity for the model defined by the action (55).

In order to integrate on quantum fields in (55) in a d---covariant manner
we use background--quantum decompositions of the action with respect to
normal locally adapted coordinates along autoparallels
 defined by Christoffel d--symbols (39). We emphasize the
multiconnection character of locally anisotropic spaces; every geometric
construction with a fixed d--connection structure (from some purposes
considered as a simple or more convenient one) can be transformed, at least
locally, into a another similar one for a corresponding d--connection by
using deformations of connections (40) (or (3.2) and (3.3) if we are
interested in na--map deformations). Let
$$
\frac{\delta ^2X^{<\alpha >}}{\partial s^2}+\{\frac{<\alpha >}{<\beta
><\gamma >}\}\frac{\delta X^{<\beta >}}{\partial s}\frac{\delta X^{<\gamma >}%
}{\partial s}=0,
$$
where $X^{<\alpha >}(s=0)=X^{<\alpha >},X^{<\alpha >}(s=1)=X^{<\alpha >}+\pi
^{<\alpha >}$ and $\pi ^{<\alpha >}$ are quantum fluctuations with respect
to background $X^{<\alpha >}.\,$Covariant quantum fields $\zeta ^{<\alpha >}$
(''normal fields'') are defined as%
$$
\zeta ^{<\alpha >}=\frac{\delta X^{<\alpha >}}{\partial s}\mid _{s=0}\equiv
\zeta ^{<\alpha >}(s)\mid _{s=0}.
$$
We shall use covariantized in a $\sigma $--model manner derivations in (1,0)
higher order anisotropic s--space
$$
{\cal D}_{\ddot A}\equiv ({\cal D}_{+},{\cal D}_{=})=\nabla _{\ddot
A}+\Gamma _{\ddot A},\Gamma _{\ddot A<\beta >}^{<\alpha >}\equiv \{\frac{%
<\alpha >}{<\gamma ><\beta >}\}\nabla _{\ddot A}X^{<\gamma >},
$$
on properties of derivation $\nabla _{\ddot A}$ see (47) and (48), into
distinguished autoparallel (in a $\sigma $--model manner) d--covariant
derivation with properties
$$
D\left( s\right) T_{<\alpha >...}=\zeta ^{<\beta >}{\cal D}_{<\beta
>}T_{<\alpha >...},D(s)\zeta ^{<\alpha >}(s)=0.
$$

The derivation ${\cal D}_{\ddagger }$ is defined as
$$
2i{\cal D}_{\ddagger }\equiv \{{\cal D}_{+},{\cal D}_{+}\};
$$
we note that ${\cal D}_{\ddagger }\neq \nabla _{\ddagger }^{cov}.$ One holds
the next relations:%
$$
{\cal D}_{+}\zeta ^{<\beta >}=\nabla _{+}\zeta ^{<\beta >}+\{\frac{<\beta >}{%
<\tau ><\sigma >}\}\nabla _{+}X^{<\sigma >}\zeta ^{<\tau >},
$$
$$
{\cal D}_{=}\zeta ^{<\beta >}=\nabla _{=}\zeta ^{<\beta >}+\{\frac{<\beta >}{%
<\tau ><\sigma >}\}\nabla _{=}X^{<\sigma >}\zeta ^{<\tau >},
$$
$$
{\cal D}_{\ddagger }\zeta ^{<\beta >}=\nabla _{\ddagger }^{cov}\zeta
^{<\beta >}-\frac i2R_{~<\tau ><\sigma ><\nu >}^{<\beta >}\nabla
_{+}X^{<\sigma >}\nabla _{+}X^{<\nu >}\zeta ^{<\tau >},
$$
$$
D(s){\cal D}_{\ddot A}\zeta ^{<\beta >}(s)=\zeta ^{<\nu >}(s)\zeta ^{<\tau
>}(s)R_{~<\tau ><\sigma ><\nu >}^{<\beta >}\nabla _{\ddot A}X^{<\sigma >},
$$
where
$$
\nabla _{\ddagger }^{cov}\zeta ^{<\beta >}=\nabla _{\ddagger }\zeta ^{<\beta
>}+\{\frac{<\beta >}{<\tau ><\sigma >}\}\nabla _{\ddagger }X^{<\tau >}\zeta
^{<\sigma >}.
$$

For heterotic fermions the d--covariant and gauge--covariant formalism of
computation of quantum--background decomposition of action can be performed
by using the prescription%
$$
\Psi _{(-)}^{|I|}(s): \Psi _{(-)}^{|I|}(0)=\Psi _{(-)}^{|I|}, \Psi
_{(-)}^{|I|}(s=1)=\Psi _{(-)}^{|I|}+\Delta _{(-)}^{|I|}, D_{(-)}^2\Psi
_{-}^{|I|}(s)=0,
$$
where $\Delta _{-}^{|I|}$ are quantum fluctuations with respect to
background $\Psi ;$ functions $\Psi _{(-)}^{|I|}(s)$ interpolate in a
gauge--covariant manner with $\Psi _{(-)}^{|I|}+\Delta _{-}^{|I|}$ because
of definition of the operator $D(s):$%
$$
D(s)\Psi _{(-)}^{|I|}=[\delta ^{|I||J|}\frac \delta {\partial s}+A_{<\alpha
>}^{|I||J|}\frac{\delta X^{<\alpha >}}{\partial s}]\Psi _{(-)}^{|J|}=\frac{%
\delta \Psi _{(-)}^{|I|}}{\partial s}+A_{<\alpha >}^{|I||J|}\zeta ^{<\alpha
>}\Psi _{(-)}^{|J|}.
$$

As d--covariant and gauge--covariant quantum s--fields we use spinors
$$
\chi _{(-)}^{|J|}\equiv D(s)\Psi _{(-)}^{|J|}(s)\mid _{s=0}
$$
satisfying conditions%
$$
D(s)\chi _{(-)}^{|J|}(s)=0.
$$
We also define the next derivation in (1,0) higher order anisotropic
s--superspace
$$
({\cal D}_{+}\Psi _{(-)})^{|I|}\equiv (\delta _{|J|}^{|I|}\nabla
_{+}+A_{|J|<\alpha >}^{|I|}\nabla _{+}X^{<\alpha >})\Psi _{(-)}{}^{|J|}
$$
for which one holds identities%
$$
D(s){\cal D}_{+}\Psi _{(-)}{}^{|I|}={\cal D}_{+}\chi
_{(-)}{}^{|I|}+F_{|J|<\alpha ><\beta >}^{|I|}\zeta ^{<\alpha >}\nabla
_{+}X^{<\beta >}\Psi _{(-)}{}^{|I|},
$$
$$
D(s)F_{|J|<\alpha ><\beta >}^{|I|}=\zeta ^{<\gamma >}{\cal D}_{<\gamma
>}F_{|J|<\alpha ><\beta >}^{|I|},
$$
where ${\cal D}_{<\gamma >}$ is both d-covariant and gauge invariant
derivation and the strength d--tensor of gauge fields is defined by using
la--derivation operators,%
$$
F_{|J|<\alpha ><\beta >}^{|I|}=\delta _{<\alpha >}A_{|J|<\beta
>}^{|I|}-\delta _{<\beta >}A_{|J|<\alpha >}^{|I|}+A_{|K|<\alpha
>}^{|I|}A_{|J|<\beta >}^{|K|}-A_{|K|<\beta >}^{|I|}A_{|J|<\alpha >}^{|K|}.
$$

The action (55) consist from four groups of terms of different nature:%
$$
I_{HS}=I^{(1)}+I^{(2)}+I^{(3)}+I^{(4)}.\eqno(57)
$$

The first term
$$
I^{(1)}=-\frac i{4\pi \alpha ^{\prime }}\int d^3z^{-}Eg_{<\alpha ><\beta
>}(X)\nabla _{+}X^{<\alpha >}\nabla _{=}X^{<\beta >}
$$
is associated to the higher order anisotropic gravitational sector and has
the next background--quantum decomposition
$$
I^{(1)}[X+\pi (\zeta )]=I_0^{(1)}+I_1^{(1)}+I_2^{(1)}+...,
$$
where
$$
I_b^{(1)}=\frac 1{b!}\frac{d^bI^{(1)}}{ds^b}\mid _{s=0};b=0,1,2,....
$$
Thes terms are computed in a usual (but distinguished to the N--connection
structure) manner:%
$$
I_1^{(1)}=-\frac i{2\pi \alpha ^{\prime }}\int d^3z^{-}Eg_{<\alpha ><\beta
>}({\cal D}_{+}\zeta ^{<\alpha >})\nabla _{=}X^{<\beta >},\eqno(58)
$$
$$
I_2^{(1)}=-\frac i{4\pi \alpha ^{\prime }}\int d^3z^{-}E\{g_{<\alpha ><\beta
>}{\cal D}_{+}\zeta ^{<\alpha >}{\cal D}_{=}\zeta ^{<\beta >}+
$$
$$
R_{<\lambda ><\sigma ><\tau ><\nu >}\nabla _{+}X^{<\sigma >}\nabla
_{=}X^{<\nu >}\zeta ^{<\lambda >}\zeta ^{<\tau >}\},
$$
$$
I_3^{(1)}=-\frac i{4\pi \alpha ^{\prime }}\int d^3z^{-}E\{\frac
23(R_{<\lambda ><\sigma ><\tau ><\nu >}\nabla _{+}X^{<\sigma >}{\cal D}%
_{=}\zeta ^{<\nu >}+
$$
$$
R_{<\lambda ><\sigma ><\tau ><\nu >}\nabla _{=}X^{<\sigma >}{\cal D}%
_{+}\zeta ^{<\nu >})\zeta ^{<\lambda >}\zeta ^{<\tau >}+
$$
$$
\frac 13{\cal D}_{<\mu >}R_{<\lambda ><\sigma ><\tau ><\nu >}\nabla
_{+}X^{<\sigma >}\nabla _{=}X^{<\nu >}\zeta ^{<\mu >}\zeta ^{<\lambda
>}\zeta ^{<\tau >},
$$
$$
I_4^{(1)}=-\frac i{4\pi \alpha ^{\prime }}\int d^3z^{-}E\{\frac 14({\cal D}%
_{<\mu >}R_{<\lambda ><\sigma ><\tau ><\nu >}\nabla _{+}X^{<\sigma >}{\cal D}%
_{=}\zeta ^{<\nu >}+
$$
$$
{\cal D}_{<\mu >}R_{<\lambda ><\sigma ><\tau ><\nu >}\nabla _{=}X^{<\sigma >}%
{\cal D}_{+}\zeta ^{<\nu >})\zeta ^{<\mu >}\zeta ^{<\lambda >}\zeta ^{<\tau
>}+
$$
$$
\frac 13R_{<\lambda ><\sigma ><\tau ><\nu >}\zeta ^{<\lambda >}\zeta ^{<\tau
>}{\cal D}_{+}\zeta ^{<\sigma >}{\cal D}_{=}\zeta ^{<\nu >}+
$$
$$
[\frac 13R_{<\lambda ><\sigma ><\tau ><\nu >}R_{<\delta ><\varepsilon >\
\cdot \ <\gamma >}^{\qquad \quad <\nu >}+
$$
$$
\frac 1{12}{\cal D}_{<\delta >}{\cal D}_{<\gamma >}R_{<\lambda ><\sigma
><\varepsilon ><\tau >}]\nabla _{+}X^{<\sigma >}\nabla _{=}X^{<\varepsilon
>}\zeta ^{<\lambda >}\zeta ^{<\tau >}\zeta ^{<\delta >}\zeta ^{<\gamma >},
$$
$$
I_5^{(1)}=-\frac i{4\pi \alpha ^{\prime }}\int d^3z^{-}E\times
$$
$$
\{\frac 16({\cal D}_{<\mu >}R_{<\lambda ><\sigma ><\nu ><\tau >}{\cal D}%
_{+}\zeta ^{<\sigma >}{\cal D}_{=}\zeta ^{<\nu >}\zeta ^{<\mu >}\zeta
^{<\lambda >}\zeta ^{<\tau >}+...\},
$$
$$
I_6^{(1)}=-\frac i{4\pi \alpha ^{\prime }}\int d^3z^{-}E\{\frac 1{20}{\cal D}%
_{<\alpha >}{\cal D}_{<\mu >}R_{<\lambda ><\sigma ><\nu ><\tau >}+
$$
$$
\frac 2{45}R_{<\lambda ><\gamma ><\nu ><\tau >}R_{<\alpha ><\sigma >\ \cdot
\ <\mu >}^{\qquad \quad <\gamma >}{\cal D}_{+}\zeta ^{<\sigma >}{\cal D}%
_{=}\zeta ^{<\nu >}\zeta ^{<\alpha >}\zeta ^{<\mu >}\zeta ^{<\lambda >}\zeta
^{<\tau >}+...\},
$$
where by dots, in this section, are denoted those terms (containing
multiples $\left( \nabla X\right) )$ which are not important for calculation
of anomalies, see section 11.

The second term in (57)
$$
I^{(2)}=-\frac i{4\pi \alpha ^{\prime }}\int d^3z^{-}Eb_{<\alpha ><\beta
>}(X)\nabla _{+}X^{<\alpha >}\nabla _{=}X^{<\beta >}
$$
having the next background--quantum decomposition
$$
I^{(2)}[X+\pi (\zeta )]=I_0^{(2)}+I_1^{(2)}+I_2^{(2)}+...,
$$
where
$$
I_b^{(2)}=\frac 1{b!}\frac{d^bI^{(2)}}{ds^b}\mid _{s=0};b=0,1,2,....
$$
describes a Wess--Zumino--Witten like model of interactions (in a higher
order anisotropic variant). The diagram vertexes depends only on intensity $%
H $ of antisymmetric d--tensor $b:$%
$$
H_{<\tau ><\mu ><\nu >}=\frac 32\delta _{[<\tau >}b_{<\mu ><\nu >]}=
$$
$$
\frac 12(\delta _{<\tau >}b_{<\mu ><\nu >}+\delta _{<\mu >}b_{<\nu ><\tau
>}-\delta _{<\nu >}b_{<\mu ><\tau >}).
$$
By straightforward calculations we find the coefficients:%
$$
I_1^{(2)}=-\frac i{2\pi \alpha ^{\prime }}\int d^3z^{-}E\zeta ^{<\tau
>}\nabla _{+}X^{<\alpha >}\nabla _{=}X^{<\beta >}H_{<\tau ><\alpha ><\beta
>},
$$
$$
I_2^{(2)}=-\frac i{4\pi \alpha ^{\prime }}\int d^3z^{-}E\{\zeta ^{<\tau >}%
{\cal D}_{+}\zeta ^{<\alpha >}\nabla _{=}X^{<\beta >}H_{<\tau ><\alpha
><\beta >}+
$$
$$
\zeta ^{<\tau >}\nabla _{+}X^{<\alpha >}{\cal D}_{=}\zeta ^{<\beta
>}H_{<\tau ><\alpha ><\beta >}+
$$
$$
\zeta ^{<\lambda >}\zeta ^{<\tau >}\nabla _{+}X^{<\alpha >}\nabla
_{=}X^{<\beta >}{\cal D}_{<\lambda >}H_{<\tau ><\alpha ><\beta >}\},
$$
$$
I_3^{(2)}=-\frac i{4\pi \alpha ^{\prime }}\int d^3z^{-}E\{\frac 23\zeta
^{<\tau >}{\cal D}_{+}\zeta ^{<\mu >}{\cal D}_{=}\zeta ^{<\nu >}H_{<\tau
><\mu ><\nu >}+
$$
$$
\frac 23\zeta ^{<\tau >}\zeta ^{<\rho >}\zeta ^{<\gamma >}R_{~\cdot <\gamma
><\rho >[<\delta >}^{<\mu >}H_{<\nu >]<\tau ><\mu >}\nabla _{+}X^{<\delta
>}\nabla _{=}X^{<\nu >}+
$$
$$
\frac 23\zeta ^{<\tau >}\zeta ^{<\lambda >}({\cal D}_{+}\zeta ^{<\mu
>}\nabla _{=}X^{<\nu >}+{\cal D}_{=}\zeta ^{<\nu >}\nabla _{+}X^{<\mu >})%
{\cal D}_{<\lambda >}H_{<\tau ><\mu ><\nu >}+
$$
$$
\frac 13\zeta ^{<\sigma >}\zeta ^{<\lambda >}\zeta ^{<\tau >}\nabla
_{+}X^{<\alpha >}\nabla _{=}X^{<\beta >}{\cal D}_{<\sigma >}{\cal D}%
_{<\lambda >}H_{<\tau ><\alpha ><\beta >}\},
$$
$$
I_4^{(2)}=-\frac i{4\pi \alpha ^{\prime }}\int d^3z^{-}E\{\frac 12\zeta
^{<\tau >}\zeta ^{<\lambda >}{\cal D}_{+}\zeta ^{<\mu >}{\cal D}_{=}\zeta
^{<\nu >}H_{<\tau ><\mu ><\nu >}+
$$
$$
\frac 16\zeta ^{<\lambda >}\zeta ^{<\tau >}\zeta ^{<\rho >}\zeta ^{<\gamma
>}\nabla _{+}X^{<\delta >}\nabla _{=}X^{<\nu >}{\cal D}_{<\lambda
>}(R_{~\cdot <\gamma ><\rho >[<\delta >}^{<\mu >}H_{<\nu >]<\tau ><\mu >})+
$$
$$
\frac 23\zeta ^{<\tau >}\zeta ^{<\rho >}\zeta ^{<\gamma >}R_{~\cdot <\gamma
><\rho >[<\delta >}^{<\mu >}H_{<\nu >]<\tau ><\mu >}\times (\nabla
_{+}X^{<\delta >}{\cal D}_{=}X^{<\nu >}+
$$
$$
{\cal D}_{+}X^{<\nu >}\nabla _{=}X^{<\delta >})+\zeta ^{<\lambda >}\zeta
^{<\tau >}\zeta ^{<\alpha >}\zeta ^{<\beta >}\nabla _{+}X^{<\nu >}\nabla
_{=}X^{<\rho >}\times
$$
$$
(\frac 13{\cal D}_{<\lambda >}H_{<\tau ><\mu >[<\rho >}R_{<\nu ><\alpha >\
\cdot \ <\beta >}^{\qquad \quad <\mu >}-\frac 1{12}{\cal D}_{<\alpha >}{\cal %
D}_{<\beta >}{\cal D}_{<\gamma >}H_{<\tau ><\nu ><\rho >})+
$$
$$
\frac 14\zeta ^{<\tau >}\zeta ^{<\lambda >}\zeta ^{<\gamma >}({\cal D}%
_{+}\zeta ^{<\mu >}\nabla _{=}X^{<\nu >}+
$$
$$
{\cal D}_{=}\zeta ^{<\nu >}\nabla _{+}X^{<\mu >}){\cal D}_{<\gamma >}{\cal D}%
_{<\lambda >}H_{<\tau ><\mu ><\nu >}\},
$$
$$
I_5^{(2)}=-\frac i{4\pi \alpha ^{\prime }}\int d^3z^{-}E\{\frac 12\zeta
^{<\tau >}\zeta ^{<\lambda >}\zeta ^{<\gamma >}{\cal D}_{+}\zeta ^{<\mu >}%
{\cal D}_{=}\zeta ^{<\nu >}\times
$$
$$
[\frac 15{\cal D}_{<\gamma >}{\cal D}_{<\lambda >}H_{<\tau ><\mu ><\nu
>}+\frac 2{15}R_{~\cdot <\gamma ><\lambda >[<\mu >}^{<\rho >}H_{<\nu >]<\tau
><\rho >}\}+...,
$$
$$
I_6^{(2)}=-\frac i{4\pi \alpha ^{\prime }}\int d^3z^{-}E\zeta ^{<\tau
>}\zeta ^{<\lambda >}\zeta ^{<\alpha >}\zeta ^{<\beta >}{\cal D}_{+}\zeta
^{<\rho >}{\cal D}_{=}\zeta ^{<\nu >}\times
$$
$$
[\frac 19{\cal D}_{<\lambda >}H_{<\tau ><\mu >[<\rho >}R_{<\nu >]<\alpha >\
\cdot \ <\beta >}^{\qquad \quad <\mu >}+\frac 1{18}{\cal D}_{<\lambda
>}R_{~\cdot <\gamma ><\rho >[<\delta >}^{<\mu >}H_{<\nu >]<\tau ><\mu >}+
$$
$$
\frac 1{18}{\cal D}_{<\alpha >}{\cal D}_{<\beta >}{\cal D}_{<\lambda
>}H_{<\tau ><\rho ><\nu >}]+...,
$$
where by dots are denoted terms not being important for calculation of
anomalies and operations of symmetrization ( ) and antisymmetrization are
taken without coefficients.

The third term in (57)
$$
I^{(3)}=I_0^{(3)}+I_1^{(3)}+I_2^{(3)}+...+I_s^{(3)}+...
$$
is of Fradkin--Tseitlin dilaton type with coefficients%
$$
I_0^{(3)}=-\frac 1{4\pi \alpha ^{\prime }}\int d^3z^{-}E\alpha ^{\prime
}\Sigma ^{+}\Phi ,
$$
$$
I_1^{(3)}=-\frac 1{4\pi \alpha ^{\prime }}\int d^3z^{-}E\alpha ^{\prime
}\Sigma ^{+}\zeta ^{<\alpha >}\delta _{<\alpha >}\Phi ,
$$
$$
I_2^{(3)}=-\frac 1{4\pi \alpha ^{\prime }}\int d^3z^{-}E\alpha ^{\prime
}\Sigma ^{+}\frac 1{2!}\zeta ^{<\beta >}\zeta ^{<\alpha >}{\cal D}_{<\alpha
>}{\cal D}_{<\beta >}\Phi ,...,
$$
$$
I_s^{(3)}=-\frac 1{4\pi \alpha ^{\prime }}\int d^3z^{-}E\alpha ^{\prime
}\Sigma ^{+}\frac 1{s!}\zeta ^{<\alpha _1>}...\zeta ^{<\alpha _s>}{\cal D}%
_{<\alpha _1>}...{\cal D}_{<\alpha _s>}\Phi .
$$

The forth term in (57)%
$$
I^{(4)}=-\frac 1{4\pi \alpha ^{\prime }}\int d^3z^{-}E~\Psi _{(-)}{}^{|I|}%
{\cal D}_{+}\Psi _{(-)}{}^{|I|}
$$
has the next background--quantum decomposition
$$
I^{(4)}[\Psi +\Delta (\chi ),X+\pi (\zeta
)]=I_0^{(4)}+I_1^{(4)}+I_2^{(4)}+...,
$$
with coefficients%
$$
I_1^{(4)}=-\frac 1{4\pi \alpha ^{\prime }}\int d^3z^{-}E~\{\chi
_{(-)}{}^{|I|}{\cal D}_{+}\chi _{(-)}{}^{|I|}+\Psi _{(-)}{}^{|I|}{\cal D}%
_{+}\Psi _{(-)}{}^{|I|}+\eqno(59)
$$
$$
\Psi _{(-)}{}^{|I|}F_{|I||J|<\mu ><\nu >}\zeta ^{<\mu >}\nabla _{+}X^{<\nu
>}\Psi _{(-)}{}^{|J|}\},
$$
$$
I_2^{(4)}=-\frac 1{4\pi \alpha ^{\prime }}\int d^3z^{-}E~\{\chi
_{(-)}{}^{|I|}{\cal D}_{+}\chi _{(-)}{}^{|I|}+
$$
$$
2\chi _{(-)}{}^{|I|}F_{|I||J|<\mu ><\nu >}\zeta ^{<\mu >}\nabla _{+}X^{<\nu
>}\Psi _{(-)}{}^{|J|}-
$$
$$
\frac 12\zeta ^{<\nu >}\zeta ^{<\mu >}\Psi _{(-)}{}^{|I|}\Psi _{(-)}{}^{|J|}%
{\cal D}_{<\lambda >}F_{|I||J|<\mu ><\nu >}\nabla _{+}X^{<\nu >}-
$$
$$
\frac 12\Psi _{(-)}{}^{|I|}\Psi _{(-)}{}^{|J|}F_{|I||J|<\mu ><\nu >}\zeta
^{<\mu >}{\cal D}_{+}\zeta ^{<\nu >}\},
$$
$$
I_3^{(4)}=-\frac 1{4\pi \alpha ^{\prime }}\int d^3z^{-}E~\{\chi
_{(-)}{}^{|I|}F_{|I||J|<\mu ><\nu >}\zeta ^{<\mu >}\nabla _{+}X^{<\nu >}\chi
_{(-)}{}^{|J|}+
$$
$$
\zeta ^{<\lambda >}\zeta ^{<\mu >}\chi _{(-)}{}^{|I|}{\cal D}_{<\lambda
>}F_{|I||J|<\mu ><\nu >}\nabla _{+}X^{<\nu >}\Psi _{(-)}{}^{|J|}+
$$
$$
\chi _{(-)}{}^{|I|}F_{|I||J|<\mu ><\nu >}\zeta ^{<\mu >}{\cal D}_{+}\zeta
^{<\nu >}\Psi _{(-)}{}^{|J|}-
$$
$$
\frac 13\zeta ^{<\lambda >}\zeta ^{<\nu >}\Psi _{(-)}{}^{|I|}{\cal D}%
_{<\lambda >}F_{|I||J|<\mu ><\nu >}\Psi _{(-)}{}^{|J|}{\cal D}_{+}\zeta
^{<\mu >}-
$$
$$
\frac 16\zeta ^{<\lambda >}\zeta ^{<\tau >}\zeta ^{<\mu >}\Psi
_{(-)}{}^{|I|}\Psi _{(-)}{}^{|J|}\nabla _{+}X^{<\nu >}\times
$$
$$
({\cal D}_{<\tau >}{\cal D}_{<\lambda >}F_{|I||J|<\mu ><\nu >}-R_{<\lambda
>]<\nu >\ \cdot \ <\tau >}^{\qquad \quad <\gamma >}F_{|I||J|<\mu ><\nu >}),
$$
$$
I_4^{(4)}=-\frac 1{4\pi \alpha ^{\prime }}\int d^3z^{-}E~\{\frac 12\chi
_{(-)}{}^{|I|}F_{|I||J|<\mu ><\nu >}\chi _{(-)}{}^{|J|}\zeta ^{<\nu >}{\cal D%
}_{+}\zeta ^{<\mu >}-
$$
$$
\frac 12\zeta ^{<\lambda >}\zeta ^{<\mu >}\chi _{(-)}{}^{|I|}({\cal D}%
_{<\lambda >}F_{|I||J|<\mu ><\nu >})\chi _{(-)}{}^{|J|}\nabla _{+}X^{<\nu
>}+
$$
$$
\frac 23\zeta ^{<\lambda >}\zeta ^{<\mu >}\chi _{(-)}{}^{|I|}({\cal D}%
_{<\lambda >}F_{|I||J|<\mu ><\nu >})\Psi _{(-)}{}^{|J|}{\cal D}_{+}\zeta
^{<\nu >}+\frac 13\zeta ^{<\gamma >}\zeta ^{<\lambda >}\zeta ^{<\mu >}\chi
_{(-)}{}^{|I|}\times
$$
$$
({\cal D}_{<\gamma >}{\cal D}_{<\lambda >}F_{|I||J|<\mu ><\nu >}+R_{<\gamma
><\nu >\ \cdot \ <\lambda >}^{\qquad \quad <\rho >}F_{|I||J|<\mu ><\rho
>})\nabla _{+}X^{<\nu >}\Psi _{(-)}{}^{|J|}-
$$
$$
\frac 1{24}\zeta ^{<\lambda >}\zeta ^{<\mu >}\zeta ^{<\gamma >}\Psi
_{(-)}{}^{|I|}\Psi _{(-)}{}^{|J|}(3{\cal D}_{<\mu >}{\cal D}_{<\lambda
>}F_{|I||J|<\gamma ><\nu >}+
$$
$$
R_{<\lambda ><\nu >\ \cdot \ <\gamma >}^{\qquad \quad <\rho >}F_{|I||J|<\tau
><\rho >}){\cal D}_{+}\zeta ^{<\nu >}-\frac 1{24}\zeta ^{<\tau >}\zeta
^{<\lambda >}\zeta ^{<\mu >}\zeta ^{<\gamma >}\Psi _{(-)}{}^{|I|}\Psi
_{(-)}{}^{|J|}\times
$$
$$
[{\cal D}_{<\gamma >}{\cal D}_{<\tau >}{\cal D}_{<\lambda >}F_{|I||J|<\mu
><\nu >}+({\cal D}_{<\gamma >}R_{<\lambda ><\nu >\ \cdot \ <\mu >}^{\qquad
\quad <\rho >})F_{|I||J|<\tau ><\rho >}+
$$
$$
3({\cal D}_{<\gamma >}F_{|I||J|<\tau ><\rho >})R_{<\lambda ><\nu >\ \cdot \
<\mu >}^{\qquad \quad <\rho >}]\nabla _{+}X^{<\nu >}\},
$$
$$
I_5^{(4)}=-\frac 1{4\pi \alpha ^{\prime }}\int d^3z^{-}E~\{-\frac 13\zeta
^{<\mu >}\zeta ^{<\lambda >}\chi _{(-)}{}^{|I|}
$$
$$
({\cal D}_{<\lambda >}F_{|I||J|<\mu ><\nu >})\chi _{(-)}{}^{|J|}{\cal D}%
_{+}\zeta ^{<\mu >}+...\},
$$
$$
I_6^{(4)}=-\frac 1{4\pi \alpha ^{\prime }}\int d^3z^{-}E~\{-\frac 1{24}\zeta
^{<\gamma >}\zeta ^{<\mu >}\zeta ^{<\lambda >}\chi _{(-)}{}^{|I|}\chi
_{(-)}{}^{|J|}\times
$$
$$
(3{\cal D}_{<\gamma >}{\cal D}_{<\lambda >}F_{|I||J|<\mu ><\nu
>}+F_{|I||J|<\gamma ><\rho >}R_{<\lambda ><\nu >\ \cdot \ <\mu >}^{\qquad
\quad <\rho >}){\cal D}_{+}\zeta ^{<\nu >}\}+...
$$

The kinetic terms for quantum fields $\zeta ^{<\mu >}$ and $\chi
_{(-)}{}^{|J|}$ in the decompositions (2.14) and (2.15) define the
propagators $\left( 2\pi \alpha ^{\prime }=1\right) $%
$$
<\zeta ^{<\mu >}(\ddot u)\zeta ^{<\nu >}(\ddot u^{\prime })>=g^{<\mu ><\nu >}%
\frac{D_{+}}{\Box }\delta _{(-)}^3(\ddot u,\ddot u^{\prime })=
$$
$$
g^{<\mu ><\nu >}\frac 1{(2\pi )^2}\int d^{d_2}p\frac
1{(-p^2)}D_{+}[e^{ip(z-z^{\prime })}\delta _{(-)}(\theta -\theta ^{\prime
})],
$$
$$
<\chi _{(-)}{}^{|I|}(\ddot u)\chi _{(-)}{}^{|J|}(\ddot u^{\prime })>=i\delta
^{|I||J|}\frac{\partial _{=}D_{+}}{\Box }\delta _{(-)}^3(\ddot u,\ddot
u^{\prime })=
$$
$$
\frac{\delta ^{|I||J|}}{(2\pi )^2}\int d^{d_2}p\frac{p_{=}}{p^2}%
D_{+}[e^{ip(z-z^{\prime })}\delta _{(-)}^3(\ddot u,\ddot u^{\prime })].
$$
Finally, we remark that background--quantum decompositions of the action
(55) for heterotic string define the Feynman rules (vertixes and
propagators) for the corresponding generalization of the two--dimensional
sigma model which are basic for a perturbation quantum formalism in higher
order anisotropic spaces.

\section{Green--Schwarz Action in DVS--Bundles}

The Green--Scwarz covariant action ( GS--action ) for superstrings can be
 considered as a two dimensional $\sigma $--model with Wess--Zumino--Wit\-ten
term and flat, dimension $d=10,$ s--space as the tangent space \cite
{gs,howe1,howe2,mar}. The GS--action was generalized for the curved
background {\sf N=1}, $d=10$ of the superspace under the condition that
motion equations hold \cite{witten} and under similar conditions for {\sf N=2}%
, $d=10$ supergravity \cite{grisaru}.

The GS--action in dimensions $d=3,4,6,10$ can be represented as
$$
I=\frac 12\int d^2z\sqrt{-\gamma }\gamma ^{\ddot e\ddot \imath }\partial
_{\ddot e}u^{<\alpha >}\partial _{\ddot \imath }u^{<\beta >}(l_{<\alpha >}^{<%
\underline{\alpha }>}l_{<\beta >}^{<\underline{\beta }>})\widehat{\eta }_{<%
\underline{\alpha }><\underline{\beta }>}+\eqno(60)
$$
$$
\frac 12\int d^2z\varepsilon ^{\ddot e\ddot \imath }\partial _{\ddot
e}u^{<\alpha >}\partial _{\ddot \imath }u^{<\beta >}B_{<\alpha ><\beta >}
$$
by using of the flat  vielbein $l_{<\alpha >}^{<\underline{\alpha }>}$
and 2--form
$$
B=\frac 12\delta u^{<\alpha >}\Lambda \delta u^{<\beta >}B_{<\alpha ><\beta
>}
$$
in the flat $d=10$ s--space with coordinates $u^{<\alpha >}.$ An important
role in formulation of the GS--action plays the fact that 3--form $H=dB$ is
closed, $dH=0.$ Because $H$ is s--invariant the 2--form $B$ changes on
complete derivation under s--transforms%
$$
\delta ^{\star }H=0,\delta ^{\star }dB=d\delta ^{\star }B=0,\delta ^{\star
}B=d\Lambda ,
$$
where $\delta ^{\star }$ is dual to $d,$ which ensures the s--invariance of
the GS--action.

We generalize the action (60) for higher order anisotropic s--spaces by
changing the flat vielbein $l_{<\alpha >}^{<\underline{\alpha }>}$ into the
locally anisotropic, $\widehat{E}_{<\alpha >}^{<\underline{\alpha }>},$
with a possible dependence of the Lagrangian on scalar fields (see \cite
{robb} for locally isotropic spaces),%
$$
I=\frac 12\int d^2z[\sqrt{-\gamma }\gamma ^{\ddot e\ddot \imath }\partial
_{\ddot e}u^{<\alpha >}\partial _{\ddot \imath }u^{<\beta >}(E_{<\alpha >}^{<%
\underline{\alpha }>}E_{<\beta >}^{<\underline{\beta }>})\widehat{\eta }_{<%
\underline{\alpha }><\underline{\beta }>}+\eqno(61)
$$
$$
\varepsilon ^{\ddot e\ddot \imath }\partial _{\ddot e}u^{<\alpha >}\partial
_{\ddot \imath }u^{<\beta >}E_{<\alpha >}^{<\underline{\alpha }>}E_{<\beta
>}^{<\underline{\beta }>}B_{<\underline{\alpha }><\underline{\beta }>}+
$$
$$
VV_{\underline{\ddot e}}^{\ddot e}\overline{\Psi }^{|\underline{I}|}\gamma ^{%
\underline{\ddot e}}(\partial _{\ddot e}\delta _{|\underline{I}||\underline{J%
}|}+E_{\ddot e}^{<\alpha >}A_{|\underline{I}||\underline{J}|<\alpha >})\Psi
^{|\underline{J}|}],
$$
where $P$ is a scalar function, $\gamma _{\ddot e\ddot \imath }=V_{\ddot e}^{%
\underline{\ddot e}}V_{\ddot \imath }^{\underline{i}}\gamma _{\underline{%
\ddot e}\underline{\ddot \imath }},~V=\det (V_{\ddot e}^{\underline{\ddot e}%
}),u^{<\alpha >}$ are coordinates of the higher order anisotropic s--space
and $\Psi ^{|\underline{J}|}$ are two dimensional (heterotic) MW--fermions
in the fundamental representation of the interior symmetry group $Gr$.

As background s--fields we shall consider
$$
E^{<\underline{\alpha }>}=d\widehat{u}^{<\beta >}E_{<\beta >}^{<\underline{%
\alpha }>}(z),~B=\frac 12E^{<\underline{\alpha }>}E^{<\underline{\beta }%
>}B_{<\underline{\alpha }><\underline{\beta }>}(z),
$$
$$
A_{|\underline{I}||\underline{J}|}=A_{|\underline{I}||\underline{J}|<%
\underline{\alpha }>}E^{<\underline{\alpha }>},
$$
where%
$$
E_{\ddot a}^{<\underline{\alpha }>}\equiv \partial _{\ddot a}u^{<\alpha
>}E_{<\alpha >}^{<\underline{\alpha }>}=(\widehat{E}_{\ddot a}^{<\underline{%
\alpha }>},E_{\ddot a}^{<\underline{\alpha }>}=E_{\ddot a}^{\underline{\ddot
a}}).
$$
S--fields $A_{|\underline{I}||\underline{J}|}$ belong to the adjoint
representation of the interior symmetry group $Gr$.

The action (61) is invariant under transforms\\ $\left( \triangle E^{<%
\underline{\alpha }>}\equiv \triangle u^{<\alpha >}E_{<\alpha >}^{<%
\underline{\alpha }>}\right) :$%
$$
\triangle \widehat{E}^{<\underline{\alpha }>}=0,\triangle E^{\ddot
e}=2(\Gamma _{<\underline{\alpha }>})^{\ddot e\ddot o}\widehat{E}_{\ddot
e}^{<\underline{\alpha }>}V_{\underline{\ddot e}}^{\ddot e}k_{\ddot o}^{%
\underline{\ddot o}},\eqno(62)
$$
$$
\triangle \Psi ^{|\underline{I}|}=-(\triangle E^{<\underline{\alpha }>})A_{<%
\underline{\alpha }>}^{|\underline{I}||\underline{J}|}\Psi ^{|\underline{J}%
|},\triangle V_{\ddot e}^{\underline{\ddot e}}=-\frac 12(\gamma _{\ddot
e\ddot \imath }+\varepsilon _{\ddot e\ddot \imath })M^{\underline{\ddot o}%
\ddot \imath }k_{\underline{\ddot o}}^{\underline{\ddot e}},
$$
where $\varepsilon _{\ddot e\ddot \imath }$ is defined as a two--dimensional
tensor, parameter $k_{<\alpha >}^{\underline{\ddot e}}$ is anti--self--dual
as a two--vector, juggling of indices of $d$ --dimensional Dirac matrices is
realized by using the $d$--dimensional matrix of charge conjugation and ,
for simplicity, we can consider matrices $(\Gamma _{<\underline{\alpha }%
>})^{\ddot e\ddot o}$ as symmetric; we shall define below the value $%
M^{<\alpha >\ddot \imath }.$

The variation of action (61) under transforms (62) can be written as%
$$
\triangle I=\int d^2z\frac 12[e^P\triangle (V\gamma ^{\ddot e\ddot \imath })%
\widehat{E}_{\ddot e}^{<\underline{\alpha }>}\widehat{E}_{\ddot \imath }^{<%
\underline{\beta }>}\eta _{<\underline{\alpha }><\underline{\beta }>}+%
\eqno(63)
$$
$$
\triangle E^{\ddot o}(e^P(V\gamma ^{\ddot e\ddot \imath })\widehat{E}_{\ddot
e}^{<\underline{\alpha }>}\widehat{E}_{\ddot \imath }^{<\underline{\beta }%
>}\eta _{<\underline{\alpha }><\underline{\beta }>}D_{\ddot
o}P-2e^P\triangle (V\gamma ^{\ddot e\ddot \imath })E_{\ddot e}^{<\underline{%
\alpha }>}E_{\ddot \imath }^{<\underline{\beta }>}T_{<\underline{\alpha }%
>\ddot o<\underline{\beta }>}+
$$
$$
\varepsilon ^{\ddot e\ddot \imath }E_{\ddot e}^{<\underline{\alpha }%
>}E_{\ddot \imath }^{<\underline{\beta }>}H_{<\underline{\alpha }><%
\underline{\beta }>\ddot o}-V\overline{\Psi }^{|\underline{I}|}\gamma
^{\ddot e}\Psi ^{|\underline{J}|}E_{\ddot e}^{<\underline{\alpha }>}F_{<%
\underline{\alpha }>\ddot o|\underline{I}||\underline{J}|})+\triangle (VV_{%
\underline{\ddot e}}^{\ddot e})\overline{\Psi }^{|\underline{I}|}\gamma ^{%
\underline{\ddot e}}({\cal D}_{\ddot e}\Psi )^{|\underline{J}|},
$$
where ${\cal D}_{\ddot e}$ is the $Gr$--covariant derivation and the torsion
2--form $T^{<\underline{\alpha }>}$, the strength 3--form $H$ and the
supersymmetric Yang--Mills strength 2--form
$F^{|\underline{I}||\underline{J}%
|}$ are respectively defined by relations%
$$
T^{<\underline{\alpha }>}=dE^{<\underline{\alpha }>}+E^{<\underline{\beta }%
>}\Omega _{<\underline{\beta }>}^{<\underline{\alpha }>}=\frac 12E^{<%
\underline{\beta }>}E^{<\underline{\gamma }>}T_{<\underline{\beta }><%
\underline{\gamma }>}^{<\underline{\alpha }>},
$$
$$
H=dB=\frac 16E^{<\underline{\gamma }>}E^{<\underline{\beta }>}E^{<\underline{%
\alpha }>}H_{<\underline{\alpha }><\underline{\beta }><\underline{\gamma }%
>},
$$
$$
F^{|\underline{I}||\underline{J}|}=dA^{|\underline{I}||\underline{J}|}+A^{|%
\underline{I}||\underline{K}|}A^{|\underline{K}||\underline{J}|}=\frac 12E^{<%
\underline{\beta }>}E^{<\underline{\gamma }>}F_{<\underline{\beta }><%
\underline{\gamma }>}^{|\underline{I}||\underline{J}|}.
$$

The variation of action (63) under transforms (63) vanishes if and only
if there are satisfied the next conditions:

1) 3--form $H$ is closed under condition
$$
(\Gamma ^{<\underline{\alpha }>})_{\ddot e\ddot \imath }(\Gamma _{<%
\underline{\alpha }>})_{\ddot o\ddot u}+(\Gamma ^{<\underline{\alpha }%
>})_{\ddot \imath \ddot o}(\Gamma _{<\underline{\alpha }>})_{\ddot e\ddot
u}+(\Gamma ^{<\underline{\alpha }>})_{\ddot o\ddot e}(\Gamma _{<\underline{%
\alpha }>})_{\ddot \imath \ddot u}=0,
$$
which holds for dimensions $d=3,4,6,10;$

2) there are imposed constraints
$$
\widehat{T}_{\ddot e\ddot \imath }^{<\underline{\alpha }>}=-i(\widehat{%
\Gamma }^{<\underline{\alpha }>})_{\ddot e\ddot \imath },~\widehat{\eta }%
_{<\gamma >(<\alpha >}\widehat{T}_{<\beta >)\ddot a}^{<\gamma >}=\widehat{%
\eta }_{<\alpha ><\beta >}B_{\ddot a},~F_{\ddot e\ddot \imath }^{|\underline{%
I}||\underline{J}|}=0,\eqno(64)
$$
$$
F_{<\underline{\alpha }>\ddot \imath }^{|\underline{I}||\underline{J}|}=(%
\widehat{\Gamma }_{<\underline{\alpha }>})_{\ddot e\ddot \imath }w^{\ddot e|%
\underline{I}||\underline{J}|},~H_{\ddot e\ddot \imath \ddot o}=0,~H_{\ddot
e\ddot \imath <\underline{\alpha }>}=-ie^P(\widehat{\Gamma }_{<\underline{%
\alpha }>})_{\ddot e\ddot \imath },~
$$
$$
\widehat{H}_{\ddot u<\alpha ><\beta >}=2e^P(\Gamma _{<\underline{\alpha }%
>})_{\ddot o}^{\ddot e}(\Gamma _{<\underline{\beta }>})_{\ddot u}^{\ddot
o}H_{\ddot e};
$$

3) The coefficient $M^{\underline{\ddot o}\ddot \imath }$ from (62) is
taken in the form%
$$
M^{\underline{\ddot o}\ddot \imath }=4iE^{\underline{\ddot o}\ddot \imath }-4%
\widehat{E}_{<\underline{\alpha }>}^{\ddot \imath }(\widehat{\Gamma }^{<%
\underline{\alpha }>})^{\underline{\ddot o}\underline{\ddot u}}H_{\underline{%
\ddot u}}-\Psi ^{|\underline{I}|}\gamma ^{\ddot \imath }\Psi ^{|\underline{J}%
|}w_{|\underline{I}||\underline{J}|}^{\underline{\ddot o}}e^{-P},
$$
where
$$
D_{\underline{\ddot u}}P+2H_{\underline{\ddot u}}-2B_{\underline{\ddot u}%
}=0;
$$

4) The last term in (63) vanishes because of conditions of chirality,
$$
\Psi ^{|\underline{I}|}=-\gamma _5\Psi ^{|\underline{I}|}.
$$

In the locally isotropic s--gravity it is known \cite{gates1,nillson,atick}
that s--field equations of type (64) are compatible with Bianchi
identities and can be interpreted as standard constraints defining
supergravity in the superspace. Considering locally adapted to
N--connections geometric objects end equations (64) we obtain a variant of
higher order anisotropic supergravity
(see \cite{vlasg,v96jpa1,v96jpa2,v96jpa3} for details on locally anisotropic
supergravity) which for dimensions $d=n+m=10$ contain, distinguished by the
N--connection structure, motion equations of {\sf N=1} of higher order
anisotropic supergravity and super--Yang--Mills matter.

The above presented constructions can be generalized in order to obtain a
variant of higher order anisotropic {\sf N=2, }$d=10$ supergravity from
so--called IIB--superstrings \cite{grisaru} (which, in our case, will be
modified to be locally anisotropic). To formulate the model we use a locally
adapted s--vielbein 1--form $E^{<\underline{\alpha }>}=\delta u^{<\alpha
>}E_{<\alpha >}^{<\underline{\alpha }>},$ a $SO(1,9)\otimes U(1)$ connection
1--form $\Omega _{<\underline{\alpha }>}^{<\underline{\beta }>},$ a 2--form
of complex potential $A$ and one real 4--form $B.$ Strengths are defined in
a standard manner:%
$$
T^{<\underline{\alpha }>}=DE^{<\underline{\alpha }>}=\delta E^{<\underline{%
\alpha }>}+E^{<\underline{\beta }>}\Omega _{<\underline{\beta }>}^{<%
\underline{\alpha }>},\eqno(65)
$$
$$
R_{<\underline{\alpha }>}^{<\underline{\beta }>}=\delta \Omega _{<\underline{%
\alpha }>}^{<\underline{\beta }>}+\Omega _{<\underline{\alpha }>}^{<%
\underline{\gamma }>}\Omega _{<\underline{\gamma }>}^{<\underline{\beta }>},
$$
$$
F=\delta A,~G=\delta B+A\overline{F}-\overline{A}F.
$$

On the mass shell (on locally anisotropic spaces we shall consider
distinguished metrics) ds--tensors (65) are expressed in terms of one
scalar s--field $V\in SU(1,1):$%
$$
V=\left(
\begin{array}{cc}
q & s \\
\overline{u} & \overline{v}
\end{array}
\right) ,q\overline{q}-s\overline{s}=1.
$$
Excluding a scalar by using the local U(1)--invariance we can use the first
components of complex s--fields $\left( q,s\right) $ as physical scalar
fields of the theory.

The constraints defining IIB supergravity in $d=n+m=10$ higher order
anisotropic s--space contain equations (on every anisotropic ''shell'', in
locally adapted frames, they generalize constraints of IIB supergavity \cite
{howe}):%
$$
T_{\underleftarrow{b_p}\underleftarrow{c_p}}^{a_p}=T_{\underrightarrow{b_p}%
\underrightarrow{c_p}}^{a_p}=0,T_{\underleftarrow{b_p}\underrightarrow{c_p}%
}^{a_p}=-i\sigma _{\underleftarrow{b_p}\underrightarrow{c_p}}^{a_p},T_{%
\underleftarrow{b_p}c_p}^{a_p}=T_{\underrightarrow{b_p}c_p}^{a_p}=0,%
\eqno(66)
$$
$$
F_{\underleftarrow{a_p}\underleftarrow{b_p}\underleftarrow{c_p}}=F_{%
\underleftarrow{a_p}\underleftarrow{b_p}\underrightarrow{c_p}}=F_{%
\underleftarrow{a_p}\underrightarrow{b_p}\underrightarrow{c_p}}=F_{%
\underrightarrow{a_p}\underrightarrow{b_p}\underrightarrow{c_p}}=F_{a_p%
\underleftarrow{b_p}\underrightarrow{c_p}}=0,
$$
$$
F_{a_p\underleftarrow{b_p}\underleftarrow{c_p}}=-iq(\sigma _{a_p})_{%
\underleftarrow{b_p}\underleftarrow{c_p}},~F_{a_pb_p\underrightarrow{c_p}%
}=-q(\sigma _{a_pb_p})_{\underrightarrow{c_p}}^{\underrightarrow{d_p}%
}\Lambda _{\underrightarrow{d_p}},
$$
$$
F_{a_p\underrightarrow{b_p}\underrightarrow{c_p}}=-is(\sigma _{a_p})_{%
\underrightarrow{b_p}\underrightarrow{c_p}},~F_{a_pb_p\underleftarrow{c_p}%
}=s(\sigma _{a_pb_p})_{\underleftarrow{c_p}}^{\underleftarrow{d_p}}\overline{%
\Lambda }_{\underleftarrow{d_p}},
$$
$$
H_{a_p\underleftarrow{b_p}\underleftarrow{c_p}}=-i(q-\overline{s})(\sigma
_{a_p})_{\underleftarrow{b_p}\underleftarrow{c_p}},~H_{a_p\underrightarrow{%
b_p}\underrightarrow{c_p}}=-i(q-\overline{s})(\sigma _{a_p})_{%
\underrightarrow{b_p}\underrightarrow{c_p}},
$$
$$
H_{a_pb_p\underrightarrow{c_p}}=-(q-\overline{s})(\sigma _{a_pb_p})_{%
\underrightarrow{c_p}}^{\underrightarrow{d_p}}\Lambda _{\underrightarrow{d_p}%
},~H_{a_p\underrightarrow{b_p}\underrightarrow{c_p}}=-i(\overline{q}%
-s)(\sigma _{a_pb_p})_{\underleftarrow{c_p}}^{\underleftarrow{d_p}}\overline{%
\Lambda }_{\underleftarrow{d_p}},
$$
where
$$
(\sigma _{a_p}\sigma _{b_p})_{\underrightarrow{c_p}}^{\underrightarrow{d_p}%
}=(\sigma _{a_pb_p})_{\underrightarrow{c_p}}^{\underrightarrow{d_p}}+\eta
_{a_pb_p}\delta _{\underrightarrow{c_p}}^{\underrightarrow{d_p}},\eqno(67)
$$
the same formula holds for ''$\underleftarrow{}"$--underlined spinors,
spinor $\Lambda _{\underrightarrow{d_p}}$ will be used below for fixing of $%
U(1)$ gauge and 3--form $H\equiv F+\overline{F}=\delta \widetilde{B}$ is
real and closed (this condition is crucial in the construction of the
GS--action on the background of IIB--supergravity, with respect to usual
isotropic string model see \cite{grisaru}).

The action (60) can be generalized for {\sf N=2} higher order anisotropic
s--spaces in this manner:%
$$
I_S=\int d^2z\{\sqrt{-\gamma }\gamma ^{\ddot \imath \ddot u}P(q,s)\widehat{E}%
_{\ddot \imath }^{<\underline{\alpha }>}\widehat{E}_{\ddot u}^{<\underline{%
\beta }>}\widehat{\eta }_{<\underline{\alpha }><\underline{\beta }>}+\frac
12\varepsilon ^{\ddot \imath \ddot u}E_{\ddot \imath }^{<\alpha >}E_{\ddot
u}^{<\beta >}\widetilde{B}_{<\alpha ><\beta >},\eqno(68)
$$
where $P(q,s)$ is a function of scalar fields $q$ and $s,$ and
$$
E_{\ddot \imath }^{<\underline{\alpha }>}\equiv \partial _{\ddot \imath
}u^{<\alpha >}E_{<\alpha >}^{<\underline{\alpha }>}=(\widehat{E}_{\ddot
\imath }^{<\underline{\alpha }>},...,E_{\ddot \imath }^{\underleftarrow{a_p}%
},...,\overline{E}_{\ddot \imath }^{\underrightarrow{b_p}},...).
$$

The variation of the Lagrangian in (67) under respective k--transforms of
type (62) can be written as
$$
\triangle L=\sqrt{-\gamma }\gamma ^{\ddot \imath \ddot u}PE_{\ddot \imath
}^{<\gamma >}\triangle E^{<\beta >}T_{<\beta ><\gamma >}^{<\underline{\delta
}>}E_{\ddot u}^{<\underline{\beta }>}\widehat{\eta }_{<\underline{\delta }><%
\underline{\beta }>}+\eqno(69)
$$
$$
\frac 12\varepsilon ^{\ddot \imath \ddot u}E_{\ddot \imath }^{<\gamma
>}E_{\ddot u}^{<\beta >}\triangle E^{<\tau >}H_{<\tau ><\beta ><\gamma >}+
$$
$$
\frac 12[\triangle (\sqrt{-\gamma }\gamma ^{\ddot \imath \ddot u}P)+\sqrt{%
-\gamma }\gamma ^{\ddot \imath \ddot u}\triangle P]\widehat{E}_{\ddot \imath
}^{<\underline{\alpha }>}\widehat{E}_{\ddot u}^{<\underline{\beta }>}%
\widehat{\eta }_{<\underline{\alpha }><\underline{\beta }>}.
$$

Taking into account constraints (66) we can express variation (69) as
$$
\triangle L=(\{-i(\varepsilon ^{\ddot \imath \ddot u}\gamma ^{\ddot e\ddot
a}(q-\overline{s})+\gamma ^{\ddot \imath \ddot u}\varepsilon ^{\ddot e\ddot
a}P)\widehat{E}_{\ddot e}^{<\gamma >}\widehat{E}_{\ddot \imath }^{<\beta
>}(\sigma _{<\beta >}\sigma _{<\gamma >})_{\underleftarrow{\tau }}^{%
\underleftarrow{\delta }}E_{\ddot u}^{\underleftarrow{\tau }}k_{\ddot a%
\underleftarrow{\delta }}-
$$
$$
i[(-\gamma )^{-1/2}\varepsilon ^{\ddot \imath \ddot u}\varepsilon ^{\ddot
e\ddot a}(\overline{q}-s)+\sqrt{-\gamma }\gamma ^{\ddot \imath \ddot
u}\gamma ^{\ddot e\ddot a}P]\widehat{E}_{\ddot e}^{<\gamma >}\widehat{E}%
_{\ddot \imath }^{<\beta >}(\sigma _{<\beta >}\sigma _{<\gamma >})_{%
\underleftarrow{\tau }}^{\underleftarrow{\delta }}\overline{E}_{\ddot u}^{%
\underleftarrow{\tau }}k_{\ddot a\underleftarrow{\delta }}+
$$
$$
\widehat{E}_{\ddot \imath }^{<\gamma >}(\widehat{E}_{\ddot u}^{<\beta >}%
\widehat{E}_{\ddot e}^{<\alpha >}\widehat{\eta }_{<\beta ><\alpha >})(\sigma
_{<\gamma >})^{\underleftarrow{\alpha }\underleftarrow{\beta }}[\varepsilon
^{\ddot \imath \ddot u}\gamma ^{\ddot e\ddot a}(\overline{q}-s)\overline{%
\Lambda }_{\underleftarrow{\alpha }}+
$$
$$
(-\gamma )^{-1/2}\varepsilon ^{\ddot \imath \ddot u}\varepsilon ^{\ddot
e\ddot a}(q-\overline{s})\Lambda _{\underleftarrow{\alpha }}]k_{\ddot a%
\underleftarrow{\beta }}\}+h.c.)+
$$
$$
\frac 12\triangle (\sqrt{-\gamma }\gamma ^{\ddot \imath \ddot u})\widehat{E}%
_{\ddot \imath }^{<\gamma >}\widehat{E}_{\ddot u}^{<\beta >}\widehat{\eta }%
_{<\gamma ><\beta >}P+\frac 12\sqrt{-\gamma }\gamma ^{\ddot \imath \ddot u}%
\widehat{E}_{\ddot \imath }^{<\gamma >}\widehat{E}_{\ddot u}^{<\beta >}%
\widehat{\eta }_{<\gamma ><\beta >}\triangle P,
$$
where $h.c.$ denotes Hermitian conjugation.

Using relation (67) and fixing the $U(1)$--gauge as to have
$$
P=q-\overline{s}=\overline{q}-s~\mbox{ and }~\triangle P=(q-\overline{s}%
)(\triangle E^{\underleftarrow{\alpha }}\Lambda _{\underleftarrow{\alpha }%
}-\triangle \overline{E}^{\underleftarrow{\alpha }}\overline{\Lambda }_{%
\underleftarrow{\alpha }})
$$
we can obtain zero values of the coefficients before $(\sigma _{a_pb_p})$%
--terms. The rest of terms in $\triangle L$ vanish for a corresponding
fixing of the variation $\triangle (\sqrt{-\gamma }\gamma ^{\ddot \imath
\ddot u}).$

So, in this section we have constructed a model of higher order anisotropic
IIB--superstring on the background of IIB supergravity with broken chiral $%
U(1)$--subgroup of the supersymmetry $SU(1,1)$--group of automorphisms of $%
N=2,d=n+m=10$ supergravity. We omit in this works calculus for
supersymmetric $\beta$--functions.

\section{Fermi Strings in Higher Order An\-isot\-rop\-ic Spaces}

There are some types of Fermi strings in dependence of the number {\sf %
N=0,1,2,4} of supersymmetry generators (see, for instance, \cite
{hul,ket,ketn} for reviews and basic references on this classification for
locally isotropic strings). The aim of this section is to present basic
results on Fermi and heterotic strings on higher order anisotropic
backgrounds: the construction of actions and calculation of superconformal
anomalies.

We note that there are two possible interpretations of models considered
in this work. On one hand they can be considered as locally anisotropic
supersymmetric two dimensional supersymmetric nonlinear sigma models
connected with supergravity. Under quantization of such type theories the
superconformal invariance is broken; the two point functions of graviton
and gravitino, computed from the quantum effective action can became
nontrivial in result (this conclusion was made \cite{abd} for locally
isotropic sigma models and, in general, holds good for locally anisotropic
generalizations). On the other hand our models can be interpreted as Fermi
strings on higher order anisotropic background. We shall follow the second
treatment.

The effective ''off--shell'' action $\Gamma $ for a (infinite) set of locally
anisotropic fields is introduced (in a manner similar to \cite{fts}) as%
$$
\Gamma [G,H,...]=\sum_\chi e^{\widetilde{\sigma }\chi }\int [D\gamma _{\ddot
e\ddot \imath }^{(e)}][Du^{<\alpha >}]\exp (-I),
$$
$$
I=\frac 1{2\pi \alpha ^{\prime }}\int d^2z\{\frac 12\sqrt{\gamma ^{(e)}}%
\gamma _{(e)}^{\ddot a\ddot \imath }\partial _{\ddot a}\widehat{u}^{<\alpha
>}\partial _{\ddot \imath }\widehat{u}^{<\beta >}G_{<\alpha ><\beta >}(u)+
$$
$$
\varepsilon ^{\ddot a\ddot \imath }\partial _{\ddot a}\widehat{u}^{<\alpha
>}\partial _{\ddot \imath }\widehat{u}^{<\beta >}H_{<\alpha ><\beta
>}(u)+...\},
$$
where dots are used instead of possible sources (with higher order
derivations), compatible with the reparametrization invariance, of another
types of perturbations and $\gamma _{\ddot e\ddot \imath }^{(e)}$ is the
Euclid two dimensional metric. In (69) we consider in explicit form the
components of locally anisotropic graviton and antisymmetric d--tensor and,
for simplicity, omit the dilaton field and topological considerations.
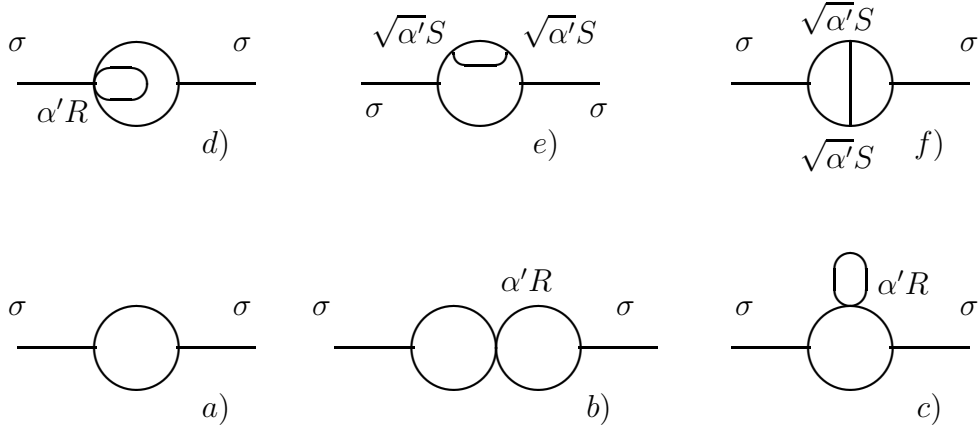
\begin{figure}[htbp]
\begin{picture}(400,180) \setlength{\unitlength}{1pt}
\thicklines

\put(70,35){\circle{30}}
\put(85,35){\line(1,0){30}}
\put(25,35){\line(1,0){30}}
\put(85,2){\makebox(30,20){$a)$}}
\put(95,40){\makebox(30,20){$\sigma$}}
\put(10,40){\makebox(30,20){$\sigma$}}

\put(190,35){\circle{30}}
\put(145,35){\line(1,0){30}}
\put(222,35){\circle{30}}
\put(237,35){\line(1,0){30}}
\put(230,2){\makebox(30,20){$b)$}}
\put(240,40){\makebox(30,20){$\sigma$}}
\put(125,40){\makebox(30,20){$\sigma$}}
\put(200,51){\makebox(35,20){${\alpha}' R$}}

\put(340,35){\circle{30}}
\put(295,35){\line(1,0){30}}
\put(355,35){\line(1,0){30}}
\put(340,61){\oval(12,20)}
\put(355,2){\makebox(30,20){$c)$}}
\put(370,40){\makebox(30,20){$\sigma$}}
\put(285,40){\makebox(30,20){$\sigma$}}
\put(345,50){\makebox(30,20){${\alpha}' R$}}

\put(70,135){\circle{30}}
\put(85,135){\line(1,0){30}}
\put(25,135){\line(1,0){30}}
\put(85,102){\makebox(30,20){$d)$}}
\put(95,140){\makebox(30,20){$\sigma$}}
\put(10,140){\makebox(30,20){$\sigma$}}
\put(64,135){\oval(20,12)}
\put(27,115){\makebox(30,20){${\alpha}' R$}}

\put(200,135){\circle{30}}
\put(155,135){\line(1,0){30}}
\put(215,135){\line(1,0){30}}
\put(210,102){\makebox(30,20){$e)$}}
\put(230,115){\makebox(30,20){$\sigma$}}
\put(145,115){\makebox(30,20){$\sigma$}}
\put(200,147){\oval(20,10)[b]}
\put(210,145){\makebox(40,20){$\sqrt{{\alpha}'} S$}}
\put(153,145){\makebox(40,20){$\sqrt{{\alpha}'} S$}}

\put(340,135){\circle{30}}
\put(295,135){\line(1,0){30}}
\put(355,135){\line(1,0){30}}
\put(340,119){\line(0,1){32}}
\put(355,102){\makebox(30,20){$f)$}}
\put(370,140){\makebox(30,20){$\sigma$}}
\put(285,140){\makebox(30,20){$\sigma$}}
\put(320,150){\makebox(30,20){$\sqrt{{\alpha}'} S$}}
\put(320,98){\makebox(30,20){$\sqrt{{\alpha}'} S$}}

\end{picture}
\caption{\it The  diagrams defining the con\-form\-al anomaly of a closed
 boson string  in a higher order an\-isot\-rop\-ic space}
\end{figure}

The problem of calculation of $\Gamma $ is split into two steps: the first
is the calculation of the effective action on an higher order anisotropic
with fixed Euler characteristic $\chi $ then the averaging on all metrics
and topologies. In order to solve the first task we shall compute%
$$
\exp \{-W[G,H,g]\}=\int [D\eta ]\exp \{-I[u+\upsilon (z),g]\}.
$$

The general structure $W$ is constructed from dimension and symmetry
considerations \cite{poly}
$$
W=-\frac{\beta (u)}\varepsilon \int R(z)\sqrt{\gamma _{(e)}}d^2z+\gamma
(u)\int (R(z)\sqrt{\gamma _{(e)}})_z\Box _{zz^{\prime }}^{-1}(R(z)\sqrt{%
\gamma _{(e)}})_{z^{\prime }}d^2zd^2z^{\prime },\eqno(73)
$$
where the dimensional regularization $(\varepsilon =2-d_2$ on world
sheet),\thinspace $\gamma _{(e)}$ is the determinant of the two dimensional
metric, $R=R_{\quad \ddot a\ddot \imath }^{\ddot a\ddot \imath },$ one holds
the relation $\beta (u)=4\gamma (u)$ for dimensionless functions (we can
computer them as perturbations on $\alpha ^{\prime })$ and note that the
second term in (73) is the Weyl anomaly which shall be computed by using
normal locally adapted to N--connection coordinates on higher order
anisotropic space.

For a two--dimensional conformal flat two--dimensional, $d_2=2,$ metric $%
\gamma _{\ddot e\ddot \imath }^{(e)}$ we find
$$
\gamma _{\ddot e\ddot \imath }^{(e)}=e^{2\sigma }\delta _{\ddot e\ddot
\imath },~R\left( u\right) =-2e^{-2\sigma }\Box \sigma ,~\gamma
_{(e)}^{\ddot a\ddot \imath }=e^{-2\sigma }\delta ^{\ddot a\ddot \imath },~%
\sqrt{\gamma _{(e)}}=e^{2\sigma }.
$$

From decomposition $\gamma _{\ddot e\ddot \imath }^{(e)}=\delta _{\ddot
e\ddot \imath }+h_{\ddot e\ddot \imath }$ with respect to first and second
order terms on $h$ we have
$$
\sqrt{\gamma ^{(e)}}\gamma _{(e)}^{\ddot a\ddot \imath }=[1+\varepsilon
\sigma (z)+\frac \varepsilon 2(d-4)\sigma ^2(z)]\delta ^{\ddot a\ddot \imath
}
$$
when $h_{\ddot e\ddot \imath }=2\sigma h_{\ddot e\ddot \imath },h\equiv
h_{\ddot a}^{\ddot a}=2\sigma d_{(2)}.$ We give similar formulas for the
frame decomposition of two metric%
$$
e^{1/2}e_{\underline{\ddot e}}^{\ddot e}=(1+\frac 12\varepsilon \sigma
)\delta _{\underline{\ddot e}}^{\ddot e},e^{-1/2}e_{\ddot e\underline{\ddot e%
}}=(1-\frac 12\varepsilon \sigma )\delta _{\ddot e\underline{\ddot e}},
$$
which are necessary for dealing with spinors in curved spaces.

Transforming the quantum field $\zeta \rightarrow \sqrt{2\pi \alpha ^{\prime
}}\zeta ,\,$where $\zeta $ is the tangent d--vector in the point $u\in {\cal %
E}^{<z>}$ of the higher order anisotropic space and using the
conformal--flat part of the two dimensional metric we obtain this effective
action necessary for further calculations%
$$
I_{int}^{eff}=\int d^2z[\frac 13(2\pi \alpha ^{\prime })^{1/2}\varepsilon
^{\ddot e\ddot \imath }H_{<\alpha ><\beta ><\gamma >}(u)\partial _{\ddot
e}\zeta ^{<\alpha >}\partial _{\ddot \imath }\zeta ^{<\beta >}\zeta
^{<\gamma >}+\eqno(74)
$$
$$
\frac{2\pi \alpha ^{\prime }}6\varepsilon \sigma (z)R_{<\alpha ><\beta
><\gamma ><\delta >}(u)\partial _{\ddot e}\zeta ^{<\alpha >}\partial _{\ddot
e}\zeta ^{<\delta >}\zeta ^{<\beta >}\zeta ^{<\gamma >}+
$$
$$
\frac 12\varepsilon \sigma (z)\partial _{\ddot e}\zeta ^{<\alpha >}\partial
_{\ddot e}\zeta ^{<\alpha >}+\frac{2\pi \alpha ^{\prime }}6R_{<\alpha
><\beta ><\gamma ><\delta >}(u)\partial _{\ddot e}\zeta ^{<\alpha >}\partial
_{\ddot e}\zeta ^{<\delta >}\zeta ^{<\beta >}\zeta ^{<\gamma >}+
$$
$$
\frac{2\pi \alpha ^{\prime }}4D_{<\alpha >}H_{<\alpha ><\beta ><\gamma
>}(u)\varepsilon ^{\ddot e\ddot \imath }\partial _{\ddot e}\zeta ^{<\beta
>}\partial _{\ddot \imath }\zeta ^{<\gamma >}\zeta ^{<\alpha >}\zeta
^{<\delta >},
$$
where $(u)$ higher order anisotropic space coordinates not depending on two
coordinates $z.$ So, the anomaly in (74) takes the form%
$$
4\vartheta \int d^2z~\sigma (z)\Box \sigma (z)\eqno(73)
$$

If $\widetilde{\sigma }$ in (70)\ is the connection on topologies constant
the first term in (17) can be absorbed by the renormalization of this
connection constant. The set of two--loop diagrams defining (73) is
illustrated in the figure 1. We note that we must take into account
tedpoles because of the compactness of the string world sheet there are not
infrared divergences.

In the one--loop approximation (figure 1) we find%
$$
\beta ^{(1)}=\frac 1{24\pi }(n+m_1+...+m_z),~\gamma ^{(1)}=\frac 1{96\pi
}(n+m_1+...+m_z);
$$
there is correspondence with classical results \cite{poly} if we consider a
trivial distinguishing of the space--time dimension $n_E=n+m_1+...+m_z$.

The two--loop terms (figure 1) from (73) are computed as
$$
b)=-\frac{\pi \alpha ^{\prime }\varepsilon ^2R}{3(2\pi )^6}\int d^2k\sigma
(k)\sigma (-k)\int d^{d_2}pd^{d_2}q\times
$$
$$
\frac{(p\cdot k-p^2)^2(q\cdot k-q^2)+(p\cdot k-p^2)^2(p\cdot q)(q\cdot k-q^2)%
}{p^2(k-p)^2q^2(k-q)^2},
$$
$$
c)=\frac{\pi \alpha ^{\prime }\varepsilon ^2R}{3(2\pi )^4}G(0)\int
d^2k\sigma (k)\sigma (-k)\int d^{d_2}p\frac{(p\cdot k-p^2)^2}{p^2(k-p)^2},
$$
$$
d)=-\frac{\pi \alpha ^{\prime }\varepsilon ^2R}{3(2\pi )^4}G(0)\int
d^2k\sigma (k)\sigma (-k)\int d^{d_2}p\frac{(p\cdot k-p^2)^2}{p^2(k-p)^2},
$$
$$
e)+f)=\frac{2\varepsilon ^2\pi \alpha ^{\prime }H_{<\alpha ><\beta ><\gamma
>}^2}{(2\pi )^6}\varepsilon ^{\ddot e\ddot \imath }\varepsilon ^{\ddot
a\ddot u}\int d^2k\sigma (k)\sigma (-k)\int d^{d_2}pd^{d_2}q\times
$$
$$
\frac 1{p^2(k-p)^2q^2(k-p-q)^2}\times
$$
$$
\{\frac{(k\cdot p-p^2)(k\cdot q-q^2)p_{\ddot e}(k-q)_{\ddot \imath
}(k-q)_{\ddot a}q_{\ddot u}}{(k-q)^2}+
$$
$$
\frac{(k\cdot p-p^2)^2q_{\ddot e}(k-q)_{\ddot \imath }q_{\ddot
a}(k-p)_{\ddot u}}{(k-p)^2}\},
$$
where
$$
\sigma (z)\equiv \frac 1{(2\pi )^2}\int d^2p\sigma (p)\exp (-ipx).
$$

The contributions of tedpole nonvanishing diagrams mutually compensate. The
sum of the rest of contributions results in the anomaly%
$$
\gamma ^{(2)}=\frac{\alpha ^{\prime }}{64\pi }(-R+\frac 13H^2)=\frac{\alpha
^{\prime }}{64\pi }(-\widehat{R}-\frac 23H^2),
$$
where $H^2=H_{<\alpha ><\beta ><\gamma >}H^{<\alpha ><\beta ><\gamma >}$ and
$\widehat{R}$ is the scalar curvature with torsion.

Computing $W$ in the leading order on $\alpha ^{\prime }$ for the closed
boson string, it is not difficult to find the effective action $\Gamma $ for
massless perturbations of the string (of the metric $G_{<\alpha ><\beta >}$
and field $H_{<\alpha ><\beta >})$ on the tree $\left( \chi =2\right) $
level. Taking into account the identity%
$$
\int (R(z)\sqrt{\gamma _{(e)}})_z\Box _{zz^{\prime }}^{-1}(R(z)\sqrt{\gamma
_{(e)}})_{z^{\prime }}d^2zd^2z^{\prime }=16\pi
$$
for the metric on sphere we find%
$$
\Gamma ^{(0)}[G,H]\sim \int \frac{\delta ^{n_E}u}{(2\pi \alpha ^{\prime
})^{n_E/2}}\sqrt{G(u)}[1+\frac{\alpha ^{\prime }}4(-R+\frac 13H^2)],%
\eqno(74)
$$
where $n_E=n+m_1+...+m_z$ is the dimension of higher order anisotropic space.
 Formula (74) generalizes for such type of spaces (scalar curvature $R$
and torsion $H$ are for, distinguished by N--connection, on la--space) of
that presented in \cite{cfmp,ckp}. The cosmological constant in (74)
arises due to the taxion modes in the spectrum of boson strings and is
absent for superstrings. From vanishing of $\beta $--functions for $\widehat{%
R}$ and  tacking into account the contributions of reparametrization
ghosts \cite{poly} into the anomaly of boson string we obtain into the
leading approximation
$$
\beta =\frac{n_E-8}{24\pi }+\frac{\alpha ^{\prime }}{16\pi }(-\widehat{R}%
-\frac 23H^2)+...,
$$
$$
\gamma =\frac{n_E-26}{96\pi }+\frac{\alpha ^{\prime }}{64\pi }(-\widehat{R}%
-\frac 23H^2)+....
$$

In consequence, the correction to the critical dimension is
$$
D_c=26+\alpha ^{\prime }H^2+{\it O}([\alpha ^{\prime }]^2).\eqno(75)
$$
We emphasize that torsion in (75) can be interpreted in a different manner
that in the case of locally isotropic theories where $H_{...}$ is considered
as an antisymmetric strength of a specific gauge field (see the
Wess--Zumino--Witten model \cite{wesz,witten}). For locally anisotropic
spaces we suggested the idea that the $H_{...}$--terms are induced by the
distinguished components of torsions of, in our case, higher order
anisotropic spaces.
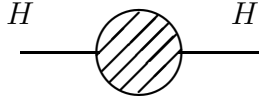
\begin{figure}[htbp]
\begin{picture}(400,70) \setlength{\unitlength}{1pt}
\thicklines

\put(220,35){\circle{30}}
\put(235,35){\line(1,0){30}}
\put(175,35){\line(1,0){30}}
\put(245,40){\makebox(30,20){$H$}}
\put(160,40){\makebox(30,20){$H$}}
\put(209,24){\line(1,1){22}}
\put(213,21){\line(1,1){21}}
\put(220,20){\line(1,1){16}}
\put(206,28){\line(1,1){21}}
\put(204,34){\line(1,1){16}}
\end{picture}
\caption{\it The  diagrams defining supergravitational and superconformal
 anomalies
in the theory of higher order anisotropic superstrings}
\end{figure}

The presented in this section constructions can be generalized for the case
of {\sf N=1 }and {\sf N=2} higher order anisotropic Fermi strings. Let
decompose action $I_S[e_{\ddot e}^{\underline{\ddot e}},u+\zeta (z),\psi
(z)] $ to within forth order on quantum fields $\zeta ^{<\alpha >}$ and $%
\psi _{\ddot e}^{<\alpha >}(z).$ After redefinition $\zeta \rightarrow \sqrt{%
2\pi \alpha ^{\prime }}\zeta ,\psi \rightarrow e^{-1/4}\sqrt{2\pi \alpha
^{\prime }}\psi $ we obtain the next additional to (72) term:%
$$
I_{F(int)}=\int d^2z[\frac i4\varepsilon \sigma \overline{\psi }^{<%
\underline{\alpha }>}\gamma ^{\ddot e}\partial _{\ddot e}\psi ^{<\underline{%
\alpha }>}-
$$
$$
\frac i2\sqrt{2\pi \alpha ^{\prime }}(1-\frac \varepsilon 2\sigma
)H_{<\alpha ><\beta ><\gamma >}\overline{\psi }^{<\alpha >}\gamma ^{\ddot
e}(\partial _{\ddot e}\zeta ^{<\gamma >})\psi ^{<\beta >}+
$$
$$
i\frac{\pi \alpha ^{\prime }}2(1+\frac \varepsilon 2\sigma )R_{<\alpha
><\beta ><\gamma ><\delta >}\overline{\psi }^{<\alpha >}\gamma ^{\ddot
e}(\partial _{\ddot e}\zeta ^{<\delta >})\psi ^{<\beta >}\zeta ^{<\gamma >}+
$$
$$
\frac{\pi \alpha ^{\prime }}{16}\widehat{R}_{<\alpha ><\beta ><\gamma
><\delta >}\overline{\psi }^{<\alpha >}(1+\gamma _5)\psi ^{<\gamma >}%
\overline{\psi }^{<\beta >}(1+\gamma _5)\psi ^{<\delta >}].
$$

Fixing the gauges
$$
{\sf N=1:~}e_{\ddot e}^{\underline{\ddot e}}=e^\sigma \delta _{\ddot e}^{%
\underline{\ddot e}}~\psi _{\ddot e}=\frac 12\gamma _{\ddot e}\lambda ,
$$
$$
{\sf N=2:~}e_{\ddot e}^{\underline{\ddot e}}=e^\sigma \delta _{\ddot e}^{%
\underline{\ddot e}}~\psi _{\ddot e}=\frac 12\gamma _{\ddot e}\lambda
,~A^{\ddot e}=\frac 12\varepsilon ^{\ddot e\ddot u}\partial _{\ddot u}\rho ,
$$
where $\lambda $ is the Maiorana (${\sf N=1}$ ) or Dirac (${\sf N=2}$ )
spinor. Because of supersymmetry it is enough \cite{fts} to computer only
the coefficient before Weyl anomaly in order to get the superconformal
anomalies. The one--loop results are
$$
{\sf N=1:}~\beta ^{(1)}=\frac{n_E}{16\pi },~\gamma ^{(1)}=\frac{n_E}{64\pi }%
,
$$
$$
{\sf N=2:}~\beta ^{(1)}=\frac{n_E}{8\pi },~\gamma ^{(1)}=\frac{n_E}{32\pi },
$$
from which, taking into account the reparametrization and superconformal
ghosts we obtain these values of critical dimension ($\gamma =0):$%
$$
{\sf N=1:}~\beta _t^{(1)}=\frac{n_E-2}{16\pi },~\gamma ^{(1)}=\frac{n_E-10}{%
64\pi },
$$
$$
{\sf N=2:}~\beta ^{(1)}=\frac{n_E}{8\pi },~\gamma ^{(1)}=\frac{n_E-2}{32\pi }%
.
$$

From formal point of view in the two--loop approximation we must consider
diagrams b)--m) from
fig. 2. By straightforward calculations by using
methods similar to those presented in \cite{fts} we conclude that all
two--loop contributions b)--m) vanish. In result we conclude that for Fermi
strings one holds the next formulas:%
$$
{\sf N=1:}~\beta =\frac{n_E-2}{16\pi }-\frac{\alpha ^{\prime }H^2}{24\pi }%
+...,~\gamma =\frac{n_E-10}{64\pi }-\frac{\alpha ^{\prime }H^2}{96\pi },
$$
$$
{\sf N=2:}~\beta =\frac{n_E}{8\pi }-\frac{\alpha ^{\prime }H^2}{24\pi }%
+...,~\gamma =\frac{n_E-2}{32\pi }-\frac{\alpha ^{\prime }H^2}{96\pi }
$$
or%
$$
{\sf N=1:}~D_c=10+\frac 23\alpha ^{\prime }H^2+...,
$$
$$
{\sf N=2:}~D_c=2+\frac 13\alpha ^{\prime }H^2...
$$
in the leading order on $\alpha ^{\prime }.$

Finally, we note that because $H^2$ contains components of N--connection and
torsion of d--connection on higher order anisotropic space we conclude that
a possible local anisotropy of space--time can change the critical dimension
of Fermi strings.

\section{Anomalies in Locally Anisotropic $\sigma $--Mo\-dels}

Anomalies in quantum field theories are considered beginning with works \cite
{adler} and \cite{bell}. Conformal and gravitational anomalies have been
analyzed in \cite{chris,duff,alvar} (see also reviews \cite
{zwz,morozov,ket,ketn,gsw}). The aim of this section is to investigate
anomalies in higher order anisotropic (1,0)--superspaces.

\subsection{One--loop calculus}

Supergravitational and conform anomalies of heterotic locally anisotropic $%
\sigma $--models connected with (1,0) higher order anisotropic supergravity
are defined by the finite (anomaly) parts of diagrams of self--energy type
(fig. 2). In order to compute anomalies it is necessary to consider all
the vertexes of the theory with no more than the linear dependence on
potentials $H_{+}^{=},H_{=}^{\ddagger }$ (see subsection 6.2 and section
9 for denotations on higher order anisotropic heterotic superstrings). We
consider this ''effective'' action (without ghosts):%
$$
I=I_0+I_{int},~I_{int}=I_0^{\prime }+I_1,\eqno(76)
$$
where
$$
I_0=-\int d^3z^{-}[iD_{+}\zeta ^{<\alpha >}\partial _{=}\zeta ^{<\beta >}%
\widehat{g}_{<\alpha ><\beta >}+\chi _{-}^{|I|}D_{+}\chi _{-}^{|I|}],
$$
$$
I_0^{\prime }=-\frac 12\int d^3z^{-}\{iD_{+}\zeta ^{<\alpha >}\partial
_{=}\zeta ^{<\beta >}(\zeta ^{<\gamma >}\widehat{A}_{<\gamma ><\alpha
><\beta >}+
$$
$$
\zeta ^{<\delta >}\zeta ^{<\gamma >}\widehat{B}_{<\delta ><\gamma ><\alpha
><\beta >}+\zeta ^{<\varepsilon >}\zeta ^{<\delta >}\zeta ^{<\gamma >}%
\widehat{{\cal D}}_{<\varepsilon ><\delta ><\gamma ><\alpha ><\beta >}+
$$
$$
\zeta ^{<\tau >}\zeta ^{<\varepsilon >}\zeta ^{<\delta >}\zeta ^{<\gamma >}%
\widehat{M}_{<\varepsilon ><\delta ><\gamma ><\tau ><\alpha ><\beta >})+\chi
_{(-)}^{|I|}\chi _{(-)}^{|J|}D_{+}\zeta ^{<\beta >}\times
$$
$$
(\zeta ^{<\alpha >}\widehat{C}_{<\alpha ><\beta >}^{|I||J|}+\zeta ^{<\gamma
>}\zeta ^{<\alpha >}\widehat{E}_{<\gamma ><\alpha ><\beta >}^{|I||J|}+\zeta
^{<\varepsilon >}\zeta ^{<\gamma >}\zeta ^{<\alpha >}\widehat{K}%
_{<\varepsilon ><\gamma ><\alpha ><\beta >}^{|I||J|}),
$$
$$
I_1=-\frac 12\int d^3z^{-}\{(\widehat{g}_{<\alpha ><\beta >}+\zeta ^{<\gamma
>}\widehat{A}_{<\gamma ><\alpha ><\beta >}+
$$
$$
\zeta ^{<\delta >}\zeta ^{<\gamma >}\widehat{B}_{<\delta ><\gamma ><\alpha
><\beta >}+\zeta ^{<\varepsilon >}\zeta ^{<\delta >}\zeta ^{<\gamma >}%
\widehat{{\cal D}}_{<\varepsilon ><\delta ><\gamma ><\alpha ><\beta >}+
$$
$$
\zeta ^{<\tau >}\zeta ^{<\varepsilon >}\zeta ^{<\delta >}\zeta ^{<\gamma >}%
\widehat{M}_{<\varepsilon ><\delta ><\gamma ><\tau ><\alpha ><\beta
>})+[iH_{+}^{=}\partial _{=}\zeta ^{<\alpha >}\partial _{=}\zeta ^{<\beta
>}+
$$
$$
\frac 12D_{+}\zeta ^{<\alpha >}(D_{+}H_{=}^{\ddagger })D_{+}\zeta ^{<\beta
>}+i(D_{+}\zeta ^{<\alpha >})H_{=}^{\ddagger }(\partial _{\ddagger }\zeta
^{<\beta >})]+
$$
$$
\chi _{(-)}^{|I|}H_{+}^{=}\partial _{=}\chi _{(-)}^{|J|}+\chi
_{(-)}^{|I|}\chi _{(-)}^{|J|}H_{+}^{=}\partial _{=}\zeta ^{<\beta >}\times
$$
$$
(\zeta ^{<\alpha >}\widehat{C}_{<\alpha ><\beta >}^{|I||J|}+\zeta ^{<\gamma
>}\zeta ^{<\alpha >}\widehat{E}_{<\gamma ><\alpha ><\beta >}^{|I||J|}+\zeta
^{<\varepsilon >}\zeta ^{<\gamma >}\zeta ^{<\alpha >}\widehat{K}%
_{<\varepsilon ><\gamma ><\alpha ><\beta >}^{|I||J|})+
$$
$$
\frac i{4\pi }\sum\limits_p^6\frac 1{p!}\zeta ^{<\alpha _1>}...\zeta
^{<\alpha _p>}\widehat{{\cal D}}_{<\alpha _1>}...\widehat{{\cal D}}_{<\alpha
_p>}\Phi (D_{+}\partial _{\ddagger }H_{=}^{\ddagger }+\partial
_{=}^2H_{+}^{=})\},
$$
where%
$$
\widehat{A}_{<\gamma ><\alpha ><\beta >}=\frac 23\widehat{H}_{<\gamma
><\alpha ><\beta >},
$$
$$
\widehat{B}_{<\delta ><\gamma ><\alpha ><\beta >}=\frac 13\widehat{R}%
_{<\delta ><\alpha ><\beta ><\gamma >}+\frac 12\widehat{{\cal D}}_{<\delta >}%
\widehat{H}_{<\gamma ><\alpha ><\beta >},
$$
$$
\widehat{{\cal D}}_{<\varepsilon ><\delta ><\gamma ><\alpha ><\beta >}=\frac
15\widehat{{\cal D}}_{<\varepsilon >}\widehat{{\cal D}}_{<\delta >}\widehat{H%
}_{<\gamma ><\alpha ><\beta >}+
$$
$$
\frac 16\widehat{{\cal D}}_{<\varepsilon >}\widehat{R}_{<\delta ><\alpha
><\beta ><\gamma >}+\frac 2{15}R_{<\delta ><\varepsilon >[<\alpha >}^{<\tau
>}\widehat{H}_{<\beta >]<\gamma ><\tau >},
$$
$$
\widehat{M}_{<\varepsilon ><\delta ><\gamma ><\tau ><\alpha ><\beta >}=\frac
1{20}\widehat{{\cal D}}_{<\varepsilon >}\widehat{{\cal D}}_{<\delta >}%
\widehat{R}_{<\gamma ><\alpha ><\beta ><\tau >}+
$$
$$
\frac 2{45}\widehat{R}_{<\beta ><\tau ><\gamma ><\vartheta >}R_{<\delta
><\varepsilon ><\alpha >}^{<\vartheta >}+\widehat{{\cal D}}_{<\varepsilon >}%
\widehat{{\cal D}}_{<\delta >}\widehat{{\cal D}}_{<\gamma >}\widehat{H}%
_{<\tau ><\alpha ><\beta >}+
$$
$$
\frac 1{18}\widehat{{\cal D}}_{<\varepsilon >}\widehat{R}_{<\gamma ><\delta
>[<\alpha >}^{<\vartheta >}\widehat{H}_{<\beta >]<\tau ><\vartheta >}+\frac
19\widehat{{\cal D}}_{<\gamma >}\widehat{H}_{<\tau ><\vartheta >[<\alpha >}%
\widehat{R}_{|<\delta ><\varepsilon >|<\beta >]}^{<\vartheta >},
$$
$$
\widehat{C}_{<\alpha ><\beta >}^{|I||J|}=-\frac 12\widehat{F}_{<\alpha
><\beta >}^{|I||J|},\widehat{E}_{<\gamma ><\alpha ><\beta >}^{|I||J|}=-\frac
13\widehat{{\cal D}}_{<\gamma >}\widehat{F}_{<\alpha ><\beta >}^{|I||J|},
$$
$$
\widehat{K}_{<\varepsilon ><\gamma ><\alpha ><\beta >}^{|I||J|}=-\frac
1{24}(3\widehat{{\cal D}}_{<\varepsilon >}\widehat{{\cal D}}_{<\gamma >}%
\widehat{F}_{<\alpha ><\beta >}^{|I||J|}+\widehat{F}_{<\varepsilon ><\tau
>}^{|I||J|}R_{<\alpha ><\gamma ><\beta >}^{<\tau >}.
$$
For simplicity, in this section we shall omit tilde ''$\widetilde{}"$ over
geometric objects (such as curvatures and torsions computed for Christoffel
distinguished symbols (39)) but maintain hats ''$\widehat{}"$ in order to
point out even components on the s--space).

We write the supergravitational anomaly in this general form (see \cite{wesz}
for locally isotropic models):%
$$
\frac 1{32\pi }\int d^3z^{-}\{\gamma _1D_{+}H_{+}^{=}\frac{\partial _{=}^4}{%
\Box }H_{+}^{=}-i\gamma _2D_{+}H_{=}^{\ddagger }\frac{\partial _{=}^3}{\Box }%
H_{=}^{\ddagger }\},
$$
where background depending coefficients will be defined from a perturbation
calculus on $\alpha ^{\prime }$ by using (76).
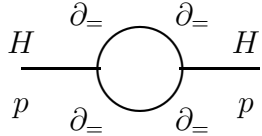
\begin{figure}[htbp]
\begin{picture}(400,70) \setlength{\unitlength}{1pt}
\thicklines

\put(220,35){\circle{30}}
\put(235,35){\line(1,0){30}}
\put(175,35){\line(1,0){30}}
\put(245,35){\makebox(30,20){$H$}}
\put(160,35){\makebox(30,20){$H$}}
\put(245,10){\makebox(30,20){$p$}}
\put(160,10){\makebox(30,20){$p$}}
\put(185,45){\makebox(30,20){${\partial}_{=}$}}
\put(225,45){\makebox(30,20){${\partial}_{=}$}}
\put(185,5){\makebox(30,20){${\partial}_{=}$}}
\put(225,5){\makebox(30,20){${\partial}_{=}$}}
\end{picture}
\caption{\it $(1,0)$ supergraf defining the one--loop anomaly}
\end{figure}

For computation of supergrafs we use a standard techniques \cite{gates} of
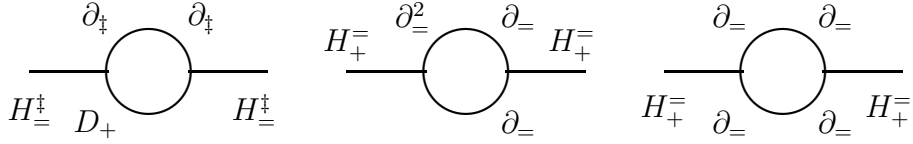
\begin{figure}[htbp]
\begin{picture}(400,70) \setlength{\unitlength}{1pt}
\thicklines

\put(70,35){\circle{30}}
\put(85,35){\line(1,0){30}}
\put(25,35){\line(1,0){30}}
\put(95,10){\makebox(30,20){$H^{\ddagger}_{=}$}}
\put(10,10){\makebox(30,20){$H^{\ddagger}_{=}$}}
\put(35,45){\makebox(30,20){${\partial}_{\ddagger}$}}
\put(75,45){\makebox(30,20){${\partial}_{\ddagger}$}}
\put(35,5){\makebox(30,20){${D}_{+}$}}

\put(190,35){\circle{30}}
\put(205,35){\line(1,0){30}}
\put(145,35){\line(1,0){30}}
\put(215,35){\makebox(30,20){$H^{=}_{+}$}}
\put(130,35){\makebox(30,20){$H^{=}_{+}$}}
\put(155,45){\makebox(30,20){${\partial}_{=}^2$}}
\put(195,45){\makebox(30,20){${\partial}_{=}$}}
\put(195,5){\makebox(30,20){${\partial}_{=}$}}

\put(310,35){\circle{30}}
\put(325,35){\line(1,0){30}}
\put(265,35){\line(1,0){30}}
\put(335,10){\makebox(30,20){$H^{=}_{+}$}}
\put(250,10){\makebox(30,20){$H^{=}_{+}$}}
\put(275,45){\makebox(30,20){${\partial}_{=}$}}
\put(315,45){\makebox(30,20){${\partial}_{=}$}}
\put(275,5){\makebox(30,20){${\partial}_{=}$}}
\put(315,5){\makebox(30,20){${\partial}_{=}$}}

\end{picture}
\caption{\it $(1,0)$ supergrafs defining the one--loop dilaton contribution
 to the anomaly}
\end{figure}
reducing to integrals in momentum space which is standard practice in
quantum field theory. For instance, the one loop diagram corresponding to
figure 3 is computed%
\begin{figure}[htbp]
\begin{picture}(400,70) \setlength{\unitlength}{1pt}
\thicklines

\put(220,35){\circle{30}}
\put(235,35){\line(1,0){30}}
\put(175,35){\line(1,0){30}}
\put(245,10){\makebox(30,20){$H_{=}^{\ddagger}$}}
\put(160,10){\makebox(30,20){$H^{=}_{+}$}}
\put(209,24){\line(1,1){22}}
\put(213,21){\line(1,1){21}}
\put(220,20){\line(1,1){16}}
\put(206,28){\line(1,1){21}}
\put(204,34){\line(1,1){16}}
\end{picture}
\caption{\it The  diagrams  being unessential for calculation of anomalies}
\end{figure}
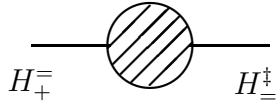

$$
\frac 1{(2\pi )^2}\int d^2k\frac{k_{=}^2(k_{=}+p_{=})^2}{k^2(k+p)^2}=-\frac
i{24\pi }\frac{p_{=}^4}{p^2}.
$$
Diagrams of type illustrated on fig. 4 give rise only to local
contributions in the anomaly and are not considered because of dimensional
considerations.
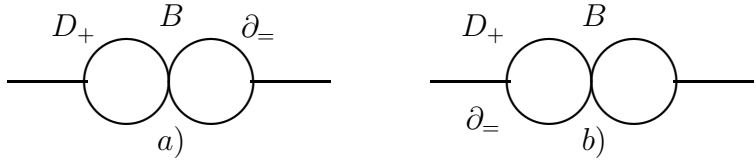
\begin{figure}[htbp]
\begin{picture}(400,70) \setlength{\unitlength}{1pt}
\thicklines

\put(90,35){\circle{30}}
\put(122,35){\circle{30}}
\put(137,35){\line(1,0){30}}
\put(45,35){\line(1,0){30}}
\put(55,45){\makebox(30,20){$D_{+}$}}
\put(92,50){\makebox(30,20){$B$}}
\put(125,45){\makebox(30,20){${\partial}_{=}$}}
\put(92,2){\makebox(30,20){$a)$}}

\put(250,35){\circle{30}}
\put(282,35){\circle{30}}
\put(297,35){\line(1,0){30}}
\put(205,35){\line(1,0){30}}
\put(210,45){\makebox(30,20){$D_{+}$}}
\put(252,50){\makebox(30,20){$B$}}
\put(210,10){\makebox(30,20){${\partial}_{=}$}}
\put(252,2){\makebox(30,20){$b)$}}
\end{picture}
\caption{\it $B$--depending two--loop corrections to the anomaly}
\end{figure}

We note that comparing with similar locally isotropic results \cite{ketn}
the torsions $H_{\cdot }^{\cdot }$ are generated by components of the
distinguished torsion of the higher order anisotropic background.

\subsection{Two--loop calculus}

The two--loop $B$--depending corrections to the anomaly are defined by
diagrams illustrated in fig. 6. 
The one--loop results including ghosts (56) are similar to locally
isotropic ones \cite{gates}. Thus we present a brief summary (we must take
into account the splitting of dimensions in higher order anisotropic spaces):%
$$
W_{eff}^{1-loop}=\frac 1{96\pi }\int d^3z^{-}\{(n+m_1+...+m_z-26+\frac{N_E}%
2)D_{+}H_{+}^{=}\frac{\partial _{=}^4}{\Box }H_{+}^{=}-
$$
$$
\frac{3i}2(n+m_1+...+m_z-10)D_{+}H_{=}^{\ddagger }\frac{\partial _{\ddagger
}^3}{\Box }H_{=}^{\ddagger }\}.\eqno(77)
$$
If the action (77) is completed by local conterterms
$$
H_{+}^{=}\Box H_{=}^{\ddagger },~S\partial _{=}^2H_{+}^{=},~S\partial
_{\ddagger }D_{+}H_{=}^{\ddagger },~S\partial _{=}D_{+}S
$$
and conditions $N_E-(n+m_1+...+m_z)=22$ and $\gamma _1=\gamma _2\equiv
\gamma $ are satisfied, we obtain from (77) a gauge invariant action:%
$$
W_{eff}+W_{loc}=\frac 1{16\pi }(n+m_1+...+m_z-10)\int d^3z^{-}\Sigma ^{+}(%
\frac{D_{+}}{\Box })\Sigma ^{+},\eqno(78)
$$
\begin{figure}[htbp]
\begin{picture}(400,400) \setlength{\unitlength}{1pt}
\thicklines

\put(70,35){\circle{30}}
\put(85,35){\line(1,0){30}}
\put(25,35){\line(1,0){30}}
\put(70,20){\line(0,1){30}}
\put(70,5){\makebox(15,10){$D_{+}$}}
\put(95,10){\makebox(30,20){$a)$}}
\put(35,45){\makebox(30,20){${\partial}_{=}$}}
\put(70,52){\makebox(15,10){$D_{+}$}}
\put(35,5){\makebox(30,20){${\partial}_{=}$}}

\put(190,35){\circle{30}}
\put(205,35){\line(1,0){30}}
\put(145,35){\line(1,0){30}}
\put(190,20){\line(0,1){30}}
\put(215,10){\makebox(30,20){$b)$}}
\put(155,45){\makebox(30,20){${\partial}_{=}$}}
\put(195,5){\makebox(30,20){${\partial}_{=}$}}
\put(180,5){\makebox(15,10){$D_{+}$}}
\put(185,52){\makebox(15,10){$D_{+}$}}

\put(310,35){\circle{30}}
\put(325,35){\line(1,0){30}}
\put(265,35){\line(1,0){30}}
\put(310,20){\line(0,1){30}}
\put(335,10){\makebox(30,20){$c)$}}
\put(275,45){\makebox(30,20){${\partial}_{=}$}}
\put(300,47){\makebox(30,20){$D_{+}$}}
\put(275,5){\makebox(30,20){$D_{+}$}}
\put(315,5){\makebox(30,20){${\partial}_{=}$}}
\put(70,135){\circle{30}}
\put(85,135){\line(1,0){30}}
\put(25,135){\line(1,0){30}}
\put(70,120){\line(0,1){30}}
\put(100,110){\makebox(15,10){$d)$}}
\put(57,153){\makebox(15,10){${D}_{+}$}}
\put(75,145){\makebox(30,20){${\partial}_{=}$}}
\put(35,105){\makebox(30,20){${D}_{+}$}}
\put(70,105){\makebox(30,20){${\partial}_{=}$}}
\put(190,135){\circle{30}}
\put(205,135){\line(1,0){30}}
\put(145,135){\line(1,0){30}}
\put(190,120){\line(0,1){30}}
\put(225,110){\makebox(15,10){$e)$}}
\put(155,145){\makebox(30,20){${D}_{+}$}}
\put(195,145){\makebox(30,20){${\partial}_{=}$}}
\put(195,105){\makebox(30,20){${D}_{+}$}}
\put(155,105){\makebox(30,20){${\partial}_{=}$}}
\put(310,135){\circle{30}}
\put(325,135){\line(1,0){30}}
\put(265,135){\line(1,0){30}}
\put(310,120){\line(0,1){30}}
\put(345,115){\makebox(15,10){$f)$}}
\put(275,145){\makebox(15,10){${D}_{+}$}}
\put(290,150){\makebox(30,20){${\partial}_{=}$}}
\put(290,100){\makebox(30,20){${\partial}_{=}$}}
\put(325,105){\makebox(15,10){$D_{+}$}}
\put(70,235){\circle{30}}
\put(85,235){\line(1,0){30}}
\put(25,235){\line(1,0){30}}
\put(70,220){\line(0,1){30}}
\put(105,210){\makebox(15,10){$i)$}}
\put(35,245){\makebox(30,20){${D}_{+}$}}
\put(75,245){\makebox(30,20){${\partial}_{=}$}}
\put(50,203){\makebox(30,20){${\partial}_{=}$}}
\put(85,210){\makebox(15,10){$D_{+}$}}
\put(190,235){\circle{30}}
\put(205,235){\line(1,0){30}}
\put(145,235){\line(1,0){30}}
\put(190,247){\oval(20,10)[b]}
\put(215,205){\makebox(30,20){$j)$}}
\put(175,250){\makebox(30,20){${\partial}_{=}$}}
\put(155,245){\makebox(15,10){$D_{+}$}}
\put(177,223){\makebox(30,20){${\partial}_{=}$}}
\put(205,240){\makebox(30,20){$D_{+}$}}
\put(310,235){\circle{30}}
\put(325,235){\line(1,0){30}}
\put(265,235){\line(1,0){30}}
\put(310,247){\oval(20,10)[b]}
\put(335,205){\makebox(30,20){$k)$}}
\put(275,245){\makebox(30,20){$D_{+}$}}
\put(315,245){\makebox(30,20){$D_{+}$}}
\put(272,232){\makebox(30,20){${\partial}_{=}$}}
\put(297,223){\makebox(30,20){${\partial}_{=}$}}
\put(115,335){\circle{30}}
\put(130,335){\line(1,0){30}}
\put(70,335){\line(1,0){30}}
\put(115,347){\oval(20,10)[b]}
\put(140,310){\makebox(30,20){$l)$}}
\put(90,350){\makebox(15,10){$D_{+}$}}
\put(99,328){\makebox(20,10){${\partial}_{=}$}}
\put(120,350){\makebox(15,10){$D_{+}$}}
\put(130,336){\makebox(20,10){${\partial}_{=}$}}
\put(265,335){\circle{30}}
\put(280,335){\line(1,0){30}}
\put(220,335){\line(1,0){30}}
\put(265,347){\oval(20,10)[b]}
\put(290,310){\makebox(30,20){$m)$}}
\put(243,350){\makebox(15,10){${D}_{+}$}}
\put(270,350){\makebox(15,10){${D}_{+}$}}
\put(227,331){\makebox(30,20){${\partial}_{=}$}}
\put(273,331){\makebox(30,20){${\partial}_{=}$}}

\end{picture}
\caption{\it $A^2$--depending two--loop corrections to the anomaly}
\end{figure}
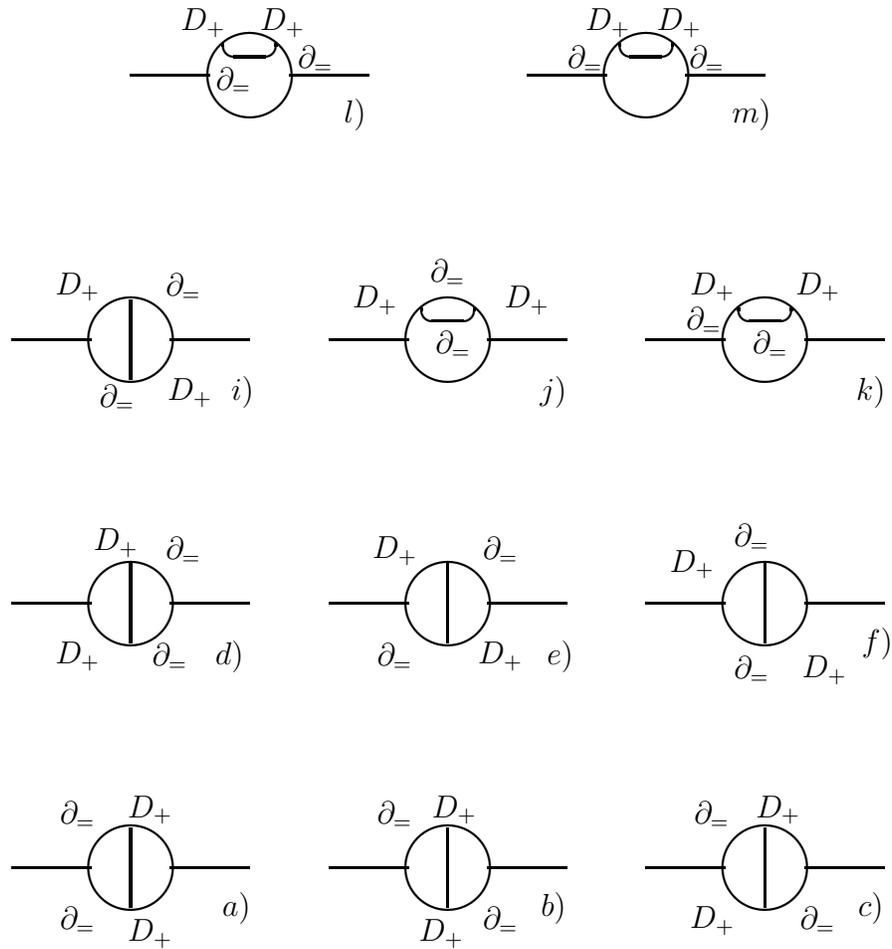

The action (78) is the conformal anomaly of our model ($\Sigma ^{+}$
depends on $S).$ The critical parameters of the higher order anisotropic
heterotic string
$$
n+m_1+...+m_z=10~\mbox{and } ~N_E=32
$$
make up the conditions of cancelation of it anomalies (the original locally
isotropic result was obtained in \cite{ghmr}). We can also compute and add
the one--loop dilaton (see fig. 5) contribution to (78):
$$
W_{eff}^{(1,\Phi )}=\frac{{\cal D}^2\Phi }{128\pi ^2}\int
d^3z^{-}[iD_{+}H_{=}^{\ddagger }\frac{\partial _{\ddagger }^3}{\Box }%
H_{=}^{\ddagger }-D_{+}H_{+}^{=}\frac{\partial _{=}^4}{\Box }H_{+}^{=}],
$$
$$
\gamma ^{(1,\Phi )}=-\frac 1{4\pi }{\cal D}^2\Phi .
$$
computed
$$
a)=I_1(p)=\int \frac{d^2kd^2q}{16\pi ^4}\frac{%
q_{=}^2(q_{=}+p_{=})(k_{=}+p_{=})}{(q^2-\mu ^2)[(q+p)^2-\mu ^2]}\times
$$
$$
\frac{k^2}{(k^2-\mu ^2)[(k+p)^2-\mu ^2]}=\frac 1{64\pi ^2}\frac{p_{=}^4}{p^2}%
+{\it O}\left( \mu ^2\right) ,
$$
$$
b)=I_2(p)=\int \frac{d^2kd^2q}{16\pi ^4}\frac{q_{=}(q_{=}+p_{=})}{(q^2-\mu
^2)[(q+p)^2-\mu ^2]}\times
$$
$$
\frac{k^2(k_{=}+p_{=})^2}{(k^2-\mu ^2)[(k+p)^2-\mu ^2]}=-\frac 1{32\pi ^2}%
\frac{p_{=}^4}{p^2}+{\it O}\left( \mu ^2\right) ,
$$
where the mass parameter $\mu ^2$ is used as a infrared regulator.

Two--loop A$^2$--dependent diagrams are illustrated in fig. 7. The
supergrafs f),i) and j) are given by the momentum integral $I_f=I_i=I_j=I(p)
:$
$$
I(p)=\int \frac{d^2kd^2q}{(2\pi )^4}\frac{k_{=}q_{=}(k_{=}+q_{=}+p_{=})^2}{%
k^2q^2(k+q+p)^2}=\frac{p_{=}^4}{96\pi ^2p^2};%
$$
to this integral there are also proportional the anomaly parts of diagrams
a)-e),k),l) and m). The rest of possible $A^2$--type diagrams do not
contribute to the anomaly part of the effective action. After a
straightforward computation of two loop diagrams we have
$$
\gamma ^{2-loop}=\frac 1{16\pi }(-\widetilde{R}+\frac 13H^2);
$$
there are not dilaton contributions in the two--loop approximation.

The anomaly coefficient $\gamma $ is connected with the central charge of
Virasoro superalgebra of heterotic string on the background of massless
modes (under the conditions of vanishing of $\beta $-functions or,
equivalently, if the motion equations are satisfied, see \cite
{ts86,ts87n,shore,osborn,curci,nied,brus}):%
$$
2\gamma +\alpha ^{\prime }\left( {\cal D}_{<\beta >}\Phi \right) ^2=%
\widetilde{\beta }^\Phi \equiv \beta ^\Phi -\frac 14\beta _{<\alpha ><\beta
>}^gg^{<\alpha ><\beta >}=
$$
$$
\frac{n+m_1+...+m_z-10}2+\frac{\alpha ^{\prime }}2L_{eff},
$$
$$
<T>_{-}=\frac 1{4\pi }\widetilde{\beta }^\Phi \Sigma ^{+}+...,
$$
where $<T>_{-}$is the averaged supertrace, $\beta _{<\alpha ><\beta >}^g$ is
the metric $\beta $-function, $\widetilde{\beta }^\Phi $ is the dilaton $%
\beta $-function and by dots there are denoted the terms vanishing on the
motion equations. We note that the $\beta $-functions and effective
Lagrangian are defined in the string theory only with the exactness of
redefinition of fields \cite{brus}. From a standard calculus according the
perturbation theory on $\alpha ^{\prime }$ we have
$$
\beta _{<\alpha ><\beta >}^g=\alpha ^{\prime }(\widetilde{R}_{<\alpha
><\beta >}-H_{<\alpha ><\beta >}^2)
$$
and
$$
S_{eff}^{(0)}\equiv \int d^{10}X\sqrt{|g|}L_{eff}^{(0)}=
$$
$$
\frac 12\int d^{10}X\sqrt{|g|}\{-\widetilde{R}-4{\cal D}^2\Phi +4({\cal D}%
_{<\alpha >}\Phi )^2+\frac 13H^2\},
$$
where%
$$
H_{<\alpha ><\beta >}^2\equiv H_{<\alpha ><\gamma ><\delta >}H_{<\beta
>}^{\quad <\gamma ><\delta >},
$$
which is a higher order anisotropic generalization of models developed in
\cite{berg,chapline,bellucci}.

\section{Discussion and Conclusions}

We have explicitly constructed a new class of superspaces with higher order
anisotropy. The status of the results in this work and the relevant open
questions are discussed as follows.

From the generally mathematical point of view it is possible a definition of
a supersymmetric differential geometric structure imbedding both type of
supersymmetric extensions of Finsler and Lagrange geometry as well various
Kaluza--Klein superspaces. The first type of superspaces, considered as
locally anisotropic, are  characterized by nontrivial nonlinear connection
structures and corresponding distinguishing of geometric objects and basic
structure equations. The second type as a rule is associated to trivial
nonlinear connections and higher order dimensions. A substantial interest
for further considerations presents the investigations of physical
consequences of models of field interactions on higher and/or lower
dimensional superspaces provided with N--connection structure.

It worth noticing that higher order derivative theories are one of currently
central division in modern theoretical and mathematical physics. It is
necessary a rigorous formulation of the geometric background for developing
of higher order analytic mechanics and corresponding extensions to
classical quantum field theories. Our results not only contain a
supersymmetric extension of higher order fiber bundle geometry, but also
propose a general approach to the ''physics'' with local anisotropic
interactions. The elaborated in this work formalism of distinguished
vector superbundles highlights a scheme by which supergravitational and
superstring theories with higher order anisotropy can be constructed.

To develop in a straightforward manner self--consistent physical theories,
 define local conservation laws,give a corresponding treatment of geometrical
objects and so on, on different extensions on Finsler spaces with nonlinear
structure of metric form and of connections, torsions and curvatures is a
highly conjectural task. Only the approach on modeling of geometric models of
the mentioned type (super)spaces on vector (super)bundles provided with
compatible nonlinear and distinguished connections and metric structures
make ''visible'' the possibility (see, for instance, \cite
{ma87,ma94,vjmp,vg,vb295,v295a,v295b,vsp96,v96jpa3}), manner of elaboration,
as well common features and differences of models of fundamental physical
fields with generic locally anisotropic interactions. From viewpoint of the
string theory fundamental ideas only some primarily changes in established
material have been introduced in this work. But we did not try to a
simple straightforward repetition of standard material in context of some
sophisticate geometries. Our main purposes were to illustrate that the higher
order  anisotropic supergravity is also naturally contained in the framework
of low energy superstring dynamics and to develop a corresponding geometric
and computational technique for supersymmetric sigma models in locally
anisotropic backgrounds.

The above elaborated methods of perturbative calculus of anomalies of
hetrotic sigma models in higher order anisotropic superspaces, as a matter
of principle, can be used in every finite order on $\alpha ^{\prime }$ (for
instance, by using the decomposition (76) for effective action we can, in a
similar manner as for one- and two--loop calculations presented in section
10, find corrections of anomalies up to fifth order inclusive) and are
compatible with the well known results for locally isotropic strings and
sigma models. We omit such considerations in this work.

\end{document}